\documentclass[12pt]{article} 
\usepackage{jheppub}
\usepackage[dvipsnames]{xcolor}
\usepackage{amsmath,amsfonts,amsbsy,amssymb,array,accents,dsfont}
\usepackage{enumerate}
\usepackage[shortlabels]{enumitem}
\usepackage{graphicx,verbatim} 
\usepackage{caption} 
\usepackage{subcaption} 
\usepackage{mathrsfs}
\usepackage{bm}
\usepackage{slashed}
\usepackage{ytableau}
\usepackage{enumitem}
\setlist[itemize,1]{label=\textbullet}

\usepackage{float}
\usepackage{hyperref}
\usepackage{overpic}
\usepackage[vcentermath]{youngtab}

\allowdisplaybreaks

\newcommand{\bee}{\begin{equation}}
\newcommand{\eee}{\end{equation}}

\usepackage{bbm}

\newcommand{\R}{{\mathbb{R}}}
\newcommand{\Z}{{\mathbb{Z}}}

\newcommand{\ee}{{\mathrm{e}}}
\newcommand{\ii}{{\mathrm{i}}}
\def\tL{\text{\normalfont\tiny{L}}}
\def\tR{\text{\normalfont\tiny{R}}}
\newcommand{\rep}[1]{{\boldsymbol{{#1}}}}
\newcommand{\repconj}[1]{{\overline{\boldsymbol{{#1}}}}}
\newcommand{\bp}{{\boldsymbol{p}}}
\newcommand{\be}{{\boldsymbol{e}}}
\newcommand{\bell}{{\boldsymbol{\ell}}}
\newcommand{\tn}{{\text{n}}}
\newcommand{\tw}{{\text{w}}}
\newcommand{\ket}[1]{{\left|{#1}\right\rangle }}
\newcommand{\NS}{{\texttt{{NS}}}}
\newcommand{\tNS}{{\widetilde{\texttt{{NS}}}}}
\newcommand{\Ram}{{\texttt{{R}}}}
\newcommand{\tRam}{{\widetilde{\texttt{{R}}}}}
\newcommand{\NSRam}{{\texttt{{S}}}}
\def\BPS{\text{\normalfont\tiny{BPS}}}

\title{String probes, simple currents, and the no global symmetries conjecture}

\author{Guglielmo Lockhart, Luca Novelli and Yann Proto}

\affiliation{
Bethe Center for Theoretical Physics, Universit\"at Bonn, D-53115, Germany
}

\emailAdd{glockhar}
\emailAdd{lnovelli}
\emailAdd{yproto@uni-bonn.de}

\abstract{Center one-form symmetries in consistent quantum gravity theories are expected to be either broken or gauged, thereby determining the global form of the gauge group. We shed light on this expectation from the perspective of distinguished extended objects, which we denote by faithful string probes, for which the gauge symmetry is realized as a holomorphic current algebra. We argue that the worldsheet counterpart of gauged center one-form symmetries is the existence of chiral simple currents extending the current algebra. Accordingly, we show that the consistency condition for such extensions reproduce and generalize known field theoretic and geometric obstructions to the gauging of center one-form symmetries in six and eight dimensions. We verify this picture in a number of examples arising from heterotic string compactifications, and apply it to infer the gauge group topology of a recently identified class of six-dimensional models with no known F-theoretic realization. In 6d supergavity, our results also clarify an observation of Kim and Vafa on the existence of BPS particles required for consistency upon circle reduction: these particles arise from worldsheet simple currents whose existence is dictated by the presence of a gauged center one-form symmetry. Intriguingly, in several models across different dimensions we find composite higher-spin chiral currents beyond the current algebra, hinting at a stringy generalization of center one-form symmetries in the bulk.}

\begin{document}

\maketitle

\section{Introduction and basic concepts}
The extended objects of a quantum field theory---that is, the dynamical brane probes that belong to its spectrum of higher-dimensional excitations---are an important tool to study non-perturbative aspects of the theory. This applies in particular to quantum field theories coupled to gravity, where their existence is expected based on the swampland conjectures. In particular, the cobordism conjecture~\cite{McNamara:2019rup} requires them in order to trivialize the cobordism group of quantum gravity, and the charge completeness conjecture~\cite{Polchinski:2003bq,Banks:2010zn} requires the existence of charged, higher-dimensional objects that act as sources for higher-form fields. The existence of brane probes has in turn been employed to test important features of quantum gravity theories, including to perform checks of the weak gravity conjecture~\cite{Lee:2018urn,Lee:2018spm}, obtain new consistency constraints going beyond perturbative anomalies \cite{Kim:2019vuc,Kim:2019ths,Katz:2020ewz,Angelantonj:2020pyr,Tarazi:2021duw,Hamada:2024oap,Kim:2024eoa,Hamada:2025vga}, argue that supergravities in $d>6$ spacetime dimensions can always be obtained by compactification of string theory~\cite{Hamada:2021bbz,Bedroya:2021fbu}, and conversely, to find evidence for the existence of 6d theories with eight supercharges that do not presently admit a known string theory embedding~\cite{Lockhart:2025lea,Lockhart:2026xml}.

The usefulness of extended objects in probing various aspects of quantum field theories can be traced back to the fact that they do not exist in isolation, but rather they interact with other sectors of the theory---including the perturbative sector---in a way which is dictated by the couplings of the bulk theory. This implies, among other things, that gauge symmetries in the bulk theory are realized as global symmetries of the brane probes. This correspondence is particularly stringent when considering two-dimensional probes, where in the infrared it implies the existence of current algebras on the worldsheet associated with bulk gauge symmetries.\\

In this paper we enlarge the dictionary between the symmetries of the bulk theory and the properties of string probes by clarifying how the choice of global form for the bulk gauge group $G$ is encoded on the worldsheet. Closely related to the choice of global form is the center one-form symmetry of the bulk theory, which is associated to the center $\mathcal{Z}(\widetilde{G})$ of the universal cover $\widetilde{G}$ of the gauge group, and in principle is either global, broken, or gauged. If a subgroup $\Gamma$ of the one-form symmetry group is gauged, the bulk theory realizes the gauge symmetry $G =\widetilde{G}/\Gamma$. In order to make contact with worldsheet physics, we introduce the notion of a \emph{faithful string probe} for a gauge symmetry $G$, whose worldsheet is governed by a conformal field theory (CFT) carrying a current algebra for $G$. We will elaborate on this notion in section~\ref{subsec:faithful} below.

When faithful string probes are part of the spectrum of a theory, which is the case for a variety of supergravity theories in different dimensions, the absence of global center one-form symmetries is required for consistency of the string worldsheet CFT. Specifically, if the bulk theory possesses a faithful string probe for its gauge symmetry we point out that the global form of the bulk gauge group is directly related to extensions of the Kac--Moody (KM) algebra on the worldsheet CFT by holomorphic simple current operators $J$. The basic properties of simple currents and holomorphic simple current extensions, with an emphasis on Kac--Moody algebras, are summarized in section~\ref{sec:KMsimplecurrents}. Simple currents are primary operators whose fusion rule with any other primary field $\mathcal{O}_{\rep{r}}(z,\bar{z})$, in a given representation $\rep{r}$, is of the form
\begin{align}
[J] \times [\mathcal{O}_{\rep{r}}] = [\mathcal{O}_{\text{g}(\rep{r})}]~,
\end{align}
where $\text{g}$ denotes a permutation of the primary representations. In the context of Kac-Moody algebras, these holomorphic (left-moving) operators are associated to the elements of the center of $\widetilde{G}$. This technology is applied to faithful string probes in section~\ref{sec:center1formsymmetries} where after reviewing general aspects of center one-form symmetries and anomalies in supergravity, we show that known obstructions to gauging center one-form symmetries in 6d and 8d supergravity theories (as well as geometric constraints for the existence of Mordell--Weil torsion in 8d) can be rephrased as integrality constraints for the conformal weight of simple current operators. There, we also show that modular invariance of the worldsheet CFT, independent of the spacetime dimension, requires either of the following two possibilities to be realized:
\begin{enumerate}
\item the entire spectrum of a faithful string probe is neutral with respect to the center symmetry, in which case the spectrum of the string necessarily includes the simple current operator, and the left-moving (chiral) Kac--Moody algebra is extended by this simple current;
\item  at least one operator in the spectrum of the worldsheet CFT is charged under the center, in which case the simple current cannot belong to the spectrum and the chiral algebra is not extended.
\end{enumerate}
We will interpret these two options as the worldsheet counterpart for the fact that the center one-form symmetry in the bulk is either gauged or broken. In particular, any bulk one-form symmetry which is not broken corresponds to possibility 1; we argue that in this case the simple current extension of the KM algebra, which is required by modularity of the worldsheet CFT, corresponds to gauging of the unbroken center one-form symmetry. It is interesting to note that our findings are not confined to supergravity theories, and in particular extend to any quantum field theory that possesses faithful strings. For instance, we will see that E-strings~\cite{Ganor:1996mu,Seiberg:1996vs,Klemm:1996hh} and M-strings~\cite{Haghighat:2013gba}, which are examples of faithful string probes that can occur in 6d $\mathcal{N}=(1,0)$ SCFTs decoupled from gravity, display properties that reflect the global structure of the 6d gauge and flavour symmetry.\\

In later sections of the paper we employ these ideas to study a number of examples, both old and new. Section~\ref{sec:heteroticcompactifications} is devoted to the study of heterotic string compactifications, including constructions involving the CHL string as well as more general asymmetric orbifolds. In this case the fundamental string itself serves as the faithful probe, allowing us to perform checks of our claims through explicit computations in the heterotic worldsheet CFT. In particular, we will be able to understand a number of known results regarding the global form of gauge groups in $d\geq 8$ supergravities with sixteen supercharges in the language of simple current extensions, and also provide novel details about the global properties of 6d asymmetric orbifold models which have been the object of renewed interest in recent years \cite{Baykara:2023plc,Baykara:2024vss}.

Section \ref{sec:6dBPSstrings}, on the other hand, focuses on $\mathcal{N}=(1,0)$ supergravity models in six dimensions. Our main goal here is to understand the global form of the gauge symmetry in a class of exotic theories without tensor multiplets that were recently studied in~\cite{Lockhart:2025lea,Lockhart:2026xml}. Although these models do not have any known canonical realization within string theory (in particular, they violate the conditions required for a model to admit a large-volume realization within F-theory, and they do not possess any perturbative string in their spectrum of extended operators), they are free from gauge and gravitational anomalies, and they possess faithful string probes whose properties can be analyzed in detail and pass nontrivial consistency checks. If these models are indeed consistent, the expectation is that they can be connected to geometric theories (specifically, the model corresponding to the elliptic fibration over $\mathbb{P}^2$ in F-theory) by Higgsing transitions which are only available at stringy volume of the base $\mathbb{P}^2$ and cannot be reproduced by a classical geometric transition. In spite of the lack of a canonical realization for these models, our methods allow us to determine the global form of the 6d gauge group by identifying the corresponding pattern of simple current extensions, which was one of the main motivations behind this project.

An intriguing observation which applies to these models as well as to other solvable models in various dimensions is that, beside extensions by KM simple currents, the holomorphic chiral algebra of the worldsheet CFT of their faithful string probes also appears to enjoy additional extensions by higher-spin chiral operators, which combine KM currents with additional simple currents that are unrelated to the current algebra. This points to a possible stringy generalization of center one-form symmetries being at play in this class of models, which by the considerations presented in our paper must also be either broken or gauged.

Furthermore, in the context of 6d $\mathcal{N}=(1,0)$ supergravities, our perspective sheds further light on an interesting connection which was drawn in~\cite{Kim:2024tdh} between the gauging of center one-form symmetries and the existence of BPS particles that arise, upon reduction to 5d on a circle, as excitations of wrapped strings. The existence of these particles was posited on the basis that they lead to corrections to the 5d prepotential which would otherwise be inconsistent. These BPS particles turn out to be none other than the string excitations corresponding to the KM simple currents, and our results indeed confirm that the presence of an unbroken center one-form symmetry in the bulk necessarily requires them to be part of the string spectrum.
We present our conclusions in section~\ref{sec:concl} along with a discussion of future directions of research, while the appendices contain more technical aspects of the CHL models (appendix~\ref{app:Narain}) and asymmetric orbifold models (appendix~\ref{app:modularorbits}) considered in the paper. \\

Having outlined the main ideas and results of this paper, let us now discuss in more detail what we mean by faithful string probes and the contexts in which we expect them to be relevant.

\subsection{Faithful string probes}
\label{subsec:faithful}
For definiteness, let us consider a bulk quantum field theory $\mathscr{T}$ in $d\geq 4$ dimensions, possibly coupled to gravity. We assume that the theory satisfies the following
{\begin{center}
\textbf{Key assumptions:}
\end{center}}
\begin{itemize}
\item[--]\emph{Existence of higher-form fields.} The spectrum of $\mathscr{T}$ includes at least one $\mathrm{U}(1)$ two-form field $B_2$; we denote by $\widetilde{B}_{d-4}$  its magnetic dual $(d-4)$-form field.
\item[--]\emph{Brane probes.} The spectrum of branes of $\mathscr{T}$ includes a stable string $\mathscr{S}$ and a dual $(d-5)$-brane $\mathscr{D}$ charged respectively under $B_2$ and $\widetilde{B}_{d-4}$ and carrying minimal charges consistent with Dirac quantization.
\end{itemize}

\noindent
These assumptions are broad: in particular, in any supergravity theory in $d \geq 5$ the gravity multiplet automatically contains a two-form field $B_2$ (or equivalently its electromagnetic dual $\widetilde{B}_{d-4}$). In $d=4$, our first assumption is equivalent to the requirement that the spectrum includes a scalar field $\widetilde{B}_0$. The existence of stable branes charged under $B_2$ and $\widetilde{B}_{d-4}$, on the other hand, follows in supergravity theories from the assumption of completeness of brane charge spectrum.

We say that $\mathscr{S}$ is a faithful string probe with respect to the bulk gauge symmetry $G$, which we take to be (locally) a product of simple and abelian factors $\prod_\alpha G_\alpha$, if the following criteria are met.\footnote{ We stress that we view our notion of faithful string probe as a working definition, and more general families of string probes may play a similar useful role to the ones we consider in this paper.
}
\begin{enumerate}
\item For each $G_\alpha$ the bulk theory possesses a Chern--Simons (CS) coupling
\begin{align}
&b_\alpha \int \widetilde{B}_{d-4}\wedge \text{Tr }F_\alpha\wedge F_\alpha~,&
b_\alpha &>0~,
\label{eq:cplg}
\end{align}
between $\widetilde{B}_{d-4}$ and the field strength for the $G_\alpha$ gauge field.
\item The worldsheet theory of $\mathscr{S}$ is described by a two-dimensional conformal field theory
\bee
\mathscr{T}_{\mathscr{S}} = \mathscr{T}_{\text{com}}\otimes\widetilde{\mathscr{T}}_{\mathscr{S}}~,
\label{eq:comf}
\eee
where $\mathscr{T}_{\text{com}}$ is a decoupled center of mass theory (which is a singular CFT if the spacetime of $\mathscr{T}$ is noncompact), and $\widetilde{\mathscr{T}}_{\mathscr{S}}$ is a compact, unitary, modular invariant CFT. The latter admits a left-moving current algebra for each factor of the gauge group. When $G_\alpha$ is a simple Lie group, we denote by $k_\alpha$ the corresponding level for the current algebra on the worldsheet.
\end{enumerate}

The CS-type couplings~\eqref{eq:cplg} are commonplace both in gravitational and non-gravitational theories. For instance, in perturbative heterotic string compactifications, we can take $B_2$ to be the Kalb--Ramond two-form field; in this case $\widetilde{B}_{d-4}$ couples to every gauge field in the bulk theory, which reflects the fact that the full gauge symmetry is realized as a current algebra on the worldsheet of the heterotic string. In six dimensional $\mathcal{N}=(1,0)$ supergravity and superconformal theories, the presence of couplings~\eqref{eq:cplg} is required to  cancel gauge anomalies via the Green--Schwarz mechanism.  Moreover, in 4d $\mathcal{N}=1$ theories the coupling~\eqref{eq:cplg} is just the familiar coupling between gauge fields to axions. While gravitational theories without such axionic couplings do exist---such as abelian Kaluza–Klein theory~\cite{Heidenreich:2021yda} or compactifications on rigid Calabi–Yau threefolds~\cite{Cecotti:2018ufg}---their presence appears to be a widespread feature of theories with chiral matter.\footnote{ We thank T. Rudelius for helpful comments on this point.}

In the presence of Chern--Simons couplings of the form \eqref{eq:cplg}, gauge invariance of the bulk theory coupled to a string probe~\cite{Freed:1998tg,Harvey:1998bx,Kim:2012wc} is dependent on anomaly inflow---the string worldsheet theory possesses anomalous global symmetries, with the correct 't Hooft anomalies to compensate for anomalous gauge transformations in the bulk. In particular, the most obvious way to cancel anomalies is to require the theory on the string worldsheet to possess a $G_\alpha$ global symmetry with a 't Hooft anomaly contribution to the 2d anomaly polynomial given by
\bee
k_\alpha\operatorname{Tr}(F_\alpha\wedge F_\alpha)~,
\label{eq:anom}
\eee
where $F_\alpha$ is the field strength for the global symmetry. However, it is important to note that more general possibilities are in principle allowed to occur\footnote{See \cite{Hsin:2018vcg} for related remarks.} as a consequence of nontrivial dynamics in contexts where supersymmetry or other considerations do not prevent this. Our definition of faithful string probes explicitly avoids this possibility, but in cases where it may be realized it would be very interesting to understand if a suitable extension of our definition can be found that can lead to insights into the physics of the bulk theory.\\

At this point it is useful to point out some concrete settings in supergravity where faithful string probes play a role. A primary example of a faithful string is provided by the heterotic string. In ten dimensions, its worldsheet theory carries a left-moving $c_\tL=16$ current algebra associated with $\mathrm{E}_8\times\mathrm{E}_8$ or $\mathrm{Spin}(32)/\Z_2$. Upon compactification to lower dimensions, this gives rise to a vast landscape of theories with diverse gauge symmetries, including non-simply laced and non-simply connected gauge groups, all realized on the heterotic worldsheet through current algebras, in some cases at higher level~\cite{Font:1990uw}. In perturbative heterotic compactifications, the correspondence between simple current extensions on the worldsheet and gauged one-form symmetries in spacetime becomes especially transparent, since the spacetime spectrum itself is constructed from the GSO-projected heterotic CFT. From this perspective, our framework does not so much provide a new method for determining the global structure of the gauge group as it reveals a common underlying structure present in all models with non-simply connected non-abelian gauge symmetry. More broadly, faithful string probes arise in settings that are intrinsically non-perturbative, as well as in supergravity models for which no string-theoretic realization is presently known. In such situations, the worldsheet CFT of the probe string can furnish genuinely new information about the topology of the gauge group: the appearance or absence of higher-spin simple currents directly reflects whether the corresponding center one-form symmetries are gauged in the bulk theory. This is relevant in supergravity theories in dimensions $4\leq d \leq 6$, where one encounters a zoo of two-dimensional brane probes in addition to the fundamental strings, which couple to two-form fields that reside in multiplets other than the gravity multiplet. In theories that have a known realization via geometric compactification in M- or F-theory, these classes of strings can be realized by wrapping respectively M5 or D3 branes on suitable holomorphic cycles in the internal geometry, and their low energy behavior is again described by two-dimensional theories with conformal symmetry. Although in this paper we focus exclusively on examples in $d\geq 6$ with at least eight supercharges, it is worth mentioning that faithful string probes (possibly generalized in a suitable way) may be a useful concept in a broader setting. Here we will content ourselves with mentioning a few possibilities. First of all, five-dimensional minimal supergravity theories possess supergravity strings which are described by (0,4) CFTs in the infrared and have been studied in \cite{Katz:2020ewz,Kaufmann:2024gqo}. At a generic point on the Coulomb branch the left-moving sector of the strings CFT supports abelian current algebras corresponding to the Cartan of the gauge group, which enhances to non-abelian KM algebras at special points in the moduli space of the CFT corresponding to gauge symmetry enhancements in the bulk. The properties of string probes have also been investigated in four-dimensional theories with as little as $\mathcal{N}=1$ supersymmetry \cite{Lee:2019tst,Lee:2020gvu,Xu:2020nlh,Grieco:2025bjy,Casas:2025qxz}, and have been found to be in line with expectations from the swampland conjectures. \\

Let us also remark that when some of the levels $k_\alpha$ vanish, the string $\mathscr{S}$ is no longer a faithful probe;
in this case, while the existence of gauged center one-form symmetries for the bulk theory still leads on the string worldsheet to extensions of the current algebras with non-zero levels, the information carried by the string worldsheet is no longer sufficient to uniquely reconstruct exactly which subgroup of the center one-form symmetry is gauged. We will illustrate this point in section~\ref{subsec:faithful6d} with an explicit example.

\subsubsection*{Remarks on conformal symmetry}
Our definition of faithful string probe relies on the crucial assumption that its worldsheet is described by a well-defined conformal field theory. In practice this condition requires some care.\footnote{We are thankful to Kai Xu for helpful comments on this topic.} In the remainder of this section we comment on related subtleties. Let us begin by considering heterotic string theory on a manifold of special holonomy $X$. Compactification is modeled on the worldsheet by introducing a non-linear sigma model with target space $X$. The $\beta$-function of the sigma model depends, among other things, on the metric of the compactification space; requiring it to vanish coincides to choosing a valid vacuum configuration for the bulk fields, and in particular imposing that the internal space metric satisfy the $\alpha'$-corrected Einstein equations. For example, consider heterotic string theory on a K3 manifold, which gives rise to a $\mathcal{N}=(1,0)$ supergravity in 6d. This admits a dual realization as F-theory on an elliptic fibration over a Hirzebruch surface. The heterotic string maps in this frame to a D3 brane wrapping the rational fiber in the Hirzebruch surface. It is described by a (0,4) quantum field theory which is not scale invariant but in the infrared flows to the worldsheet CFT of the heterotic string. More generally, one can consider the infrared conformal field theories associated to 6d BPS strings, which in F-theory settings also arise from D3 branes wrapping curves in the base of an elliptic fibration. This involves some well-known subtleties, which we will illustrate by focusing on the case of BPS strings in 6d $\mathcal{N}=(1,0)$ theories \cite{HaghighatMurthyVafaVandoren2015,Kim:2019vuc}.

The QFT describing a string of given charge possesses in general a complicated moduli space of classical vacua consisting of multiple connected branches. At the quantum level, different branches can become disconnected and are described by distinct conformal field theories, potentially with very different features (including different superconformal R-symmetry and central charges). For instance, branches may describe multi-string configurations where the BPS string splits into different components, or single-string configurations describing a bound state of the constituent strings. Moreover, some branches may correspond to non-compact instanton moduli spaces, while others branches may originate from compact moduli spaces. It is the CFTs of single-string branches of the latter type, which are expected to arise in the infrared description of supergravity strings, which our faithful probe strings are modeled after.  In particular, for these class of strings the bulk gauge symmetries are expected to give rise to left-moving current algebras at integral level.\footnote{This should be contrasted to instanton strings, for which the symmetry acts on bosonic zero-modes and gives rise to a negative-level KM algebra acting on the BPS spectrum \cite{DelZotto:2018tcj}.} Even in this restricted setting, it is still in principle possible that the CFTs associated to this type of branches may develop singular behavior \cite{HaghighatMurthyVafaVandoren2015}, which a priori may invalidate some of the requirements for it to be a faithful string. This is an issue which is certainly worthy of further investigation but is beyond the scope of this paper. On the other hand, there do exist many classes of string probes (of which we will encounter several examples in the text) whose classical moduli space of vacua consists of a unique compact branch. In such cases we can be confident that the infrared CFTs are well behaved, and indeed we will be able to extract highly nontrivial information on the symmetries of the bulk theory from them.

\section{The center symmetries of current algebras}
\label{sec:KMsimplecurrents} 
In this section we review some standard notions about simple currents and chiral algebra extensions in two-dimensional CFTs, with a focus on theories which possess a holomorphic Kac--Moody algebra.

\subsection{Simple currents}
Consider a unitary two-dimensional CFT with central charges $c_\tL$ and $c_\tR$. For the moment, we assume that the theory is rational: its left- and right-moving chiral algebras---consisting of the Virasoro algebra possibly enlarged by additional (anti-)holomorphic fields---admit only a finite number of primary representations. We denote the primary fields by $\phi_i(z,\bar{z})$, where the index $i$ labels the distinct representations, and $z,\bar{z}$ are complex coordinates on the Euclidean worldsheet.
The fusion rules of the CFT encode the conformal families that
appear in the operator product expansion; they take the form
\begin{align}
[\phi_i]\times[\phi_j] = \sum_k \mathcal{N}_{ij}{}^k [\phi_k]~,
\end{align}
where the fusion coefficients $\mathcal{N}_{ij}{}^k$ are non-negative integers. A \emph{simple current} is a primary operator $J$ whose fusion with any other primary yields a single primary. Such an operator acts as a permutation on the set of representations:
\begin{align}
[J]\times [\phi_i] = [\phi_{J(i)}]~,
\end{align}
where, by standard abuse of notation, $J(i)$ denotes the permutation of the primary field labels induced by fusion with $J$. Every CFT contains a trivial simple current, the identity operator $\mathbbm{1}$. 
The presence of a non-trivial simple current $J$ has several direct consequences. Let us summarize the main features; additional details may be found in~\cite{Schellekens:1989am,Schellekens:1990xy}. First, there exists a primary operator $J^{\text{c}}$, called the conjugate of $J$, such that
\begin{align}
[J]\times [J^{\text{c}}] = [\mathbbm{1}]~.
\label{eq:conjugatecurrent}
\end{align}
The fusion $[J]\times [J]$ of the simple current $J$ with itself produces a new simple current, distinct from $J$.\footnote{The primary field generated by $[J]\times[J]$ must be a simple current due to the associativity of the fusion rules. It is distinct from $J$ since a fusion product of the form $[J]\times [J] = [J]$ would be incompatible with~\eqref{eq:conjugatecurrent}.} By iterating this process, one obtains a sequence
\begin{align}
[J]\times [J^\ell] = [J^{\ell+1}]~,
\end{align}
with $J^1=J$ and $J^0=\mathbbm{1}$. Because the number of primary fields is finite, there exists a smallest positive integer $n$ such that
\begin{align}
J^n = \mathbbm{1}~.
\end{align}
This integer $n$ is called the order of the simple current. The set $\{J^1,\dots,J^n\}$, equipped with the fusion product, is isomorphic to the cyclic group $\Z_n$. Evidently, the conjugate simple currents are given by $(J^\ell)^{\text{c}} = J^{n-\ell}$. 
More generally, the collection of all simple currents of a rational CFT forms a finite abelian group
\begin{align}
\mathcal{Z} \simeq \Z_{n_1}\times \dots\times \Z_{n_d}~,
\end{align}
often referred to as the center of the CFT. Choosing a set of generators $J_1,\dots,J_d$ with respective orders $n_1,\dots,n_d$, any simple current can be uniquely expressed as
\begin{align}
J =(J_1)^{\ell_1}\dots (J_d)^{\ell_d}~,
\end{align}
where $(\ell_1,\dots,\ell_d) \in (\Z/n_1\Z)\times\dots\times(\Z/n_d\Z)$. Fusion of simple currents corresponds to the group multiplication in the CFT center $\mathcal{Z}$, while current conjugation corresponds to taking the inverse.\\

The preceding discussion applies equally well to the chiral sector of a CFT. Restricting attention to the left-moving algebra generated by the stress tensor $T(z)$ and possibly additional holomorphic fields, one can define simple currents with respect to this chiral algebra alone. In this context, the primary fields $\phi_i$ correspond to representations of the left-moving algebra, independently of the right-moving sector. In particular, the notion of simple currents remains meaningful even for non rational CFTs, whenever a chiral subalgebra possesses a finite set of representations. This situation naturally arises when the CFT has a decomposable energy-momentum tensor
\begin{align}
T = T_1+T_2~,
\label{eq:decomposableCFT}
\end{align}
with $T_1$ and $T_2$ generating mutually commuting Virasoro algebras. One may then focus on the chiral algebra associated with $T_1$ (possibly enhanced by additional holomorphic fields that are neutral with respect to $T_2$). If this subalgebra admits a finite set of primary representations, a well-defined notion of simple currents follows, and the full CFT inherits the associated algebraic properties. Such structure occurs in CFTs with a holomorphic current algebra, where the decomposition~\eqref{eq:decomposableCFT} arises from the Sugawara construction.

\subsection{Extensions by chiral currents}
\label{subsec:simplecurrentextension}
We now restrict our attention to CFTs possessing purely left-moving simple currents, and examine the structure they induce on the chiral algebra. We consider the theory on a genus-one worldsheet with complex structure parameter $\tau$. Let us denote the left-moving primary fields by $\phi_i$, with conformal weights $h_i$. The multiplicities of non-null descendants in the conformal family $[\phi_i]$ are encoded in the character $\chi_i(\tau)$, which transforms under $\mathrm{SL}(2,\mathbb{Z})$ as
\begin{align}
\chi_i(\tau+1) &= \ee^{2\pi\ii (h_i-c_\tL/24)} \, \chi_i(\tau)~,&
\chi_i(-1/\tau) &= \sum_j \mathcal{S}_{ij} \, \chi_j(\tau)~,
\end{align}
where $\mathcal{S}$ is the modular matrix. This matrix is unitary and symmetric, and its square implements charge conjugation. When the chiral algebra admits a non-trivial simple current $J$, the modular matrix satisfies additional relations. Specifically, for primaries belonging to the same orbit under the action of $J$, one has~\cite{Schellekens:1989dq,Intriligator:1989zw}
\begin{align}
\mathcal{S}_{J(i)j} = \ee^{2\pi\ii Q(J)_j} \mathcal{S}_{ij}~.
\label{eq:modularmatrixorbit}
\end{align}
Here, $Q(J)_i$ denotes the monodromy charge of the primary $\phi_i$ with respect to $J$, defined by
\begin{align}
Q(J)_i = h_{J}+h_{i}-h_{J(i)} \quad\text{mod }1~,
\end{align}
which captures the monodromy of the OPE of $\phi_i$ with $J$.\\

If the CFT admits a purely left-moving simple current $J$ with integer conformal weight $h_J$, then $J$ can be interpreted as a local operator, since it carries integer spin. When $J$ appears in the spectrum, the CFT is said to be extended by $J$, and the holomorphic chiral algebra can be enlarged accordingly. Although the resulting extended algebra may be intricate, its primary representations are readily determined from their decomposition in terms of the primaries of the original
chiral algebra. Consequently, the CFT spectrum must organize into representations of the extended algebra.

More precisely, we say that the CFT is \emph{extended by the simple current} $J$ when the following conditions hold:
\begin{itemize}
\item[\emph{i}.] the simple current $J$ is present in the spectrum;
\item[\emph{ii}.] all primaries appearing in the spectrum have trivial monodromy charge with respect to $J$;
\item[\emph{iii}.] primaries belonging to the same orbit under the action of $J$ appear in the spectrum with equal multiplicity.
\end{itemize}
These conditions are not independent; in a modular invariant CFT they are, in fact, equivalent. To see this, consider for simplicity a rational theory in which the left- and right-moving primary representations are labeled by $i$ and $\bar{\imath}$, respectively. The genus $1$ partition function $Z$ can be expressed as
\begin{align}
Z(\tau,\bar{\tau}) = \sum_{i,\bar{\imath}} M_{i\bar{\imath}} \,\chi_{i}(\tau) \overline{\chi}_{\bar{\imath}}(\bar{\tau})~,
\label{eq:CFTpartitionfunction}
\end{align}
and the spectrum of the CFT is specified by the multiplicities $M_{i\bar{\imath}}$, which are non-negative integers. In particular, the identity operator $\mathbbm{1}$, labeled by $0$, satisfies $M_{0\bar{0}}=1$. The above conditions translate into
\begin{subequations}
\renewcommand{\theequation}{\theparentequation\emph{\roman{equation}}}
\begin{align}
M_{J(0)\bar{0}}>0~,\\
\ee^{2\pi\ii Q(J)_i}M_{i\bar{\imath}}=M_{i\bar{\imath}}~,
\label{eq:trivialmonodromy}
\\
M_{J(i)\bar{\imath}}=M_{i\bar{\imath}}~.
\label{eq:orbitmultiplicity}
\end{align}
\end{subequations}
Assuming condition (\emph{i}), the holomorphic field $J$ appears (at least once) in the spectrum. Single-valuedness of correlation functions then requires every local operator $\mathcal{O}_{i\bar{\imath}}$ to be local with respect to $J$. Their OPE with $J$ exhibits the monodromy
\begin{align}
\mathcal{O}_{i\bar{\imath}}(\ee^{2\pi\ii}z,\ee^{-2\pi\ii}\bar{z}) J(0) = \ee^{-2\pi\ii Q(J)_i} \mathcal{O}_{i\bar{\imath}}(z,\bar{z})J(0)~.
\end{align}
Absence of branch cuts therefore requires $Q(J)_i=0$, establishing condition (\emph{ii}). In turn, using the property~\eqref{eq:modularmatrixorbit} of the modular matrix together with modular invariance of the partition function, condition~\eqref{eq:trivialmonodromy} implies~\eqref{eq:orbitmultiplicity}, so that operators related by the action of $J$ appear with identical multiplicities. Since the holomorphic current $J$ lies in the same orbit as the identity $\mathbbm{1}$, specializing~\eqref{eq:orbitmultiplicity} to $i=\bar{\imath}=0$ yields
\begin{align}
M_{J(0)\bar{0}} = 1~,
\end{align}
showing that the spectrum contains a single copy of the current $J$.\\

This structure is elegantly reflected in the modular properties of characters. Let $I$ denote a set of representatives for the orbits of $J$ with trivial monodromy charge.\footnote{In other words, $I$ is a maximal subset of left-moving primaries such that $Q(J)_i=0$ for all $i\in I$, and no two elements of $I$ belong to the same $J$-orbit.} For each orbit, define the extended character
\begin{align} 
\mathcal{X}_i(\tau) = \sum_{j\in\{i,J(i),\dots\}} \chi_j(\tau)~. 
\end{align}
Because of property~\eqref{eq:modularmatrixorbit}, these extended characters furnish a representation of $\mathrm{SL}(2,\mathbb{Z})$: they satisfy
\begin{align}
\mathcal{X}_i(\tau+1) &= \ee^{2\pi\ii (h_i-c_\tL/24)} \, \mathcal{X}_i(\tau)~,&
\mathcal{X}_i(-1/\tau) &= \sum_{j\in I} \widetilde{\mathcal{S}}_{ij} \, \mathcal{X}_j(\tau)~.
\end{align}
The first relation reflects that primaries within the same orbit and with trivial monodromy charge share the same conformal weight modulo $1$. The matrix $\widetilde{\mathcal{S}}$ appearing in the second relation is determined from the modular matrix of the unextended theory. If the CFT is extended by the integral spin simple current $J$, then its partition function~\eqref{eq:CFTpartitionfunction} can be written as
\begin{align}
Z(\tau,\bar{\tau}) = \sum_{i\in I,\bar{\imath}} M_{i\bar{\imath}}  \,\mathcal{X}_{i}(\tau) \overline{\chi}_{\bar{\imath}}(\bar{\tau})~,
\end{align}
in terms of extended characters.

Although we have restricted the discussion to rational CFTs for ease of presentation, the construction extends to decomposable theories whose left-moving chiral algebra factorizes into two subalgebras, one of which possessing finitely many integrable representations. If the spectrum contains a left-moving simple current $J$ of this rational subalgebra, the entire spectrum organizes into representations of the corresponding extended algebra: only operators local with respect to $J$ are allowed, and operators within the same orbit under the action of $J$ appear with identical multiplicities.

\subsection{Kac--Moody algebras}
\label{subsec:KMcenter}
Having reviewed the general CFT structures associated to simple currents, we now specialize to unitary theories with a Kac--Moody algebra. We first consider a left-moving current algebra based on a simple Lie algebra $\mathfrak{g}$. By the Sugawara construction, the stress tensor decomposes as $T=T_{\mathfrak{g}}+T_{\text{res}}$, where $T_{\mathfrak{g}}$, constructed from bilinears in the KM currents, has central charge
\begin{align}
c_\mathfrak{g} = \frac{k\operatorname{dim}\mathfrak{g}}{k+h^\vee_\mathfrak{g}}~.
\end{align}
Here, $h^\vee_\mathfrak{g}$ is the dual Coxeter number of $\mathfrak{g}$, and $k$ denotes the KM level, which must be a strictly positive integer in a unitary CFT.

The current algebra admits a finite set of integrable highest weight representations, labeled by irreducible representations of $G$, the simply connected compact Lie group associated with $\mathfrak{g}$. Let $\lambda_\rep{r}$ be the highest weight of the representation $\rep{r}$, and let $\gamma_\mathfrak{g}$ denote the highest root of $\mathfrak{g}$. Then the primary representations of the $\mathfrak{g}$ current algebra at level $k$ are characterized by the condition
\begin{align}
\lambda_\rep{r}\cdot\gamma_\mathfrak{g} \leq k~.
\end{align}
Here, $\cdot$ denotes the Killing form, normalized so that long roots have length
squared $2$. Since the CFT is decomposable, the left-moving conformal dimensions of operators can be written as $h_\tL = h_\rep{r}+h_\tL^{\text{res}}$. The Sugawara contribution is given by
\begin{align}
h_\rep{r} = \frac{\lambda_\rep{r}\cdot(\lambda_\rep{r}+2\rho_{\mathfrak{g}})}{2(k+h^\vee)}~,
\end{align}
with $\rho_{\mathfrak{g}}$ the Weyl vector of the Lie algebra---the half-sum of all positive roots. The residual contribution $h_\tL^{\text{res}}$ is non-negative by unitarity.\\

Simple currents of KM algebras were classified in~\cite{Fuchs:1990wb}. They are closely related to the center symmetries of $G$,  the simply connected compact Lie group whose Lie algebra is $\mathfrak{g}$. More precisely, there is a one-to-one correspondence\footnote{There is a single exception: the $\mathfrak{e}_8$ current algebra at level $2$, which admits a non-trivial simple current even though $\mathrm{E}_8$ has trivial center. This subtlety will not affect the present discussion, whose focus is on the realization of center symmetries in the current algebra. However, the resulting $\mathbb{Z}_2$ symmetry will reappear in section \ref{sec:heteroticcompactifications} in examples arising from the CHL string.}
\begin{align*}
\{\text{simple currents of the $\mathfrak{g}$ current algebra}\}\;\longleftrightarrow\;\{\text{central elements of $G$}\}~.
\end{align*}
To make this correspondence explicit, we introduce some notation. Let $\mathcal{Z}(G)$ denote the center of $G$. This is a finite abelian group, which is cyclic provided $G$ is distinct from $\mathrm{Spin}(4l)$; we assume this for simplicity. We label its elements by $\ell \in \{0,1,\dots,n-1\}$, where $n$ is the order of $\mathcal{Z}(G)$. By Schur’s lemma, in any irreducible representation $\rep{r}$ of the group $G$, a central element $\ell$ must be represented as
\begin{align}
\exp({2\pi\ii\, Q_\rep{r}^{\ell}} )\,\mathrm{id}_\rep{r}~,
\end{align}
where the charges $Q_\rep{r}^{\ell}$ take values in $\tfrac{1}{n}\Z$. Correspondingly, for any central element $\ell$ one finds a simple current $J_{\mathfrak{g}}^\ell$ among the integrable representations of the KM algebra. These currents act on primary fields $\phi_\rep{r}$ via fusion as
\begin{align}
[J_{\mathfrak{g}}^{\ell}] \times [\phi_\rep{r}] &= [\phi_{\rep{r}'}]~,
\end{align}
where $\rep{r}'=\text{g}(\rep{r})$ is related to $\rep{r}$ by an outer automorphism $\text{g}$ of the extended Dynkin diagram of the Lie algebra $\mathfrak{g}$---see~\cite{DiFrancesco:1997nk} for additional details. Importantly, the monodromy charge of the KM primary field $\phi_\rep{r}$ with respect to $J_{\mathfrak{g}}^\ell$ is given by
\begin{align}
Q(J_{\mathfrak{g}}^\ell)_\rep{r} = Q_\rep{r}^{\ell }~,
\label{eq:monodromycharge=centralcharge}
\end{align}
and thus coincides with the charge of the representation under the center symmetry.\\

Let us illustrate these considerations with an example. Consider the $\mathfrak{a}_r$ current algebra at level $k$. It admits $\binom{k+r}{k}$ primary representations, labeled by highest weight representations $\rep{r}$ of $\mathfrak{su}(r+1)$, with Dynkin labels
\begin{align}
\rep{r} = [l_1l_2\cdots l_r]
\end{align}
subject to the constraint $l_1+\dots+l_r\leq k$. The center of the $\mathfrak{su}(r+1)$ current algebra---like that of $\mathrm{SU}(r+1)$---is isomorphic to $\Z_{r+1}$. It is generated by a simple current $J_{\mathfrak{a}_r}$ associated with the representation
\begin{align}
\rep{r} &= [k\,0\cdots0]
\nonumber\\
&=\underbrace{\,\footnotesize\yng(2)\cdots \yng(2)\,}_{k}~,
\label{eq:q2}
\end{align}
whose dimension is $\operatorname{dim}\rep{r}=\binom{k+r}{r}$. Its conformal weight is equal to
\begin{align}
h(J_{\mathfrak{a}_r}) &= \frac{k r}{2(r+1)}~.
\end{align}
The fusion of $J_{\mathfrak{a}_r}$ with a generic primary field acts by a cyclic permutation of the affine Dynkin labels: we have
\begin{align}
[J_{\mathfrak{a}_r}]\times[\phi_{[l_1l_2\cdots l_r]}] = [\phi_{[l_0l_1\cdots l_{r-1}]}]~,
\end{align}
where $l_0 = k-l_1-\dots -l_r$ is the affine label, associated with the affine node of the extended Dynkin diagram. In particular, the full set of non-trivial simple currents $J_{\mathfrak{a}_r}^\ell$, with $\ell=1,\dots,r$, corresponds to representations of the form $\rep{r}=[0\cdots0\,k\,0\cdots0]$, where the entry $k$ appears in the $\ell$-th position.\\

For any simple Lie algebra $\mathfrak{g}$ associated with a simply connected Lie group $G$ with non-trivial center, the conformal dimensions of simple currents for the corresponding $\mathfrak{g}$ current algebra at level $k$ are given by\footnote{For even $r$, the $\mathfrak{d}_r$ current algebra---like $\mathrm{Spin}(2r)$---has center $\Z_2\times\Z_2$. Accordingly, its simple currents are labeled by $\ell_1,\ell_2 \in \{0,1\}$. The preceding discussion extends straightforwardly to this case, with the obvious modifications, such as monodromy charges $Q^{\ell_1,\ell_2}_\rep{r}$.}
\begin{align}
\mathfrak{a}_r &= \mathfrak{su}(r+1)~:&
h(J_{\mathfrak{a}_r}^{\ell}) &= \frac{k \ell(r+1-\ell)}{2(r+1)}~,~\ell=1,\dots,r~,
\nonumber\\
\mathfrak{b}_r &= \mathfrak{spin}(2r+1)~:&
h(J_{\mathfrak{b}_r}) &= \frac{k }{2}~,
\nonumber\\
\mathfrak{c}_r &= \mathfrak{sp}(r)~:&
h(J_{\mathfrak{c}_r}) &= \frac{k r}{4}~,
\nonumber\\
\mathfrak{d}_r &= \mathfrak{spin}(2r)~,~ r\text{ even}~:&
h(J_{\mathfrak{d}_r}^{1,0}) &=h(J_{\mathfrak{d}_r}^{0,1})= \frac{k r}{8}~,\quad h(J_{\mathfrak{d}_r}^{1,1})= \frac{k }{2}~,
\nonumber\\
\mathfrak{d}_r &= \mathfrak{spin}(2r)~,~ r\text{ odd}~:&
h(J_{\mathfrak{d}_r}) &=h(J_{\mathfrak{d}_r}^3)= \frac{k r}{8}~,\quad h(J_{\mathfrak{d}_r}^2)= \frac{k }{2}~,
\nonumber\\
\mathfrak{e}_6 &~:&
h(J_{\mathfrak{e}_6}) &=h(J_{\mathfrak{e}_6}^2)= \frac{2k}{3}~,
\nonumber\\
\mathfrak{e}_7 &~:&
h(J_{\mathfrak{e}_7}) &= \frac{3k}{4}~.
\label{eq:KMconformaldimensions}
\end{align}

The generalization to the semi-simple case is straightforward. Consider a left-moving current algebra based on a Lie algebra
\begin{align}
\mathfrak{g} = \mathfrak{g}_1 + \dots + \mathfrak{g}_N~,
\end{align}
written as a direct sum of simple factors, with corresponding KM levels $k_1,\dots,k_N$. Integrable representations of the current algebra are then labeled by highest weight representations $\rep{r} = (\rep{r}_1,\dots,\rep{r}_N)$ of the simply-connected Lie group $G=G_1\times\dots\times G_N$. The center of $G$ is a finite abelian group of the form
\begin{align}
\mathcal{Z}(G) \simeq \Z_{n_1} \times \dots \times \Z_{n_d}~.
\end{align}
Each central element $\ell=(\ell_1,\dots,\ell_d)$ gives rise to a simple current $J^\ell_{\mathfrak{g}}$. Its action on the primary representation $\rep{r}$ factorizes across the simple components. The associated monodromy charge is equal to the charge $Q^\ell_\rep{r}$ of the representation $\rep{r}$ under the center symmetry.

\subsubsection*{Enhanced $\mathfrak{u}(1)$ currents}
Many features of center symmetries in KM algebras admit a natural analogue for $\mathfrak{u}(1)$ current algebras, provided the latter are suitably enhanced by higher-spin chiral fields. Let us consider a CFT with a left-moving $\mathfrak{u}(1)$ current algebra generated by a holomorphic spin-$1$ operator $\mathcal{J}(z)$. A local operator $\mathcal{O}(z,\bar{z})$ is a primary of charge $Q$ if
\begin{align}
\mathcal{J}(z)\mathcal{O}(0)\simeq\frac{Q \,\mathcal{O}(0)}{z}~.
\end{align}
The spectrum organizes into $\mathfrak{u}(1)$ primaries and their descendants. The current algebra is characterized by
\begin{align}
\mathcal{J}(z)\mathcal{J}(0) \simeq \frac{r}{z^2}~,
\end{align}
where $r>0$ in a unitary theory. In general, this constant $r$ is not physical---it can be set to one by a rescaling of the current. However, it becomes meaningful when the CFT has an integer charge spectrum. More precisely, if the ratio of charges of any two charged operators is rational, one can canonically normalize $\mathcal{J}$ so that all charges are integral and primitive. In this normalization, $r$ is fixed and in fact integer, as follows, for instance, from spectral flow arguments~\cite{Melnikov:2019tpl}.

Upon bosonizing the current as $\mathcal{J}=\ii\sqrt{r}\partial H$, the chiral scalar $H$ may be used to construct vertex operators. In particular, the holomorphic chiral fields
\begin{align}
\Omega^+ &= :\ee^{\ii\sqrt{r}H}:~,&
\Omega^- &= :\ee^{-\ii\sqrt{r}H}:~,
\end{align}
are mutually local with all operators in the theory. Their presence signals an extension of the left-moving chiral algebra.\footnote{Such enhancements frequently arise in (2,2) and (0,2) SCFTs, where $\mathcal{J}$ is the R-symmetry current of the N=2 superconformal algebra. We thank I. Melnikov for enlightening discussions on this point.} These operators carry charges $Q=\pm r$ and conformal weight $h_\tL=r/2$. We restrict to even $r$, so that the extended left-moving algebra is bosonic. Writing
\begin{align}
k = \frac{r}{2} ~,
\end{align}
we interpret $k$ as the level of the enhanced $\mathfrak{u}(1)$ algebra. The latter admits a finite set of representations labeled by $s\in\{0,1,\dots,2k-1\}$, with conformal weights
\begin{align}
h_s =
\begin{cases}
\frac{s^2}{4k} & s=0,\dots,k-1~, \\
\frac{(s-2k)^2}{4k} & s=k,\dots,2k-1~.
\end{cases}
\label{eq:u(1)conformalweights}
\end{align}
Each conformal family contains infinitely many $\mathfrak{u}(1)$ primary representations with charges $Q = s\text{ mod }2k$, and is encoded in the character
\begin{align}
\chi^{\mathfrak{u}(1)}_s(\tau) = \frac{1}{\eta(\tau)}\sum_{l\in\Z} q^{k\left(l+\frac{s}{2k}\right)^2}~.
\end{align}
The fusion product is
\begin{align}
[s]\times [s'] = [s+s'\text{ mod }2k]~.
\end{align}
Every extended primary representation is a simple current of the enhanced $\mathfrak{u}(1)$ algebra---the center of the extended algebra is isomorphic to $\Z_{2k}$. The monodromy charge of a primary with respect to the simple current labeled by $s$ reproduces its charge under the $\mathrm{U}(1)$ transformation $\ee^{-2\pi\ii\frac{s}{2k}}$. Equivalently, the simple currents of the extended $\mathfrak{u}(1)$ current algebra are in one-to-one correspondence with the discrete subgroup
\begin{align}
\Z_{2k}\subset\mathrm{U}(1)~.
\end{align}

\section{String probes and center one-form symmetries}
\label{sec:center1formsymmetries}
This section contains the main conceptual claims of the paper, which in later sections will be applied to explicit examples in various dimensions. We begin in section~\ref{subsec:centeranomalies} by reviewing the consistency conditions for gauge groups with nontrivial topology, which can be obstructed by mixed anomalies involving the local two-form fields of the theory. In section~\ref{subsec:simplecurrentanomalycorrespondence} we explain how the global structure of the gauge group is encoded in the worldsheet CFT of a faithful string through extensions of the current algebra by chiral operators of integral spin, and relate the corresponding integrality conditions on their conformal dimensions to obstructions to gauging the bulk one-form symmetry. In section~\ref{subsec:noglobalcenter}, we apply the general discussion of simple current extensions from section~\ref{subsec:simplecurrentextension} to the CFT of a faithful probe string, and show that the theory either admits a simple current extension or contains states that explicitly break the center symmetry. We argue that these two possibilities correspond in the bulk to the center one-form symmetry being respectively gauged or broken. We also comment on the existence of faithful strings in non-gravitational six-dimensional theories, as well as on generalizations involving abelian gauge factors. Finally, in section~\ref{subsec:MWtorsion} we review the relation between center one-form symmetries and Mordell--Weil torsion in F-theory: specifically, we show that in eight dimensions the consistency conditions for simple current extensions on the probe string reduce to the integrality conditions satisfied by the Mordell--Weil group of elliptically fibered K3 surfaces when the gauge group has maximal rank, and generalize them to lower rank models involving frozen singularities.

\subsection{Review of center one-form symmetries and anomalies}
\label{subsec:centeranomalies}
Gauge theories are specified not only by their local degrees of freedom, but also by global data, foremost among which is the global structure of the gauge group. Even in pure Yang--Mills theory, distinct theories may share the same gauge algebra while differing in their spectrum of extended operators. To make this distinction precise, we begin with a compact simple Lie group $G$; extensions to semi-simple groups and the inclusion of abelian factors will be discussed later. The group $G$ need not be simply connected. Writing $\widetilde{G}$ for its universal cover, it may be described as a quotient
\begin{align}
G = \widetilde{G}/\Gamma~,
\end{align}
where  $\Gamma\subset\mathcal{Z}(\widetilde{G})$ is a subgroup of the center, and hence $\pi_1(G)\simeq \Gamma$.

Consider a quantum field theory with gauge algebra $\mathfrak{g}$ in which all matter fields transform in representations of $G$. At the level of local observables, the choice between $G$ and its universal cover $\widetilde{G}$ is invisible: the two theories share the same Lagrangian description and correlation functions. The difference lies in the spectrum of electric and magnetic probes of the two theories. Wilson line operators are labeled by representations of $G$, so only those representations of $\widetilde{G}$ that are neutral under the subgroup $\Gamma$ are allowed. On the other hand, passing to a non-simply connected gauge group enlarges the spectrum of admissible `t Hooft line operators relative to the theory with gauge group $\widetilde{G}$. A non-trivial global structure of a gauge group $G$ can be understood as arising from gauging a subgroup $\Gamma \subset \mathcal{Z}(\widetilde{G})$, viewed as a one-form center symmetry of the theory with simply connected gauge group~\cite{Kapustin:2014gua,Gaiotto:2014kfa}. Such one-form symmetries act on extended operators---namely Wilson lines---and may be coupled to background two-form gauge fields; gauging them amounts to summing over these backgrounds and produces the theory with gauge group $G=\widetilde{G}/\Gamma$.

In a consistent theory of quantum gravity  global symmetries are widely believed to be absent. This applies to discrete and higher form symmetries as well as ordinary continuous zero-form symmetries \cite{Banks:2010zn, Harlow:2018tng,McNamara:2019rup}. Accordingly, if a theory that couples to gravity contains gauge fields for a non-abelian group and its spectrum is neutral under a subgroup of the center symmetry, the one-form symmetry that corresponds to the center should be gauged, implying that the correct global form of the gauge group is non-simply connected. This picture is further reinforced by the completeness hypothesis~\cite{Polchinski:2003bq}, which asserts that any gauge theory coupled to quantum gravity must contain objects transforming in every finite-dimensional irreducible representation of the gauge group---a spectrum satisfying this condition is said to be complete. Our aim will be to provide additional corroboration for these ideas by studying how the presence of center one-form symmetries is reflected in the worldsheet CFTs of faithful probe strings, which appear naturally in various classes of quantum field theories coupled to gravity.

\subsubsection*{Instanton fractionalization}
The global form of the gauge group determines the topological sectors which are summed over in the path integral, corresponding to discrete choices of gauge bundle topology. This structure is particularly transparent in four-dimensional gauge theories. For a simply connected group $G$, the topology of a principal $G$-bundle $\mathcal{E}$ over a Euclidean four-manifold $M$ is characterized by a single invariant, the instanton number,
\begin{align}
\nu(\mathcal{E}) = \frac{1}{16\pi^2 h^\vee_{\mathfrak{g}}}\int_M \operatorname{Tr}_{\text{\textbf{adj}}}(F\wedge F)~,
\end{align}
which is integer-valued when $M$ is compact and without boundary.\footnote{The normalization is such that a minimal ``one-instanton'' configuration has $\nu=1$. The trace is taken in the adjoint representation of $G$ and $h^\vee_{\mathfrak{g}}$ denotes the dual Coxeter number. In the rest of the text we use the shorthand $\operatorname{Tr}=\operatorname{Tr}_{\text{\textbf{adj}}}/(16\pi^2 h^\vee_{\mathfrak{g}})$.} When $G$ is not simply connected, additional topological sectors appear, corresponding to $G$-bundles that do not lift to bundles of the universal cover $\widetilde{G}$. The obstruction to such a lift is measured by characteristic classes, often referred to as generalized Stiefel--Whitney classes. For a simple group with $\pi_1(G)\simeq \mathbb{Z}_n$ cyclic, this obstruction is encoded in a class
\begin{align}
w_2\in H^2(M,\Z_n)~,
\end{align}
which must vanish for the bundle to lift. In the case of $\widetilde{G}=\mathrm{Spin}(4 l)$, whose center is $\Z_2\times\Z_2$, one instead has two classes
\begin{align}
w_2^{(1)},w_2^{(2)}\in H^2(M,\Z_2)~.
\end{align}
Bundles with non-trivial Stiefel--Whitney classes can carry fractional instanton number. More precisely, the instanton number satisfies
\begin{align}
\nu(\mathcal{E}) = \mathcal{\alpha}_{\mathfrak{g}} \int_M \mathfrak{P}[w_2(\mathcal{E})] \text{ mod }1~,
\label{eq:fractionalinstanton}
\end{align}
where $\mathfrak{P}$ denotes the Pontryagin square, mapping $H^2(M,\Z_n)$ to $H^4(M,\Z_{2n})$ for even $n$ and to $H^4(M,\Z_n)$ for odd $n$. The coefficient $\alpha_{\mathfrak{g}}$, determined in~\cite{Aharony:2013hda,Cordova:2019uob}, is a rational number that quantifies the fractionalization of the instanton number. Its values for compact simple Lie groups with non-trivial center are collected in table~\ref{tab:mixedanomalycoefficients}.
\begin{table}[h!]
\centering
\def\arraystretch{1.2}
\begin{tabular}{c|ccccccc}
$\mathfrak{g}$ & $\mathfrak{a}_r$ & $\mathfrak{b}_r$ & $\mathfrak{c}_r$ & $\mathfrak{d}_r$ & $\mathfrak{d}_r$ & $\mathfrak{e}_6$ & $\mathfrak{e}_7$ \\[-8pt]
&&&&  \scriptsize ($r$ even) & \scriptsize ($r$ odd)\\ \hline
$\alpha_{\mathfrak{g}}$ & \large $\frac{r}{2(r+1)}$ & \large $\frac{1}{2}$ & \large $\frac{r}{4}$ & \large $(\frac{r}{8},\frac{1}{2})$ & \large $\frac{r}{8}$ & \large $\frac{2}{3}$ & \large $\frac{3}{4}$ \\
\end{tabular}
\caption{Values of the fractional instanton coefficients $\alpha_{\mathfrak{g}}$}
\label{tab:mixedanomalycoefficients}
\end{table}
Let us remark that, for $\mathrm{Spin}(2r)$ with even $r$, two independent coefficients appear, $\alpha_{\mathfrak{d}_r}=r/8$ and $\alpha'_{\mathfrak{d}_r}=1/2$, associated with the $\Z_2\times\Z_2$ center. In this case, the relation~\eqref{eq:fractionalinstanton} is replaced by
\begin{align}
\nu = \mathcal{\alpha}_{\mathfrak{d}_r} \int_M \mathfrak{P}[w_2^{(1)}+w_2^{(2)}] + \mathcal{\alpha}_{\mathfrak{d}_r}' \int_M w_2^{(1)} \cup w_2^{(2)} \text{ mod }1~,
\end{align}
where $\cup$ denotes the cup product. Fractional instantons play an important role in four-dimensional gauge theories with non-simply connected gauge group, where they modify, in particular, the periodicity of the $\theta$-angle~\cite{Aharony:2013hda}.

\subsubsection*{Gauged center one-form symmetries and anomalies}
The fractionalization of instanton number for non-simply connected gauge groups can obstruct the gauging of center one-form symmetries. This effect arises in theories containing two-form fields that transform non-trivially under the gauge symmetry, and has been studied in particular in minimal higher-dimensional supergravity~\cite{Apruzzi:2020zot,Cvetic:2020kuw,BenettiGenolini:2020doj}. We briefly review the mechanism, following~\cite{Apruzzi:2020zot}. In a theory with gauge group $G$ whose spectrum is invariant under a subgroup of center symmetries, passing to the theory with non-simply connected gauge group amounts to gauging this one-form symmetry. Concretely, this requires introducing a background field $C=w_2$ for the center symmetry and summing over its possible configurations. The obstruction arises because the theory may fail to be well defined in a non-trivial $C$ background.

In supergravity theories with sixteen supercharges, the massless spectrum contains a local two-form field $B$, which transforms non-trivially under gauge transformations. As a result, its field strength $H$ satisfies a Bianchi identity of the form
\begin{align}
\text{d} H = b\operatorname{Tr}(F\wedge F) +\dots~,
\end{align}
where $b$ is an integer coefficient, and we ignored possible additional contributions involving the spacetime curvature. Equivalently, one may dualize $B$ to a local $(d-4)$-form field, which couples through
\begin{align}
b \int \widetilde{B}\wedge\operatorname{Tr}(F\wedge F)~.
\end{align}
The field $\widetilde{B}$ is only locally defined, and its descriptions on different patches are related by higher-form gauge transformations. Among these are large gauge transformations
\begin{align}
\widetilde{B} \quad\to \quad\widetilde{B} + \widetilde{B}_{\text{flat}} ~,
\label{eq:flatshifts}
\end{align}
where $\widetilde{B}_{\text{flat}}$ is a closed $(d-4)$-form with integer periods. Although flat, such a form may still be topologically non-trivial. In the absence of a background for the center one-form symmetry, these transformations shift the action by an integer and therefore leave the partition function invariant.

The situation changes in the presence of a non-trivial background $w_2=C$. Under the transformation~\eqref{eq:flatshifts}, the action shifts by
\begin{align}
b\,\mathcal{\alpha}_{\mathfrak{g}} \int \widetilde{B}_{\text{flat}}\,\cup \mathfrak{P}[C] \quad\text{mod }1~.
\label{eq:mixedanomaly}
\end{align}
When this quantity is not integral, the partition function acquires a non-trivial phase and becomes ambiguous. This signals a mixed anomaly between the center one-form symmetry and the large gauge transformations of the dual $(d-4)$-form field. Since the latter are gauge redundancies, the former cannot be consistently gauged whenever the phase encoded in~\eqref{eq:mixedanomaly} is non-trivial.

These obstructions were analyzed in detail in~\cite{Cvetic:2020kuw} for minimal supergravity theories in eight dimensions. Such theories preserve sixteen supercharges and are strongly constrained by quantum consistency, while string constructions provide a large class of explicit realizations---including models with non-simply laced and non-simply connected gauge groups---where these conditions can be tested~\cite{Font:2021uyw,Cvetic:2021sjm}. In this setting, the dual field is a local four-form $\widetilde{B}$, which couples to the Yang--Mills field strengths as
\begin{align}
\sum_{\alpha=1}^N b_\alpha \int \widetilde{B}\wedge\operatorname{Tr}(F_\alpha\wedge F_\alpha)~,
\end{align}
where $\alpha=1,\dots,N$ labels the simple factors $G_\alpha\subset G$ of the non-abelian gauge group. The integers $b_\alpha$ coincide with the levels of the corresponding current algebras in perturbative string realizations, where the fundamental string provides a faithful probe carrying primitive charge. For a center element $(\ell_1,\dots,\ell_N)\in\mathcal{Z}(\widetilde{G}_1)\times\dots\times\mathcal{Z}(\widetilde{G}_N)$ of the simply connected cover of the gauge group, the corresponding one-form symmetry is anomaly-free precisely when~\cite{Cvetic:2020kuw}
\begin{align}
\sum_{\alpha=1}^N b_\alpha \ell_\alpha^2 \alpha_{\mathfrak{g}_\alpha} \quad\in\Z~.
\label{eq:8dmixedanomaly}
\end{align}
Here we assume for simplicity that all simple factors have cyclic center. When one of the factors is $\mathrm{Spin}(2r)$ with even $r$, its $\Z_2\times\Z_2$ center symmetries, labeled by $(\ell_1,\ell_2)$, contribute instead
\begin{align}
b_{\mathfrak{d}_r}\left((\ell_1+\ell_2)^2\alpha_{\mathfrak{d}_r}+\ell_1\ell_2\alpha_{\mathfrak{d}_r}'\right)
\label{eq:mixedanomalySpin}
\end{align}
to the mixed anomaly.

\subsubsection*{Charge quantization for coupled strings}
For an anomalous center one-form symmetry, the fractional value of the anomaly~\eqref{eq:mixedanomaly} admits an equivalent interpretation in terms of charge quantization for objects electrically charged under the local two-form fields~\cite{Hsieh:2020jpj,Apruzzi:2020zot}. Let us illustrate this in six-dimensional $\mathcal{N}=(1,0)$ supergravity. Such a theory contains one self-dual two-form field from the gravity multiplet and $T$ anti-self-dual two-form fields from the tensor multiplets, which we collectively denote by $B^\mu$, with $\mu=0,1,\dots,T$. Their couplings to the Yang--Mills sector are completely determined by anomaly cancellation. In a theory free of perturbative anomalies, the eight-form anomaly polynomial factorizes as
\begin{align}
I_8 = \Omega_{\mu\nu}X_4^\mu X_4^\nu~,
\end{align}
where $\Omega_{\mu\nu}$ is the pairing of signature $(1,T)$ on the charge lattice. The four-forms $X_4$ are given by
\begin{align}
X_4^\mu = \frac{1}{2}a^\mu \operatorname{tr}(R\wedge R) + \sum_{\alpha=1}^N \frac{2}{\lambda_\alpha} b^\mu_\alpha \operatorname{tr}(F_\alpha\wedge F_\alpha)~,
\label{eq:6danomalyX}
\end{align}
in terms of the anomaly coefficients $a$ and $b_\alpha$. Here, the trace is taken in the fundamental representation, while the constants $\lambda_\alpha$ are chosen so that an minimal embedded $\mathrm{SU}(2)$ instanton has unit topological charge. Once this factorization holds, the anomaly is canceled through the generalized Green--Schwarz mechanism. The action then contains the coupling
\begin{align}
\sum_{\alpha=1}^N \Omega_{\mu\nu}\,b_\alpha^\mu \int B^\nu\wedge\operatorname{Tr}(F_\alpha\wedge F_\alpha)~,
\end{align}
which is precisely the type of term responsible for the mixed anomaly of center one-form symmetries discussed above.

Six-dimensional supergravity theories admit non-critical strings, which act as sources for the fields $B^\mu$. A string of charge $Q^\mu$, whose worldsheet spans a surface $\Sigma\subset M$, couples through
\begin{align}
\Omega_{\mu\nu}\, Q^\mu\int_\Sigma B^\nu~.
\label{eq:6dstringcoupling}
\end{align}
Even when no local counterterm exists to cancel the mixed anomaly~\eqref{eq:mixedanomaly}, the anomalous phase in the partition function associated to the one-form symmetry may still be compensated by the contribution of such probe strings. However, this is possible only if the corresponding coupling does not violate charge quantization. Consider first a theory invariant under the full center symmetry of a simple simply connected gauge group $G$. Turning on a background configuration $C$ for the center symmetry induces the anomalous phase discussed above. Canceling it by the string coupling~\eqref{eq:6dstringcoupling} is consistent with charge quantization only when~\cite{Apruzzi:2020zot}
\begin{align}
\alpha_\mathfrak{g}\,Q\cdot b \quad\in \Z~,
\end{align}
where we use the shorthand $Q\cdot b = \Omega_{\mu\nu} Q^\mu b^\nu$. For a semi-simple gauge group, and for a center symmetry labeled by $(\ell_1,\dots,\ell_N)$, this condition generalizes to
\begin{align}
\sum_{\alpha=1}^N \alpha_{\mathfrak{g}_\alpha}\ell_\alpha^2\,Q\cdot b_\alpha  \quad\in \Z~.
\label{eq:6dDiracquantization}
\end{align}
Here we again assume, for simplicity of notation, that each simple factor has cyclic center. When one factor is $\mathrm{Spin}(2r)$ with even $r$, whose center is $\Z_2\times\Z_2$, the corresponding condition is modified exactly as in~\eqref{eq:mixedanomalySpin}, with the replacement $k_{\mathfrak{d}_r}\to Q\cdot b_{\mathfrak{d}_r}$.

\subsection{Holomorphic simple currents and gauge group topology}
\label{subsec:simplecurrentanomalycorrespondence}

In this section we consider faithful string probes that couple to a non-abelian gauge group $G$, and show that the obstructions for gauging the center one-form symmetry in the bulk exactly match with the conditions for the chiral algebra to possess holomorphic simple currents of integer conformal weight. On the worldsheet CFT of faithful string probes the gauge symmetry is realized as a left-moving current algebra. As a result, all worldsheet states organize into representations of the affine algebra, and hence into irreducible representations of the corresponding Lie algebra. If the gauge group is written as $G=\widetilde{G}/\Gamma$, with $\widetilde{G}$ the simply connected cover and $\Gamma\subset\mathcal{Z}(\widetilde{G})$ a subgroup of its center, then only representations neutral under the center symmetry $\Gamma$ are allowed in the physical spectrum. The current algebra already severely constrains which representations of $\widetilde{G}$ may appear: only finitely many can arise as primary representations, depending on the KM level. Requiring that only the $\Gamma$-neutral subset actually appears in the CFT spectrum is non-trivial. As reviewed in section~\ref{subsec:KMcenter}, each element of the center symmetry determines a simple current of the affine algebra, which we denote by
\begin{align}
J_{\mathfrak{g}}(z)~.
\end{align}
Its monodromy charge with a primary field in a representation $\rep{r}$ of $\widetilde{G}$ coincides with the charge of $\rep{r}$ under the center symmetry. Thus, if the bulk gauge group is $G=\widetilde{G}/\Gamma$, every local operator on the worldsheet must be neutral under $\Gamma$, and therefore must be local with respect to the corresponding simple current. For such a current to belong to the spectrum of a consistent modular-invariant CFT, it must define a local operator, and hence has integer spin. Since
\begin{align}
h_\tL &= h(J_{\mathfrak{g}})~,&
h_\tR &=0~,
\end{align}
this requires the KM conformal weight $h(J_{\mathfrak{g}})$---listed in~\eqref{eq:KMconformaldimensions} for simple Lie algebras---to be an integer. Importantly, this quantity depends only on the levels of the simple factors of the gauge algebra and on group-theoretic data. It therefore provides a direct constraint on the possible global form of the gauge group: if the simple current associated with a putative quotient $G=\widetilde{G}/\Gamma$ has non-integer conformal weight, then that global form is incompatible with the existence of faithful probe strings.

Consider for example a gauge theory with algebra $\mathfrak{c}_r=\mathfrak{sp}(r)$, and suppose that the theory admits a faithful probe string whose worldsheet realizes this symmetry as a $\mathfrak{c}_r$ KM algebra at level $k$. We ask whether the gauge group can consistently be
\begin{align}
G = \mathrm{Sp}(r)/\Z_2~.
\end{align}
The $\mathfrak{c}_r$ KM algebra at level $k$ has a $\Z_2$ symmetry generated by a simple current $J_{\mathfrak{c}_r}$. Its monodromy charges coincide with the charges under the $\Z_2$ center of $\mathrm{Sp}(r)$. If the bulk theory is defined with gauge group $\mathrm{Sp}(r)/\Z_2$, all worldsheet states must be neutral under this symmetry, and hence local with respect to $J_{\mathfrak{c}_r}$. The conformal weight of the simple current is
\begin{align}
h(J_{\mathfrak{c}_r}) = \frac{k r}{4}~.
\end{align}
This quantity must be integral for $J_{\mathfrak{c}_r}$ to define a local operator of the CFT. Therefore, once the level $k$ for the faithful probe string is fixed, the allowed global form of the gauge group is constrained. For instance, if the gauge $\mathfrak{sp}(r)$ symmetry is realized at level $k=1$, the non-simply connected form is allowed only when $r$ is a multiple of 4.

Similar constraints on the allowed global form of the gauge group were derived in 8d supergravity theories with sixteen supercharges from the absence of mixed anomalies for center one-form symmetries~\cite{Cvetic:2020kuw}. In particular, the inconsistency of certain $\mathrm{Sp}(r)/\Z_2$ theories found in~\cite{Cvetic:2020kuw} leads to exactly the same condition on the rank. This agreement in fact extends to arbitrary semi-simple gauge. For a simple Lie algebra $\mathfrak{g}$, the relation takes the simple form
\begin{align}
h(J_{\mathfrak{g}}) = k \,\alpha_{\mathfrak{g}}~.
\label{eq:simplecurrentanomalycorrespondence}
\end{align}
Here, $k$ is the level of the worldsheet current algebra, $J_{\mathfrak{g}}$ is the simple current generating the center symmetry, and $\alpha_{\mathfrak{g}}$ is precisely the fractional instanton coefficient introduced in the previous section and listed in table~\ref{tab:mixedanomalycoefficients}. The condition that the simple current has integer conformal weight is therefore equivalent to the absence of a mixed anomaly for the corresponding center one-form symmetry. From the spacetime perspective, a non-integral value of $k\,\alpha_{\mathfrak{g}}$ signals a mixed anomaly between the center one-form symmetry and large gauge transformations of higher-form fields, obstructing the gauging of the one-form symmetry. From the worldsheet perspective, the same non-integrality means that the operator $J_{\mathfrak{g}}$ has fractional spin, and therefore cannot consistently extend the left-moving current algebra. The obstruction to gauging the center symmetry is thus encoded directly in the chiral algebra of faithful probe strings.

For groups with non-cyclic center, the story is completely analogous. In particular, for the $\mathfrak{d}_r$ algebra with even rank, the center is $\Z_2\times\Z_2$, and the current algebra admits simple currents labeled by $\ell_1,\ell_2\in\{0,1\}$. Their conformal weights satisfy
\begin{align}
h(J^{\ell_1,\ell_2}_{\mathfrak{d}_r}) = k\left((\ell_1+\ell_2)^2 \alpha_{\mathfrak{d}_r}+\ell_1\ell_2\alpha_{\mathfrak{d}_r}'\right)\text{ mod 1}~,
\end{align}
in terms of the two coefficients $\alpha_{\mathfrak{d}_r}$ and $\alpha_{\mathfrak{d}_r}'$ given in table~\ref{tab:mixedanomalycoefficients}. The correspondence also extends immediately to semi-simple gauge groups. For $\mathfrak{g}= \mathfrak{g}_1+\dots+\mathfrak{g}_N$, and a center symmetry labeled by $(\ell_1,\dots,\ell_N)$, one finds
\begin{align}
h(J^\ell_{\mathfrak{g}}) = \sum_{\alpha=1}^N k_\alpha\ell_\alpha^2 \alpha_{\mathfrak{g}_\alpha}\text{ mod 1}~,
\label{eq:conformaldimensionJsemisimple}
\end{align}
again with the appropriate modifications for factors with $\Z_2\times\Z_2$ center.\\

Faithful probe strings therefore provide a criterion for the existence of obstructions to gauging center one-form symmetries which is purely based on the existence of a consistent probe string CFT, independent of spacetime dimension. Applied to eight-dimensional supergravity theories, it reproduces the anomaly cancellation conditions derived in~\cite{Cvetic:2020kuw}. In the context of six-dimensional $\mathcal{N}=(1,0)$ supergravity, it gives a worldsheet interpretation for the integrality constraints of~\cite{Apruzzi:2020zot} that were reviewed in section \ref{subsec:centeranomalies}. Indeed, the low-energy worldsheet theory of a faithful probe string of charge $Q$ coupled to the bulk theory through~\eqref{eq:6dstringcoupling} contains a left-moving current algebra for each simple factor $\mathfrak{g}_\alpha$, with level given by~\cite{Kim:2019vuc}
\begin{align}
k_\alpha = Q\cdot b_\alpha~,
\end{align}
The condition~\eqref{eq:6dDiracquantization} for the gauged one-form symmetry to be compatible with charge quantization is then exactly equivalent to the condition that the corresponding simple current extension of the KM algebra be consistent on the worldsheet.

\subsection{Faithful strings and the absence of global center one-form symmetries}
\label{subsec:noglobalcenter}
We now argue that the absence of obstructions to gauging a center one-form symmetry corresponds automatically to a simple current extension of the current algebra on the worldsheet of faithful probes. We analyze the implications for quantum gravity theories as well as certain non-gravitational theories, where we argue that the extension signifies that the corresponding center one-form symmetry is gauged. We also describe how this picture can be modified to allow for additional abelian gauge factors.

Suppose the bulk theory has a non-abelian gauge algebra $\mathfrak{g}$. On the worldsheet theory of a faithful probe string, this gauge symmetry is realized as a left-moving current algebra for $\mathfrak{g}$, so the CFT states organize into a finite set of integrable representations $\rep{r}$ of the affine algebra. By the Sugawara construction, the left-moving stress tensor decomposes as $T = T_{\mathfrak{g}}+T_{\text{res}}$, where $T_{\mathfrak{g}}$ is the KM contribution and $T_{\text{res}}$ (possibly enlarged by additional integer-spin currents) generates the residual left-moving algebra. Local operators are therefore labeled by a representation $\rep{r}$ of the current algebra, together with labels $i$ and $\bar{\imath}$ for representations of the residual left-moving algebra and of the right-moving algebra, respectively. The genus-one partition function takes the form
\begin{align}
Z_{\text{probe}}(\tau,\bar{\tau}) = \sum_{\rep{r},i,\bar{\imath}} M_{\rep{r}i\bar{\imath}}\, \chi^{\mathfrak{g}}_{\rep{r}}(\tau)\,\chi_i^{\text{res}}(\tau) \,\bar{\chi}_{\bar{\imath}}(\bar{\tau})~,
\end{align}
where the multiplicities $M_{\rep{r}i\bar{\imath}}$ are non-negative integers. The identity operator appears uniquely, so that $M_{\rep{1}0\bar{0}}=1$. Modular invariance imposes strong constraints on these multiplicities; in particular,
\begin{align}
M_{\rep{r}i\bar{\imath}} = \sum_{\rep{s},j,\bar{\jmath}}\mathcal{S}^{\mathfrak{g}}_{\rep{r}\rep{s}}\,\mathcal{S}^{\text{res}}_{ij}\,\bar{\mathcal{S}}_{\bar{\imath}\bar{\jmath}} \,M_{\rep{s}j\bar{\jmath}}~,
\label{eq:probestringmodularinvariance}
\end{align}
where $\mathcal{S}^{\mathfrak{g}}$, $\mathcal{S}^{\text{res}}$ and $\bar{\mathcal{S}}$ are modular matrices of the left- and right-moving chiral algebras. Revisiting the general discussion of simple current extensions from section~\ref{subsec:simplecurrentextension}, we now derive the consequences of modular invariance in the presence of a center one-form symmetry.

Consider a center symmetry of the gauge group, characterized by charges $Q_{\rep{r}}$ assigned to representations of $\mathfrak{g}$. The probe string is invariant under this symmetry precisely when only neutral representations appear, namely when
\begin{align}
\ee^{2\pi\ii Q_{\rep{r}}}M_{\rep{r}i\bar{\imath}} = M_{\rep{r}i\bar{\imath}}~.
\end{align}
As reviewed in section~\ref{subsec:KMcenter}, every such center symmetry is associated with a simple current of the affine algebra, corresponding to a distinguished primary representation that we denote by $\rep{r}_{\mathrm{g}}$. Its fusion action on KM primaries takes the form
\begin{align}
[\rep{r}_{\text{g}}] \times [\rep{s} ] = [\text{g}(\rep{s})]~,
\end{align}
where $\text{g}\,:\,\rep{r}\mapsto\text{g}(\rep{r})$ is a certain permutation of the KM primary representations, induced by an outer automorphism of the extended Dynkin diagram~\cite{DiFrancesco:1997nk}. In particular, the simple current representation itself is $\rep{r}_{\text{g}}=\text{g}(\rep{1})$. The modular matrix of the current algebra satisfies
\begin{align}
\mathcal{S}^{\mathfrak{g}}_{\text{g}(\rep{r})\rep{s}} = \ee^{2\pi\ii Q_{\rep{s}}} \mathcal{S}^{\mathfrak{g}}_{\rep{r}\rep{s}}~.
\end{align}
This relation, combined with modular invariance~\eqref{eq:probestringmodularinvariance}, implies the following constraint on the CFT spectrum:
\begin{align}
M_{\text{g}(\rep{r})i\bar{\imath}} = M_{\rep{r}i\bar{\imath}}~.
\label{eq:outerautomorphismspectrum}
\end{align}
Thus every local operator transforming in the representation $\rep{r}$ must be accompanied by another operator in the representation $\mathrm{g}(\rep{r})$, so that the spectrum organizes into orbits of the outer automorphism $\mathrm{g}$. Applied to the vacuum, \eqref{eq:outerautomorphismspectrum} implies in particular that the simple current itself must appear in the spectrum with multiplicity one. It is therefore realized as a purely left-moving state, transforming in the representation $\rep{r}_{\mathrm{g}}$ of the current algebra and trivially under the residual left-moving algebra.
To summarize, in a modular invariant CFT with a left-moving $\mathfrak{g}$ current algebra, there are two possibilities for a given center symmetry:
\begin{enumerate}
\item Every local operator is neutral under the center symmetry. Then the corresponding simple current is local with respect to the full operator algebra, and the left-moving chiral algebra is necessarily extended by it. The spectrum organizes into orbits of the associated outer automorphism, and in particular the simple current itself appears as a purely left-moving state of integral spin, with multiplicity one.
\item At least one local operator carries non-trivial charge under the center symmetry. In this case, the associated simple current cannot belong to the spectrum, since it would have non-trivial monodromy with that operator. In particular, this must occur whenever the conformal weight of the simple current is non-integral.
\end{enumerate}

When the CFT describes a faithful probe, the spacetime interpretation is that the corresponding center one-form symmetry is either broken or gauged. In case 1, the presence of the simple current as a holomorphic operator on the worldsheet reflects the fact that the one-form symmetry has been gauged. In case 2, the symmetry is explicitly broken by string excitations, which correspond to extended charged operators in the bulk theory. Several pieces of evidence support this picture. As discussed in section~\ref{subsec:simplecurrentanomalycorrespondence}, the consistency conditions for a simple current extension of the probe string CFT match precisely the conditions for gauging the corresponding center one-form symmetry, in supergravity settings where the latter are understood. Moreover, the presence of the simple current often has direct physical consequences in the bulk theory.
For example, as we will discuss in section~\ref{subsec:5dreduction}, in six-dimensional supergravity the existence of such an extension for a faithful BPS string constrains the Kähler cone of the five-dimensional theory obtained after circle reduction, and is required for consistency with positivity constraints~\cite{Kim:2024tdh} for the five-dimensional prepotential.\\

We can understand more explicitly the link between bulk one-form symmetries and simple current extensions on the worldsheet by considering the following setup. Consider a configuration where a $(d-5)$-brane of minimal charge dual to a faithful string is infinitely extended along $d-4$ directions in spacetime. Let us focus on the remaining four dimensions $M_4$, in which we suppose a string is supported on a 2-dimensional manifold $\Sigma$. The CS coupling \eqref{eq:cplg}
between the $(d-4)$-form field and the field strength for a gauge group $G_\alpha$ reduces to a four-dimensional axionic coupling
\begin{align}
&b_\alpha \int_{M_4} \phi\, \operatorname{Tr}(F_\alpha\wedge F_\alpha)~.
\label{eq:cplg2}
\end{align}  
Consider a path $\gamma$ which winds once around $\Sigma$ in $M_4$. Along this path, the field $\phi$ shifts as
\bee
\phi \to \phi + 2\pi \ii~,
\eee
which is crucial in the familiar derivation of Dirac quantization from single-valuedness of correlation functions involving both electric and magnetic sources. Correspondingly the coupling \eqref{eq:cplg2} shifts by a term
\begin{align}
&2\pi \ii \,b_\alpha \int_{M_4}\text{Tr }(F_\alpha\wedge F_\alpha)~,
\label{eq:shift}
\end{align}  
which signals the presence of an interface in $M_4$ which is bounded by $\Sigma$. Indeed, a faithful string probe supports a KM algebra for $G_\alpha$, which extends to a level $b_\alpha$ Chern--Simons theory on the interface and produces the required shift~\eqref{eq:shift}.\footnote{As emphasized in \cite{Hsin:2018vcg}, this is not the only logical possibility: depending on the dynamics of the theory different choices of TQFT may in principle be realized on interfaces, corresponding to different 2d RCFTs on the boundary, provided they possess the correct 't Hooft anomalies. In our definition of faithful string probes we assume that is not the case.} Although the bulk theory is dynamical, the existence of an unbroken center one-form symmetry is associated to the existence of Gukov--Witten (GW) surface operators $\mathscr{U}_\ell$ for $\ell\in\mathcal{Z}(G)$ which are topological, and in particular can be moved across interfaces as in \cite{Hsin:2018vcg}. An operator $\mathscr{U}_\ell$ that intersects the interface induces a Wilson line operator $\mathscr{W}_\ell$ in the Chern--Simons theory that is nilpotent and serves as a topological GW operator for the center one-form symmetry of the CS theory. Conversely, any such operator in the CS theory can be used to define a corresponding GW operator in the bulk theory.
These operators $\mathscr{W}_\ell$ are in turn well known to be in one-to-one correspondence with simple current operators of the boundary Kac--Moody algebra, and extending the 2d theory by simple currents is equivalent to gauging the center one-form symmetry of the interface \cite{Moore:1988ss,Moore:1989yh,Hsin:2018vcg}, which is implemented by projecting onto the spectrum of invariant line operators and identifying lines related by the action of the topological operators $\mathscr{W}_\ell$. We take this as strong evidence for the fact that the center one-form symmetry in the bulk, which is generated by the closely related operators $\mathscr{U}_\ell$, is also gauged accordingly. It would be very interesting to understand in more detail the conditions under which this claim can be rigorously justified. This would require an extension of the arguments of Section 3.2 of \cite{Hsin:2018vcg} to interfaces in dynamical theories and is beyond the scope of the present paper. In the various examples in $d\geq 6$ which we consider in the coming sections we do remark, however, that the occurrence of KM simple current extensions is in perfect agreement with the gauging of center one-form symmetries in the bulk. Whether situations can arise in lower dimensional theories where the one-form symmetry realized in the bulk deviates from the one predicted by faithful string probes is a very interesting question which we leave for future research. 

We emphasize that these results do not explicitly require the bulk theory to be coupled to gravity, but rather they are implied by the presence of faithful string probes. This is a common ingredient of supergravity theories, for which the existence of a compact CFT on the worldsheet of strings is in line with the fact that the spectrum of excitations of brane probes should be discrete~\cite{Hamada:2021bbz,Bedroya:2021fbu}, based on black hole entropy considerations.\\

Conversely, the BPS strings that arise in non-gravitational theories are generally associated to non-compact moduli spaces, which correspond to the moduli of instanton, monopole, or vortex configurations in dimensions 6, 5, or 4, respectively. Nevertheless it is worth to point out that, in cases where the non-gravitational theory does possess faithful probes, they do lead to nontrivial predictions regarding the global properties of the bulk gauge symmetry. Let us focus for definiteness on the case of six-dimensional superconformal theories and restrict our attention to strings charged with respect to a single two-form field. The Dirac self-pairing ot the strings determines the possible couplings of the two-form field to 6d gauge and matter fields. In all but two instances, the two-form field is paired to a non-trivial gauge symmetry \cite{Morrison:2012np,Heckman:2015bfa}; when this is the case the corresponding BPS strings carry instanton charge and are described by non-compact worldsheet CFTs, so they do not give rise to faithful probes. The two exceptions are the so-called E-strings and M-strings. In both cases, the CFT describing a string of minimal charge arises purely from left-moving degrees of freedom. These degrees of freedom give rise respectively to an $\mathfrak{su}(2)$ and $\mathfrak{e}_8$ level $1$ Kac--Moody algebra on the worldsheet, which may be either gauged or ungauged in the 6d theory. In the latter case we are, strictly speaking, considering faithful strings for 6d gauge and flavor symmetries, which is a slight generalization of our definition of section~\ref{subsec:faithful}. Nevertheless our worldsheet analysis of the associated center one-form symmetries still applies.\\

In the case of M-strings, the $\mathfrak{su}(2)$ current algebra at level $1$ does not possess simple current operators of integer weight, and the spectrum of the theory includes matter in the fundamental representation of $\mathrm{SU}(2)$ which confirms that the center one-form symmetry is broken. More interesting is the case of the E-strings. Depending on the way the two-form field associated to the E-string couples to other two-form fields in the SCFT, the $\mathrm{E}_8$ symmetry may be fully realized or may be broken to a maximal subalgebra. When $\mathrm{E}_8$ is unbroken, the center is trivial and correspondingly there is no simple current to extend by. The situation is different when $\mathfrak{e}_8$ is broken to a maximal subalgebra $\mathfrak{g}_1+\mathfrak{g}_2$. Let us for definiteness consider a specific example, where
\bee
\mathfrak{e}_8\;\to\;\mathfrak{su}(3)+\mathfrak{e}_6~.
\label{eq:maximalsub}
\eee
The E-string is a faithful probe for the $\mathrm{SU}(3)\times\mathrm{E}_6$ symmetry, of which each factor may be either gauged or global. The worldsheet theory admits an $\mathfrak{su}(3)+\mathfrak{e}_6$ current algebra, with both factors at level $1$. The two KM algebras possess $\mathbb{Z}_3$ simple currents, respectively of conformal dimension $1/3$ and $2/3$, which correspond to the primaries associated to, say, the $\repconj{3}$ and $\rep{27}$ representations. Their product appears in the spectrum and has conformal weight $1$. It therefore yields a simple current extension of the KM algebra, so the global form of the gauge group in 6d is in fact given by \cite{Dierigl:2020myk,Apruzzi:2020zot}
\bee
\left(\mathrm{SU}(3)\times\mathrm{E}_6\right)/\Z_3~.
\eee
 This is consistent with the fact that the Lie group $\mathrm{E}_8$ contains $\left(\mathrm{SU}(3)\times\mathrm{E}_6\right)/\Z_3$ as a subgroup, rather than the simply connected form. In this example, however, there is more happening at the level of the worldsheet: as the simple current has conformal dimension $1$, it also leads to an enlargement of the current algebra on the string, which gets enhanced back to the full $\mathfrak{e}_8$. Identical considerations apply for other breaking patterns to non-simply connected groups \cite{Dierigl:2020myk}, for instance:
\begin{align}
\mathrm{E}_8 &\supset \left(\mathrm{SU}(2)\times\mathrm{E}_7\right)/\Z_2~,&
\mathrm{E}_8 &\supset \left(\mathrm{Spin}(8)\times\mathrm{Spin}(8)\right)/(\Z_2\times\Z_2)~.
\end{align}
On the worldsheet, these correspond to conformal embeddings of the $\mathfrak{su}(2)+\mathfrak{e}_7$ and $\mathfrak{so}(8)+\mathfrak{so}(8)$ KM algebras at level one inside the $\mathfrak{e}_8$ current algebra, reflected in the following identities for the level-one characters:
\begin{align}
\chi^{\mathfrak{e}_8}_\rep{1} &= \chi^{\mathfrak{a}_1}_\rep{1}\chi^{\mathfrak{e}_7}_\rep{1} + \chi^{\mathfrak{a}_1}_\rep{2}\chi^{\mathfrak{e}_7}_\rep{56}~,&
\chi^{\mathfrak{e}_8}_\rep{1} &= \chi^{\mathfrak{d}_4}_\rep{1} \chi^{\mathfrak{d}_4}_\rep{1} + \chi^{\mathfrak{d}_4}_\rep{8^v}\chi^{\mathfrak{d}_4}_\rep{8^v} + \chi^{\mathfrak{d}_4}_\rep{8^s}\chi^{\mathfrak{d}_4}_\rep{8^s} + \chi^{\mathfrak{d}_4}_\rep{8^c}\chi^{\mathfrak{d}_4}_\rep{8^c}~.
\end{align} 

\subsubsection*{Abelian factors}
We can straightforwardly extend the previous discussion to gauge theories whose gauge group contains abelian factors. On the worldsheet of a faithful probe string, such factors are realized by a left-moving $\mathfrak{u}(1)$ current algebra. As reviewed in section~\ref{subsec:KMcenter}, its representation theory differs significantly from that of the semi-simple case: in particular, there is in principle an infinite family of primary representations, labeled by their $\mathrm{U}(1)$ charge $Q$. 

To analyze one-form symmetries, it is useful to distinguish two situations, depending on the charge spectrum. Suppose first that the worldsheet theory contains two charged operators with charges $Q_1$ and $Q_2$ such that the ratio $Q_1/Q_2$ is irrational. In that case, no non-trivial subgroup of $\mathrm{U}(1)$ acts trivially on the full CFT spectrum. In the bulk, all center one-form symmetry associated with the abelian factor are broken. The second possibility is that the ratio of charges of any two charged operators is always rational. Then the current can be normalized so that all charges are integral. In this case, as described in section~\ref{subsec:KMcenter}, the $\mathfrak{u}(1)$ current algebra admits an extension by holomorphic spin-$k$ operators, and the resulting enhanced algebra has only finitely many primary representations, labeled by $s\in\Z_{2k}$. We refer to this extended theory as the $\mathfrak{u}(1)$ current algebra at level $k$.

Each of these $2k$ primary representations is itself a simple current. Just as in the non-abelian case, these simple currents encode center symmetries: their monodromy charges measure the charge under a $\Z_{2k}\subset \mathrm{U}(1)$ subgroup.
One may then ask which of the simple currents can consistently extend the chiral algebra. Consider a subgroup
\begin{align}
\Z_n\subset\Z_{2k}
\end{align}
generated by the primary with label $s=2k/n$. For this operator to appear as a purely left-moving local operator in a modular invariant CFT, it must have integer spin. Using the conformal weight formula~\eqref{eq:u(1)conformalweights} for extended $\mathfrak{u}(1)$ primaries, this requires
\begin{align}
k = \widetilde{k} n^2~,
\end{align}
for some integer $\widetilde{k}$. In that case, the simple current has spin $\widetilde{k}$, and extending the algebra by this operator simply produces the $\mathfrak{u}(1)$ current algebra at level $\widetilde{k}$. From the spacetime point of view, this is not a particularly dramatic effect: it corresponds precisely to the situation in which all states carry charge divisible by $n$, so that a $\mathbb{Z}_n\subset \mathrm{U}(1)$ subgroup acts trivially. Gauging the corresponding one-form symmetry leads to
\begin{align}
\mathrm{U}(1)/\Z_n \simeq \mathrm{U}(1)~,
\end{align}
which amounts only to a redefinition of the fundamental unit of charge. Nevertheless, it fits naturally into the same general picture: the presence of the simple current on the worldsheet reflects the gauging of the corresponding center symmetry in spacetime.

A more interesting situation arises when the probe string current algebra contains both a semi-simple component $\mathfrak{g}$ and a $\mathfrak{u}(1)$ factor. When the abelian charge lattice is integral, one can construct simple currents that combine a $\Z_n$ center symmetry of the non-abelian sector with a discrete $\Z_n\subset\mathrm{U}(1)$ subgroup. If the resulting combined current has integer conformal weight, it can consistently appear as a local operator of the probe string CFT. The corresponding simple current extension then admits a natural spacetime interpretation: it realizes the gauging of a one-form symmetry, leading to a gauge group with global structure
\begin{align}
\left( G\times \mathrm{U}(1) \right) /\Z_n~.
\end{align}

\subsection{Relation to Mordell--Weil torsion}
\label{subsec:MWtorsion}
In this section we connect our results to geometry in the case of theories with non-simply connected gauge group that arise as compactifications of F-theory. On the geometric side, the global structure of the gauge group is encoded in the torsion part of the Mordell--Weil group, as we review following~\cite{Cvetic:2017epq,Cvetic:2018bni}. In a background with nontrivial Mordell--Weil torsion, faithful strings---such as those obtained by wrapping D3-branes on curves in the base---exhibit corresponding simple current extensions on their worldsheet theories. We will indeed see in the case of 8d supergravities with maximal gauge group rank the integrality conditions on the conformal dimensions of simple currents imply the known constraints for the existence of Mordell-Weil torsion for elliptic K3s.

In F-theory one considers compactifications on an elliptically fibered Calabi--Yau manifold
\begin{equation}
\pi\;:\;X\;\longrightarrow\;B
\label{eq:FT-ell-fib}
\end{equation}
over a K\"ahler base $B$, equipped with a zero section which we denote by $s_0$. The effective spacetime theory lives in dimension $d=12-\operatorname{dim}_{\R}X$. The elliptic fibration can be described by a Weierstrass model, in which the fiber is embedded as an algebraic curve in the weighted projective space $\mathbb{P}^2_{[2:3:1]}$, with homogeneous coordinates $[x:y:z]$. It is specified by the Weierstrass equation
\begin{equation}
y^2=x^3+f x z^4+g z^6~,
\label{eq:FT-weierstrass}
\end{equation}
with $f\in \Gamma(B,K_B^{-4})$ and $g\in \Gamma(B,K_B^{-6})$, where $K_B$ denotes the canonical bundle of the base. The elliptic fiber degenerates along the locus where the discriminant
\begin{equation}
\Delta = 4f^3 + 27g^2
\label{eq:FT-discriminant}
\end{equation}
vanishes, which defines a complex codimension-one subvariety of $B$. For each irreducible component $C_\alpha\subset \{\Delta=0\}$, the Kodaira type of the fiber singularity determines a simple subalgebra $\mathfrak{g}_\alpha$ of the non-abelian gauge algebra of the effective theory.  More precisely, over a generic point of $C_\alpha$, the singular fiber can be resolved by a sequence of blow-ups, introducing rational curves whose intersection pattern is encoded in the affine Dynkin diagram of $\mathfrak{g}_\alpha$. This yields $\operatorname{rk}\mathfrak{g}_\alpha$ linearly independent exceptional divisors for each irreducible component, that we collectively denote by $E_a$. For simplicity, we assume that the discriminant locus has a single irreducible component $C$. Over a codimension-two locus in $C$, further singularities appear and can be resolved by introducing additional rational curves. When wrapped by M2-branes, these curves give rise to charged matter. In particular, a curve $\gamma$ of this type is associated to a weight of a representation $\rep{r}$ of $\mathfrak{g}$, whose Dynkin labels $[l_1\cdots l_r]$ are determined by
\begin{align}
E_a \cdot \Gamma = l_a~.
\end{align}
Knowledge of the singular fibers is generally insufficient to determine the global form of the gauge group or to detect the presence of possible abelian factors. This information is instead encoded in the Mordell--Weil group.

The Mordell--Weil group of $X$ is the abelian group of rational sections,
\begin{equation}
\text{MW}(X)=\left\{\text{rational sections } s:B\rightarrow X\right\}~,
\label{eq:MW-def}
\end{equation}
where addition, denoted by $\boxplus$, is defined fiberwise using the elliptic curve structure, with the zero section $s_0$ serving as the identity element. By the Mordell--Weil theorem, this group is finitely generated. Its torsion subgroup 
\begin{equation}
\text{MW}(X)_{\mathrm{tor}} \subset \text{MW}(X)
\label{eq:MW-fg}
\end{equation}
encodes gauged one-form symmetries, while free generators of $\text{MW}(X)$ are associated to abelian gauge factors~\cite{ Aspinwall:1998xj, Morrison:2012ei,Mayrhofer:2014opa,Cvetic:2017epq,Cvetic:2018bni}. Torsional sections form the kernel of the Shioda map~\cite{Shioda:1989,Shioda:1990}, a group homomorphism
\begin{align}
\varphi\;:\;\text{MW}(X)\to \text{NS}(X)\otimes \mathbb{Q}~,
\end{align}
where $\text{NS}(X)$ denotes the Néron–Severi group of $X$, that is, its group of divisors modulo algebraic equivalence. The image of a rational section $s$ under the Shioda map can be decomposed as
\begin{equation}
\varphi(s) = S-S_0-\pi\big((S-S_0)\cdot S_0\big)+\sum_{a=1}^{r} \lambda_a E_a~,
\label{eq:Shioda-map}
\end{equation}
where $S$ and $S_0$ are the divisor classes associated with $s$ and with the zero section $s_0$, respectively. The rational numbers $\lambda_a$, determined by the section $s$ and by the structure of the exceptional divisors $E_a$ over the discriminant locus, can be related to center symmetries of the universal cover $\widetilde{G}$ of the gauge group $G$~\cite{Cvetic:2017epq}. Indeed, given a curve $\gamma$ associated with a weight $[l_1\cdots l_r]$ of the representation $\rep{r}$, the intersection of $\varphi(s)$ with $\gamma$ satisfies
\begin{align}
\varphi(s)\cdot\gamma - \sum_{a=1}^r \lambda_a\, l_a\quad\in \mathbb{Z}~,
\end{align}
due to the decomposition~\eqref{eq:Shioda-map}. The resulting ``charge'' of $\gamma$,
\begin{align}
Q(s)_{\rep{r}} &= \varphi(s)\cdot\gamma \quad \text{mod } 1~,
\label{eq:Shiodacharge}
\end{align}
agrees for all weights in the same representation~$\rep{r}$. When $s$ is a torsional section, the corresponding quantity $\sum_a \lambda_a\, l_a$, evaluated on an arbitrary weight $[l_1\cdots l_r]$ of $\widetilde{G}$, is identical modulo $1$ to the charge of that weight under a center symmetry of $\mathcal{Z}(\widetilde{G})$ encoded by the fractional numbers $\lambda_a$. Hence, the Mordell--Weil torsion subgroup is associated to a subgroup $\Gamma\subset\mathcal{Z}(\widetilde{G})$ of center symmetries of the effective theory. However, for a torsional section $s$ of order $n$, the homomorphism property of the Shioda map gives
\begin{align}
n\,\varphi(s) = \varphi\bigl(\underbrace{s \boxplus \cdots \boxplus s}_{n\ \text{times}}\bigr) = \varphi(s_0)= 0~.
\end{align}
Together with the fact that the divisor group is torsion-free, this implies $\varphi(s)=0$, so that for every representation $\rep{r}$ realized by curves $\gamma$ in the Calabi--Yau geometry,
\begin{align}
Q(s)_\rep{r} =0~.
\end{align}
Thus all geometrically realized representations are neutral under the corresponding center symmetry of $\widetilde{G}$, and the associated center one-form symmetry in the spacetime theory is gauged, leading to a nontrivial gauge group topology
\begin{align}
G = \widetilde{G}/\Gamma~.
\end{align}
Strictly speaking, this is verified only for the massless spectrum of the $d$-dimensional theory, but duality arguments to M-theory~\cite{Apruzzi:2020zot} indicate that it also holds for massive excitations.\\

These considerations admit a worldsheet counterpart in the presence of a faithful string probe, realized as a D3-brane wrapping a curve in the base $B$ of the Calabi--Yau geometry. In a background with non-trivial Mordell--Weil torsion, signaling a gauged center one-form symmetry, the holomorphic chiral algebra on the string must be extended by the corresponding KM simple current. In particular, the charge~\eqref{eq:Shiodacharge}, which vanishes for representations realized geometrically in the Calabi--Yau compactification, is the analogue of the monodromy charge~\eqref{eq:monodromycharge=centralcharge} associated with the corresponding simple current, which also vanishes for all local operators on the string worldsheet when the chiral algebra is extended. The consistency conditions for the chiral algebra extension should therefore have a geometric realization within the Calabi--Yau geometry itself.

Indeed, in the case of K3 compactifications, a relation between $\text{MW}(X)$ and the consistency of gauging a center one-form symmetry was already noted in~\cite{Cvetic:2020kuw}, building on observations of Miranda and Persson on the Mordell--Weil groups of K3 surfaces~\cite{Miranda:1989}. In this setting, the effective eight-dimensional $\mathcal{N}=1$ supergravity theory has an ADE-type gauge group of rank $18$ (ignoring for simplicity cases with abelian factors), determined by the singular fibers of the K3 surface. The Mordell--Weil group of an elliptic K3 surface $X$ can be presented as a quotient~\cite{Shioda:1989}
\begin{equation}
\text{MW}(X)\cong \text{NS}(X) /L~,
\end{equation}
where $\mathrm{NS}(X)$ is the Picard lattice of $X$, endowed with the intersection pairing, and $L$ is the sublattice generated by the zero section and all the irreducible components of fibers. In particular, the torsion subgroup of $\text{MW}(X)$ is isomorphic to the quotient
\begin{align}
\text{MW}(X)_{\text{tor}} \simeq (L^\perp)^\perp /L
\label{eq:MWtorsionK3}
\end{align}
with $L^\perp$ the orthogonal complement of $L$ inside $\mathrm{NS}(X)$. The lattice $L$ admits a decomposition
\begin{equation}
L = U \oplus L_{\mathfrak{g}}~,
\end{equation}
where $U$ is a rank-$2$ hyperbolic lattice generated by the zero section $S_0$ and the class of a fiber, while $L_{\mathfrak{g}}$ is the negative definite lattice generated by the exceptional curves, and is isomorphic, up to signature, to the root lattice of the ADE algebra $\mathfrak{g}$.

Relation~\eqref{eq:MWtorsionK3} shows that $\mathrm{MW}(X)_{\mathrm{tor}}$ naturally embeds into the discriminant group $L^\ast/L$, where $L^\ast$ denotes the dual lattice of $L$. Since $U$ is self-dual, the discriminant group of $L$ coincides with that of $L_{\mathfrak g}$, and therefore
\begin{align}
\text{MW}(X)_{\text{tor}} \;\subset\; L_{\mathfrak g}^\ast/L_{\mathfrak g}~.
\label{eq:MWembeddingK3lattice}
\end{align}
As a finite abelian group, $L_{\mathfrak{g}}^\ast/L_{\mathfrak{g}}$ is homomorphic to $\mathcal{Z}(\widetilde{G})$, the center of the simply connected group associated with the ADE algebra $\mathfrak{g}$. More precisely, the embedding above preserves the discriminant quadratic form. Since $L_{\mathfrak{g}}$ is even, the quotient $L_{\mathfrak{g}}^\ast/L_{\mathfrak{g}}$ carries the quadratic form
\begin{align}
q_{\mathfrak{g}}([x])=\frac{x \cdot x}{2}\mod 1~,
\label{eq:MWdiscriminantform}
\end{align}
for $x\in L_{\mathfrak{g}}^\ast$. Because $L_{\mathfrak{g}}$ is an ADE root lattice, the form $q_{\mathfrak{g}}$ can be computed explicitly in all cases. For simply laced algebras other than $\mathfrak{d}_r$ with even $r$, the center is cyclic and $L_{\mathfrak{g}}^\ast/L_{\mathfrak{g}}$ is generated by a single element $\omega$, so that any class can be written as $[x]=\ell\omega$. The discriminant form is then written as
\begin{align}
q_{\mathfrak{g}}([x]) = -\ell^2\alpha_{\mathfrak{g}}\mod 1~,
\end{align}
where the constants $\alpha_{\mathfrak{g}}$ coincide precisely with the fractional instanton coefficients\footnote{This was already noticed in~\cite{Cvetic:2020kuw} for $\mathfrak{a}_r$ algebras.} listed in table~\ref{tab:mixedanomalycoefficients}. For $\mathrm{Spin}(2r)$ with $r$ even, whose center is $\Z_2\times\Z_2$, the discriminant group instead has two generators $\omega_1,\omega_2$, and
\begin{align}
q_{\mathfrak{d}_r}([x]) = -(\ell_1+\ell_2)^2\alpha_{\mathfrak{d}_r}-\ell_1\ell_2\alpha'_{\mathfrak{d}_r}\mod 1~,
\end{align}
where $[x]=\ell_1\omega_1+\ell_2\omega_2$. The constants $\alpha_{\mathfrak{d}_r}$ and $\alpha'_{\mathfrak{d}_r}$ again agree with the fractional instanton coefficients.

Since elements of $\mathrm{MW}(X)$ arise from classes in $\mathrm{NS}(X)$, which is an even lattice, the discriminant form~\eqref{eq:MWdiscriminantform} vanishes identically on $\mathrm{MW}(X)_{\mathrm{tor}}$. The corresponding center one-form symmetry therefore satisfies the consistency condition
\begin{align}
\ell^2\alpha_{\mathfrak{g}}\in\Z~,
\label{eq:8dconsistencyADE}
\end{align}
with the obvious modification in the $\Z_2\times\Z_2$ case. This immediately generalizes to products of ADE factors, where the discriminant form
\begin{align}
q_{\mathfrak{g}} = q_{\mathfrak{g}_1}\oplus\dots \oplus q_{\mathfrak{g}_N}
\end{align}
reproduces the condition~\eqref{eq:8dmixedanomaly} for $b_\alpha=1$, the value relevant for the rank-$18$ eight-dimensional theories. The resulting constraint therefore agrees with the condition for the absence of a mixed anomaly for the center one-form symmetry in the eight-dimensional effective supergravity theory. Equivalently, it matches the consistency condition for extending the $\mathfrak{g}$ current algebra at level $1$ by the corresponding KM simple current. This result is consistent with the fact that the 8d theory admits a dual realization in terms of the heterotic string compactified on $T^2$. In this dual frame, the heterotic worldsheet CFT contains the $\mathfrak{g}$ current algebra inside its Narain sector, and the condition~\eqref{eq:8dconsistencyADE} follows directly from the structure of the associated rank-$18$ ADE sublattice~\cite{Font:2021uyw}. 

The preceding constraint~\eqref{eq:8dconsistencyADE} admits a generalization to other settings, including eight-dimensional supergravity theories with rank-$10$ gauge group (possibly non-simply laced), realized in F-theory through compactification on a K3 surface with a frozen $\mathfrak{d}_8$ singularity~\cite{Witten:1997bs}. In that case, faithful probe strings can realize the current algebras of the various gauge factors at level $b_\alpha=1$ or $b_\alpha=2$, and equation \eqref{eq:8dmixedanomaly} provides the natural generalization of the geometric condition \eqref{eq:8dconsistencyADE}, namely
\begin{align}
\sum_{\alpha=1}^N b_\alpha \ell_\alpha^2 \alpha_{\mathfrak{g}_\alpha} \quad\in\Z~.
\end{align}
It would be interesting to understand this generalized integrality constraint in geometric terms. Turning now to six-dimensional models, one generally expects faithful strings of different string charge to give rise to several constraints which depend on how various strings couple to the bulk gauge symmetry and must all be compatible with one another. Ultimately, in models that arise from F-theory such constraints must also be compatible with geometric constraints for the existence of a suitable Mordell-Weil torsion group; it would be interesting to spell out this correspondence more explicitly.

\section{Gauge group topology in heterotic compactifications}
\label{sec:heteroticcompactifications}
In this section we illustrate the general picture developed in section \ref{sec:center1formsymmetries} through a number of examples drawn from supersymmetric heterotic string compactifications. We begin with the familiar ten-dimensional heterotic theories, in which the relation between simple currents and the global structure of the gauge group can already be seen explicitly. To access more non-trivial examples, we then turn to asymmetric orbifold compactifications. Asymmetric orbifolds, introduced in~\cite{Narain:1986qm}, have since been used to produce a wide variety of models with possibly reduced supersymmetry and various non-abelian gauge groups---see~\cite{Bianchi:2022tbr,Acharya:2022shu,Fraiman:2022aik,Gkountoumis:2023fym,Faraggi:2023mkn,Baykara:2023plc,Gkountoumis:2024dwc,Baykara:2024tjr,Baykara:2024vss,Angelantonj:2024jtu,Hamada:2024cdd,Hamada:2025cpe,Aldazabal:2025zht,Baykara:2025lhl,Larotonda:2026hxy,Gkountoumis:2025btc,Reece:2026hfk,Cheng:2026abk} for some recent applications. A well-studied setting is that of the nine-dimensional CHL string~\cite{Chaudhuri:1995fk}, which may be obtained as a $\Z_2$ asymmetric orbifold of the heterotic string compactified on a circle~\cite{Chaudhuri:1995bf}.  The orbifold action uses the outer automorphism of $\mathrm{E}_8\times\mathrm{E}_8$; it leads to a reduction of the gauge group rank and, at certain points in the moduli space, to enhanced gauge symmetry with a higher level current algebra realization. In lower dimensions, CHL compactifications yield a vast landscape of theories, including examples with non-simply laced gauge algebras as well as non-simply connected gauge groups~\cite{Font:2021uyw}. Hence, they provide a natural framework to verify explicitly the correspondence between simple current extensions and gauged one-form symmetries.

We begin in section~\ref{subsec:het10d} by analyzing in detail the extension of the $\mathfrak{so}(32)$ current algebra in the ten-dimensional $\mathrm{Spin}(32)/\mathbb{Z}_2$ heterotic theory, which provides perhaps the simplest realization of the general picture described in section \ref{sec:center1formsymmetries}. Next, we recall in section~\ref{subsec:Narain} some basic features of asymmetric orbifold compactifications, with additional details relegated to appendices~\ref{app:Narain} and~\ref{app:modularorbits} in order to streamline our exposition. With this framework in place, we then turn in sections~\ref{subsec:CHLexamples} and~\ref{subsec:6dasymmetric} to examples in lower dimensions: CHL models in nine and eight dimensions, with sixteen supercharges, as well as some asymmetric orbifold models with eight supercharges in six dimensions.

\subsection{The heterotic \texorpdfstring{$\mathrm{Spin}(32)/\Z_2$}{Spin(32)/Z2} currents}
\label{subsec:het10d}
Ten-dimensional heterotic supergravity admits two consistent gauge groups, namely $\mathrm{E}_8\times\mathrm{E}_8$ and $\mathrm{Spin}(32)/\Z_2$. On the heterotic worldsheet, both are realized by a $c_\tL=16$ left-moving current algebra at level one, constructed from sixteen left-moving fermions with different chiral GSO projections. This provides one of the simplest setting in which the correspondence between worldsheet simple currents and spacetime center one-form symmetries can be seen explicitly.

Let us begin with the $\mathrm{E}_8\times\mathrm{E}_8$ theory. Since the center of $\mathrm{E}_8$ is trivial, there is no non-trivial center one-form symmetry in the ten-dimensional supergravity theory. Correspondingly, in the heterotic CFT, the $\mathfrak{e}_8+\mathfrak{e}_8$ KM algebra admits no non-trivial simple current: each $\mathfrak{e}_8$ factor, at level $1$, has a single primary representation $\rep{1}$. Its character,
\begin{align}
\chi^{\mathfrak{e}_8}_{\rep{1}}(\tau) = q^{-\frac{8}{24}}\left(1+248\,q+4124\,q^2+34752\,q^3+\dots\right)~,
\label{eq:e8character}
\end{align}
is modular invariant by itself. The situation is more interesting for the $\mathrm{Spin}(32)/\Z_2$ heterotic string. The $\mathfrak{d}_{16}=\mathfrak{so}(32)$ current algebra at level $1$ admits four integrable representations, corresponding to the trivial, vector, spinor and spinor conjugate representations of $\mathrm{Spin}(32)$. We denote them by $\rep{1}$, $\rep{v}$, $\rep{s}$ and $\rep{c}$ respectively. They have conformal weights
\begin{align}
h_\rep{1} &=0~,&
h_\rep{v} &=\tfrac{1}{2}~,&
h_\rep{s}=h_\rep{c} &=2~,
\end{align}
and the corresponding characters are
\begin{align}
\chi^{\mathfrak{d}_{16}}_{\rep{1}}(\tau) &= q^{-\frac{16}{24}}\left(1+496\,q +36984\,q^2+1066432\,q^3+\dots\right)~,
\nonumber\\
\chi^{\mathfrak{d}_{16}}_{\rep{v}}(\tau) &= q^{-\frac{16}{24}+\frac{1}{2}}\left(
32 + 4992\,q + 217280\,q^2 + 4548352\,q^3 +\dots\right)~,
\nonumber\\
\chi^{\mathfrak{d}_{16}}_{\rep{s}}(\tau) &= q^{-\frac{16}{24}+2}\left(32768 + 1048576 \,q + 17301504\,q^2 + 197132288\,q^3 +\dots\right)
\nonumber\\
&=\chi^{\mathfrak{d}_{16}}_{\rep{c}}(\tau)~.
\end{align}
Their modular transformations are encoded by the modular matrix
\begin{align}
\mathcal{S}_{\mathfrak{d}_{16}} = \tfrac{1}{2}
\begin{pmatrix}
1 & 1 & 1 & 1 \\
1 & 1 & -1 & -1 \\
1 & -1 & 1 & -1\\
1 & -1 & -1 & 1
\end{pmatrix}~.
\end{align}
The center of $\mathrm{Spin}(32)$ is $\Z_2\times\Z_2$, and accordingly all four KM primaries are simple currents. Since the spinor representations $\rep{s}$ and $\rep{c}$ have integral conformal weight, they can consistently be used to extend the chiral algebra. Choosing, for instance, the $\Z_2$ simple current $\rep{s}$, modular invariance requires projecting out the primaries $\rep{v}$ and $\rep{c}$, which have non-trivial monodromy with respect to it. The extended theory then contains a single character,
\begin{align}
\mathcal{X}^{\mathfrak{d}_{16}}_{\rep{1}} = \chi^{\mathfrak{d}_{16}}_{\rep{1}}+\chi^{\mathfrak{d}_{16}}_{\rep{s}}~,
\end{align}
which is itself modular invariant. This is precisely the character that appears in the GSO-projected heterotic partition function of the $\mathrm{Spin}(32)/\Z_2$ theory. In this simple example, we observe the  correspondence between bulk one-form symmetries and worldsheet simple currents explicitly: the $\Z_2$ simple current extension of the $\mathfrak{so}(32)$ current algebra by the spin-$2$ simple current in the spinor representation reflects the gauging the corresponding center one-form symmetry in spacetime, leading from $\mathrm{Spin}(32)$ to the non-simply connected gauge group $\mathrm{Spin}(32)/\Z_2$.

\subsubsection*{An aside on meromorphic CFTs}
As a brief aside from our discussion of gauged one-form symmetries, it is worth pausing to note that the simple current extension of $\mathfrak{so}(32)$ reviewed above is a special case of a much more general structure that plays a fundamental role in lattice CFTs. The extended character of $\mathfrak{so}(32)$---and more generally, the extended character of the $\mathfrak{so}(16l)$ current algebra at level $1$ for any $l$---can be rewritten as
\begin{align}
\chi_\rep{1}^{\mathfrak{d}_{8l}}+\chi_\rep{s}^{\mathfrak{d}_{8l}} = \frac{1}{\eta(\tau)^{8l}}\sum_{\boldsymbol{L}\in\Lambda_{8l}} q^{\frac{1}{2}\boldsymbol{L}\cdot \boldsymbol{L}}~,
\label{eq:so(16l)extension}
\end{align}
where $\eta(\tau)$ is the Dedekind function and $\Lambda_{8l}$ is the even self-dual lattice
\begin{align}
\Lambda_{8l} = \left\lbrace (x_1,\dots,x_{8l}) \in \Z^{8l}\cup (\Z+\tfrac{1}{2})^{8l} \left| x_1+\dots+x_{8l}\in 2\Z \right.\right\rbrace ~.
\end{align}
This expression makes clear that the extended character is precisely the partition function of a lattice CFT with central charge $c_\tL=8l$, and is invariant under $\mathrm{SL}(2,\Z)$ (up to an overall phase). The spectrum of this CFT, when decomposed into KM primary representations, consists of the identity operator together with a spin-$l$ current in the spinor representation of $\mathfrak{so}(16l)$. For $l=1$, the extended character~\eqref{eq:so(16l)extension} coincides with the identity character~\eqref{eq:e8character} of the $\mathfrak{e}_8$ KM algebra at level $1$: indeed, the spin-$1$ currents in $\rep{s}=\rep{128}$ enhance the affine algebra due to the conformal embedding $\mathfrak{d}_8\subset\mathfrak{e}_8$.

More broadly, integral spin simple current extensions of KM algebras are central in the classification of $c=24$ meromorphic CFTs~\cite{Schellekens:1992db}. Of these $71$ theories, $69$ possess a semi-simple current algebra---the remaining two theories are provided by the Leech CFT, a lattice theory with $\mathfrak{u}(1)^{24}$ symmetry, and the monster module, that has no KM current. The list of~\cite{Schellekens:1992db} includes $24$ lattice CFTs, that describe free chiral bosons compactified on an even self-dual Euclidean lattice $\Lambda$. The respective rank-$24$ lattices were classified by Niemeier~\cite{Niemeier:1973}---apart from the Leech lattice, they admit a unified description based on the so-called gluing construction. The lattice $\Lambda$ is obtained from the root lattice $\Lambda_{\mathfrak{g}}$ of a rank $24$ semi-simple algebra $\mathfrak{g}$, upon enlarging it by glue vectors, which are elements of the lattice dual to $\Lambda_{\mathfrak{g}}$. In the lattice CFT, the symmetry $\mathfrak{g}$ is realized as a level $1$ KM algebra, and the aforementioned glue vectors correspond precisely to KM simple currents of integral spin. The Niemeier CFT partition function is then naturally interpreted as the identity character of the extended KM algebra---exactly as in the $\mathfrak{d}_{8l}$ example reviewed above. Let us note that, for the remaining theories of~\cite{Schellekens:1992db} at higher KM level, the chiral algebra is also an extension of the KM algebra; however the extension involves more than just the center simple currents.

\subsection{Narain compactifications}
\label{subsec:Narain}
In this section we briefly review Narain compactifications of ten-dimensional supersymmetric heterotic theory which lead to a much larger class of examples with interesting global structures. Consider the $\mathrm{E}_8\times\mathrm{E}_8$ heterotic string compactified on a $D$-dimensional torus. We focus on the internal sector of the worldsheet theory, and largely follow the presentation of~\cite{Cheng:2022nso,Israel:2023tjw}. The Narain CFT is a free theory with central charges $c_\tL=D+16$ and $c_\tR=D$. At a generic point in its moduli space, the CFT exhibits a left-moving $\mathfrak{u}(1)^{D+16}$ KM symmetry and a right-moving $\mathfrak{u}(1)^D$ KM symmetry. The vertex operators organize into primary representations of this current algebra, and are labeled by points of a lattice $\Gamma$. The latter is an even self-dual lattice of signature $(D+16,D)$. We denote its elements as $\bp\in\Gamma$. The corresponding vertex operators $\mathcal{V}_\bp(z,\bar{z})$ have left- and right-moving conformal dimensions $h_\tL(\boldsymbol{p})$ and $h_\tR(\boldsymbol{p})$. Their spin, given by
\begin{align}
h_\tL(\boldsymbol{p}) - h_\tR(\boldsymbol{p}) = \tfrac{1}{2}\bp\cdot\bp~,
\end{align}
only depends on the lattice point $\bp$. The conformal weights themselves depend on the moduli of the Narain theory. The moduli space is locally parametrized by a choice of timelike $D$-plane $\Pi_\tR$ inside the linear span $\Gamma_\R\simeq \R^{D+16,D}$ of the Narain lattice, with orthogonal $\Pi_\tL=(\Pi_\tR)^\perp$. The decomposition of $\bp$ into its $\Pi_\tL$ and $\Pi_\tR$ components then determines the conformal dimensions of the operator $\mathcal{V}_\bp$. More concretely, after fixing a basis for the lattice $\Gamma$, one may parametrize $\Pi_\tR$ in terms of quantities associated to the $D$-dimensional toroidal geometry. These Narain moduli consist of a metric, a two-form field, and Wilson line parameters---see appendix~\ref{app:Narainmoduli} for details.\\

The states of the Narain CFT that are purely left-moving or purely right-moving are respectively characterized by the lattices
\begin{align}
\Lambda_\tL &= \Gamma\cap\Pi_\tL ~,&
\Lambda_\tR &= \Gamma\cap\Pi_\tR ~.
\end{align}
For generic values of the heterotic parameters, $\Lambda_\tL=\Lambda_\tR=\{0\}$ and the theory has no chiral states in addition to the $\mathfrak{u}(1)$ currents. However, there are distinguished loci in the Narain moduli space where the rank of $\Lambda_\tL\oplus\Lambda_\tR$ jumps, and the theory acquires additional (anti-)holomorphic states. A well-studied situation is where the theory acquires additional left-moving spin-$1$ states: this corresponds to an enlargement of the left-moving current algebra
\begin{align}
\mathfrak{u}(1)^{D+16} \to \mathfrak{g}+\mathfrak{u}(1)^{D+16-\operatorname{rk}\mathfrak{g}}~.
\end{align}
These additional holomorphic currents give rise to spacetime gauge bosons in the supergravity description. We note that the enhanced algebra $\mathfrak{g}$ is realized at level $1$ in the heterotic CFT. A familiar example is provided by the locus of the moduli space where Wilson line parameters vanish: this subspace generically leads to a $\mathfrak{u}(1)^D+\mathfrak{e}_8+\mathfrak{e}_8$ gauge algebra, with further enhancements for specific values of the toroidal metric and two-form field. The moduli space contains isolated points of \emph{maximal enhancement}, that is, where the enhanced algebra has maximal rank $\operatorname{rk}\mathfrak{g}=D+16$. For instance, specific values of the Narain moduli identified in~\cite{Narain:1986am,Ginsparg:1986bx} lead to a $\mathfrak{d}_{D+16}$ gauge symmetry. For a given $D$, possible symmetry enhancement points can in principle be determined using standard lattice-theoretic techniques~\cite{Nikulin:1980}, as the root lattice of $\mathfrak{g}$ must admit an embedding in the even self-dual lattice $\Gamma$.\footnote{This embedding is not always primitive: although the positive-definite lattice $\Lambda_\tL\subset\Gamma$ is primitively embedded, there are situations where $\Lambda_\tL$ is not generated by its roots. This will happen, in particular, when the spacetime gauge group is non-simply connected: the simple current associated to the gauged one-form symmetry gives a higher-spin state in $\Lambda_\tL$ that does not belong to its root sublattice.} For $D\leq 2$, maximally enhanced points have been systematically explored in~\cite{Fraiman:2018ebo,Font:2020rsk} using generalized Dynkin diagrams~\cite{Goddard:1985,Ginsparg:1986bx,Cachazo:2000ey}.

More generally, the Narain moduli space contains a distinguished set of isolated points where $\Lambda_\tL\oplus\Lambda_\tR$ is a full rank sublattice of $\Gamma$. For these special choices of heterotic parameters, the Narain CFT is rational: its partition function may be recast as a finite sum indexed by the abelian group
\begin{align}
\Gamma/(\Lambda_\tL\oplus\Lambda_\tR)~.
\end{align}
Conversely, starting from the sublattice $\Lambda_\tL\oplus\Lambda_\tR$, one may reconstruct the even self-dual lattice $\Gamma$ by a procedure called \emph{gluing}: the sublattice is enlarged by a set of vectors from the discriminant group $(\Lambda_\tL^\ast/\Lambda_\tL)\oplus(\Lambda_\tR^\ast/\Lambda_\tR)$ obeying certain consistency conditions~\cite{Conway:1999dov}. This procedure may be used to construct Narain CFTs with specific properties, such as a prescribed left-moving current algebra.\\

In the heterotic CFT, the Narain sector describes compact scalars $X^I$ valued in a $D$-dimensional torus $T^D$, together with the sixteen left-moving Weyl fermions associated to the $\mathrm{E}_8\times\mathrm{E}_8$ heterotic symmetry. In addition, the internal sector of the CFT contains anti-holomorphic Weyl fermions $\psi^I$, the right-moving superpartners of the bosons $X^I$. Asymmetric orbifolds compactifications are obtained by combining the orbifold action on the toroidal geometry with a non-trivial action on the gauge sector. In the models considered below, the orbifold symmetry $g$ acts on vertex operators $\mathcal{V}_\bp$ as
\begin{align}
g\, \mathcal{V}_{\boldsymbol{p}}\, g^{-1}= U_g(\bp) \, \mathcal{V}_{\varphi_{g}(\boldsymbol{p})}~,
\end{align}
where $U_g(\bp)\in\mathrm{U}(1)$ and $\varphi_g\,:\,\Gamma\to\Gamma$ is a lattice automorphism. The phase $U_g(\bp)$ is intertwined with the vertex operator cocycles~\cite{Cheng:2022nso}; its presence is necessary for consistency of the OPE with the orbifold action. When the lattice symmetry $\varphi_g$ is trivial, $U_g$ must be an element of $\mathrm{Hom}(\Gamma,\mathrm{U}(1))$, and may always be written as
\begin{align}
U_g(\bp) = \ee^{2\pi\ii \,s_g\cdot\boldsymbol{p}}~,
\end{align}
in terms of a so-called shift vector $s_g\in\Gamma_\R$. More generally, the asymmetric orbifold action may combine such a shift with a lattice rotation. Modular invariance places further consistency constraints on $U_g$ and $\varphi_g$~\cite{Narain:1986qm}---for pure shift orbifolds, a recent discussion can be found in~\cite{Israel:2025ouq}. Note that, if $\varphi_g$ acts non-trivially on the right-moving degrees of freedom of $X^I$, then the orbifold action should also rotate the right-moving fermions $\psi^I$ in order to preserve worldsheet supersymmetry.

\subsection{CHL compactifications and symmetry enhancements}
\label{subsec:CHLexamples}
We now turn to specific examples in dimension $d=9$ and $d=8$, obtained from the CHL orbifold, and examine the relation between the gauge group topology and simple current extensions on the string worldsheet. The nine-dimensional CHL string~\cite{Chaudhuri:1995fk} can be realized as a $\Z_2$ asymmetric orbifold of the $\mathrm{E}_8\times\mathrm{E}_8$ heterotic theory compactified on a circle~\cite{Chaudhuri:1995bf}. Geometrically, the orbifold symmetry $g$ acts by shifting the circle coordinate by a half-period while exchanging the two $\mathrm{E}_8$ factors of the heterotic string. The CHL orbifold acts solely on the Narain sector of the heterotic worldsheet CFT, with $c_\tL=17$ and $c_\tR=1$. We choose a basis of the Narain lattice in which lattice points $\bp\in\Gamma$ admit a decomposition of the form
\begin{align}
\bp = \tw\be+\tn\be^\ast +(\bell,\bell')~.
\end{align}
Here, $\tn$ and $\tw$ are integers (respectively the momentum and winding numbers), and $(\bell,\bell')$ belongs to an $\mathfrak{e}_8+\mathfrak{e}_8$ sublattice of $\Gamma$ orthogonal to the hyperbolic factor spanned by the null vectors $\be,\be^\ast$. The CHL symmetry is then characterized by the following shift vector and lattice automorphism:
\begin{align}
s_g &= \frac{\be}{2}~,&
\varphi_g(\bp) &=\tw\be+\tn\be^\ast +(\bell',\bell)~.
\end{align}
The shift action induces a phase $U_g(\bp)=(-1)^\tn$ in the transformation of vertex operators, thereby reproducing the half-shift along the circle direction. Note that this CHL action defines a symmetry of the CFT only on a subspace of the Narain moduli space. In nine dimensions, the Narain moduli consist of a radius $r$ and a Wilson line parameter $A=(a,a')$, and the compatible locus is (locally) given by $a=a'$. The twisted sector of the $\Z_2$ orbifold can be constructed explicitly~\cite{Chaudhuri:1995bf,Mikhailov:1998si}. A detailed description of the twisted Hilbert space---which we will need below to characterize holomorphic states in certain CHL models---is provided in appendix~\ref{app:CHLspectrum}.

The CHL construction readily extends to lower dimensions, in which case the worldsheet CFT is obtained as a $\Z_2$ orbifold of the heterotic string compactified on a $D$-dimensional torus. The asymmetric orbifold acts in a similar manner, combining a half-shift along one of the circle directions with an exchange of the two $\mathrm{E}_8$ factors. The resulting spacetime theory is a supergravity in $d=10-D$ dimensions preserving sixteen supercharges. The spacetime gauge group $G$ (ignoring graviphotons) arises from the holomorphic current algebra of the worldsheet CFT. One consequence of the CHL orbifold is a reduction of the rank of the gauge group to
\begin{align}
\operatorname{rk}G=D+8~.
\end{align}
As in heterotic toroidal compactifications, one finds $G=\mathrm{U}(1)^{D+8}$ at a generic point in the moduli space, while at special loci additional spin-$1$ states may become holomorphic, enhancing $G$ to a non-abelian gauge group. In particular, the moduli space contains finitely many isolated points where $G$ is a non-abelian group of maximal rank. Such maximally enhanced models have been systematically characterized in~\cite{Font:2021uyw}.\\

For $D=1$, the CHL moduli space contains $9$ points of maximal enhancements~\cite{Font:2021uyw}, at which the spacetime gauge group $G$ is enhanced to one of the following rank-$9$ simply laced Lie groups:
\begin{align}
&\mathrm{SU}(10)~,&
&\mathrm{SU}(2)\times\mathrm{SU}(9)~,&
&\mathrm{SU}(5)\times\mathrm{SU}(6)~,\nonumber\\
&\mathrm{SU}(2)\times\mathrm{SU}(3)\times\mathrm{SU}(7)~,&
&\mathrm{SU}(5)\times\mathrm{Spin}(10)~,&
&\mathrm{Spin}(18)~,\nonumber\\
&\mathrm{SU}(4)\times\mathrm{E}_6~,&
&\mathrm{SU}(3)\times\mathrm{E}_7~,&
&\mathrm{SU}(2)\times\mathrm{E}_8~.
\label{eq:9dCHLgroups}
\end{align}
In all cases, the group $G$ is simply-connected. This is shown in~\cite{Font:2021uyw} by studying overlattices of the root lattice and their embeddings in the so-called Mikhailov lattice~\cite{Mikhailov:1998si}. The same result can be readily derived by analyzing KM simple currents. Indeed, the gauge symmetry is realized in the CHL worldsheet theory as a left-moving current algebra, with each simple factors at level $2$. By the arguments in the previous section, any gauged one-form center symmetry associated with $\pi_1(G)$ must appear in the CFT as left-moving simple currents. 

Given a gauge groups in~\eqref{eq:9dCHLgroups}, one can enumerate the KM simple currents and check whether any have integral conformal dimension. For example, the $\mathfrak{su}(10)$ current algebra at level $2$ has a $\Z_{10}$ center. The corresponding simple currents, labeled by $\ell=0,1,\dots,9$, have conformal dimension
\begin{align}
h_\tL = \frac{\ell(10-\ell)}{10}~,
\end{align}
which is non-integer for $\ell$ non-zero. It follows that the $\Z_{10}$ center one-form symmetry of $\mathrm{SU}(10)$ is not gauged: it is broken by states in the CHL spectrum. Similar considerations apply to all but one of the groups in~\eqref{eq:9dCHLgroups}: their associated current algebras admit no simple currents of integral conformal dimension,\footnote{More precisely, the level $2$ current algebras associated with $\mathrm{SU}(5)\times\mathrm{Spin}(10)$, $\mathrm{Spin}(18)$, and $\mathrm{SU}(4)\times\mathrm{E}_6$ do contain spin-$1$ simple currents, which can be traced to conformal embeddings $\mathfrak{so}(10)_2\subset\mathfrak{su}(10)_1$, $\mathfrak{so}(18)_2\subset\mathfrak{su}(18)_1$, and $\mathfrak{su}(4)_2\subset\mathfrak{su}(6)_1$. However, extending the current algebra using these currents yields a KM algebra of larger rank, and is therefore incompatible with the known gauge symmetry. Thus, only simple currents with $h_\tL>1$ are relevant. Note also that the $\mathfrak{su}(2)+\mathfrak{e}_8$ current algebra admits a spin-$2$ simple current, but the corresponding $\mathbb{Z}_2$ action does not coincide with a center symmetry.} and hence the corresponding gauge groups must have trivial fundamental group. The only potentially ambiguous case is $G=\mathrm{SU}(2)\times\mathrm{SU}(9)$, since the $\mathfrak{su}(9)$ current algebra at level $2$ admits spin-$2$ simple currents associated with the $\Z_3\subset\Z_9$ subgroup of the center. However, a direct inspection of the CHL spectrum, presented in appendix~\ref{app:CHLspectrum}, shows that these currents are absent from the theory. Therefore, the analysis of KM simple currents reproduces the result of~\cite{Font:2021uyw} that all maximally enhanced gauge groups in~\eqref{eq:9dCHLgroups} are simply-connected.\\

Further compactifying the CHL string gives rise to a large set of theories in lower dimensions. For $D=2$, the CHL moduli space contains $61$ maximally enhanced points~\cite{Font:2021uyw}, yielding eight-dimensional supergravity theories with a variety of rank-$10$ gauge groups, including non-simply laced examples such as $\operatorname{Sp}(10)$. In eight dimensions, the CHL string admits a dual realization as F-theory compactified on a K3 surface with frozen singularities. The corresponding maximally enhanced gauge algebras were determined in~\cite{Hamada:2021bbz} via a classification of configurations of singular fibers in elliptic K3 surfaces~\cite{Shimada:2005yxf}, in agreement with the list of~\cite{Font:2021uyw}. Among these, one finds $29$ supergravity theories with nontrivial gauge group topology---specifically, $\pi_1(G)=\mathbb{Z}_2$ or $\pi_1(G)=\mathbb{Z}_2\times\mathbb{Z}_2$. Again, the analysis carried out for CHL models in~\cite{Font:2021uyw,Cvetic:2021sjm} is confirmed by duality with F-theory, where gauged one-form symmetries are encoded in the Mordell--Weil torsion of the elliptic K3 surface. Thus, eight-dimensional CHL models provide a well-controlled setting to test our proposal relating worldsheet simple currents to one-form symmetries. To illustrate this correspondence, we analyze a representative example below.

\subsubsection*{An eight-dimensional theory with $\pi_1(G)=\Z_2$}
Among the maximally enhanced models in~\cite{Font:2021uyw,Hamada:2021bbz}, consider the theory with gauge algebra
\begin{align}
\mathfrak{g} = \mathfrak{su}(2) + \mathfrak{su}(3) + \mathfrak{e}_7~.
\end{align}
On the worldsheet CFT of the CHL string, this gauge symmetry is realized as a left-moving current algebra with each factor at level $2$. The left-moving central charge $c_\tL=18$ is entirely saturated by the Sugawara contribution, so that left-moving primary operators are fully determined by their $\mathfrak{g}$ representation. The KM primary representations for each factor are given in table~\ref{tab:a1a2e7KMcontent}.
\begin{table}[h!]
\centering
{\renewcommand{\arraystretch}{1.2}
$\begin{array}{c|ccc}
\multicolumn{4}{c}{\mathfrak{su}(2)}\\  \hline 
\rep{r} & \rep{1} & \rep{2} & \rep{3} \\ \hline
h_\tL & 0 & \frac{3}{16} & \frac{1}{2} \\
\end{array}
\qquad\qquad
\begin{array}{c|cccccc}
\multicolumn{7}{c}{\mathfrak{su}(3)}\\ \hline 
\rep{r} & \rep{1} & \rep{3} & \repconj{3} & \rep{8} & \rep{6} & \repconj{6} \\ \hline
h_\tL & 0 & \frac{4}{15}  & \frac{4}{15}  & \frac{3}{5}  & \frac{2}{3}  & \frac{2}{3}
\end{array}
\qquad\qquad
\begin{array}{c|cccccc}
\multicolumn{7}{c}{\mathfrak{e}_7}\\ \hline 
\rep{r} & \rep{1} & \rep{56} & \rep{133} & \rep{912} & \rep{1463} & \rep{1539} \\ \hline
h_\tL & 0 & \frac{57}{80} & \frac{9}{10} & \frac{21}{16} & \frac{3}{2} & \frac{7}{5} \\
\end{array}$}
\caption{Primary representations of the $\mathfrak{su}(2)$, $\mathfrak{su}(3)$ and $\mathfrak{e}_7$ current algebras at level $2$}
\label{tab:a1a2e7KMcontent}
\end{table}

The simple currents of the left-moving KM algebra are associated with the center symmetries of $\mathrm{SU}(2)\times\mathrm{SU}(3)\times\mathrm{E}_7$. They form a $\Z_2\times\Z_3\times\Z_2$ group, generated by the primary operators in the representations $(\rep{3},\rep{1},\rep{1})$, $(\rep{1},\rep{6},\rep{1})$, and $(\rep{1},\rep{1},\rep{1463})$. In particular, there is a single nontrivial simple current with integer conformal dimension, corresponding to the representation $(\rep{3},\rep{1},\rep{1463})$. This operator, with $h_\tL=2$, could in principle be part of the CHL spectrum; its presence would signal the gauging of the one-form symmetry associated with the combined centers of $\mathrm{SU}(2)$ and $\mathrm{E}_7$.

To determine whether this spin-$2$ current is present in the theory, it suffices to count holomorphic states at the first Virasoro levels. As shown in appendix~\ref{app:CHLspectrum}, one finds a total of $30477$ purely left-moving states with $h_\tL=2$, of which $10584$ are KM descendants of the identity. The remaining $19893$ states must organize into primary representations, and there is a unique consistent decomposition, namely
\begin{align}
(\rep{3},\rep{1},\rep{1463})+(\rep{3},\rep{8},\rep{133})+(\rep{1},\rep{8},\rep{1539})~.
\end{align}
In particular, the representation $(\rep{3},\rep{1},\rep{1463})$ is present, confirming that the $\Z_2$ simple current appears in the spectrum. The remaining $h_\tL=2$ primaries organize into a single representation of the extended KM algebra.

By our arguments, the $\Z_2$ extension of the left-moving current algebra of the CHL string signals a gauged center one-form symmetry in the eight-dimensional spacetime theory. We thus obtain the global form of the gauge group:
\begin{align}
G=(\mathrm{SU}(2)\times\mathrm{E}_7)/\Z_2\times \mathrm{SU}(3)~.
\end{align}
This is in agreement with the lattice-theoretic computation of~\cite{Font:2021uyw} as well as with the F-theory dual description, for which the elliptic K3 has $\Z_2$ Mordell--Weil torsion.\footnote{The K3 geometry for this model corresponds to configuration $\#2995$ in~\cite{Shimada:2005yxf}, as identified in~\cite{Hamada:2021bbz}.}

\subsection{Six-dimensional asymmetric orbifolds}
\label{subsec:6dasymmetric}
Let us now turn to a class of six-dimensional supergravity theories, obtained from asymmetric orbifolds of the heterotic string. The internal sector of the parent heterotic theory is described by a (0,4) SCFT with $c_\tL=20$ and $c_\tR=6$. Its degrees of freedom consist of four compact bosons $X^I$ valued in a torus $T^4$, their right-moving superpartners $\psi^I$, and sixteen left-moving Weyl fermions associated with $\mathrm{E}_8\times\mathrm{E}_8$. Let us consider a $\Z_2$ orbifold symmetry $g$ whose action on the right-moving sector,
\begin{align}
g\,X_\tR^I\,g^{-1} &= - X_\tR^I~,&
g\,\psi^I g^{-1} &= - \psi^I~,
\label{eq:Kummerorbifold}
\end{align}
preserves the small N=4 superconformal algebra. Combined with a symmetric action on the left-moving components of the scalars (together with a compatible action on the holomorphic fermions), this transformation realizes the Kummer involution, and the orbifolded theory is a heterotic non-linear sigma model on a K3 surface, at a singular $T^4/\Z_2$ point in moduli space. More generally, one may choose alternative actions on the left-moving degrees of freedom which, despite the lack of a geometric interpretation, still define consistent string backgrounds. Such constructions yield six-dimensional supergravity theories with $T=1$ tensor multiplet and supersymmetry reduced from sixteen to eight supercharges.\\

Let us examine a concrete example taken from~\cite{Baykara:2023plc}. Consider the $c_\tL=20$, $c_\tR=4$ Narain theory at a rational point in moduli space, where the sublattice $\Lambda_\tL\subset\Gamma$ encoding purely left-moving vertex operators is isomorphic to the $\mathrm{SO}(8)\times\mathrm{E}_8\times\mathrm{E}_8$ root lattice. This symmetry can be realized, for instance, in a Narain compactification with vanishing Wilson lines and a suitable choice of metric and B-field for the $4$-torus. At such an enhancement point, any vector $\bp\in\Gamma$ of the Narain lattice admits a decomposition $\bp=\bp_\tL+\bp_\tR$ into time-like and space-like components, where $\bp_\tL$ can be identified with a weight of $\mathrm{SO}(8)\times\mathrm{E}_8\times\mathrm{E}_8$. We consider the asymmetric orbifold defined by the following shift vector and lattice symmetry\footnote{To see that $\varphi_g$ is a lattice automorphism, note that we can write $\varphi_g(\bp)= 2\bp_\tL-\bp$. While $\bp_\tL$ does not necessarily lie in the Narain lattice, $2\bp_\tL$ always does.}
\begin{align}
s_g &=\tfrac{1}{2}(\boldsymbol{\omega},\boldsymbol{\omega}) ~,&
\varphi_g(\bp)&= \bp_\tL-\bp_\tR~,
\end{align}
with $\boldsymbol{\omega}$ a root of the $\mathfrak{e}_8$ lattice. This orbifold data reproduces the Kummer action on right-moving scalars in~\eqref{eq:Kummerorbifold}, and is compatible with the consistency conditions of~\cite{Narain:1986qm}.

The resulting spacetime theory is a six-dimensional $\mathcal{N}=(1,0)$ supergravity model with $T=1$ tensor multiplet---the universal tensor containing the heterotic dilaton. The gauge symmetry,
\begin{align}
\mathfrak{g} = \mathfrak{su}(2)+\mathfrak{su}(2)+\mathfrak{so}(8)+\mathfrak{e}_7+\mathfrak{e}_7~,
\label{eq:6dT=1asymorb1gaugealgebra}
\end{align}
is realized in the worldsheet CFT as a left-moving KM algebra, with all components at level $1$. Indeed, the holomorphic currents all arise in the untwisted sector from those of the un-orbifolded theory that survive the projection: each $\mathfrak{e}_8$ factor gives rise to $\mathfrak{su}(2)+\mathfrak{e}_7$, while the $\mathfrak{so}(8)$ currents from the $4$-torus remain invariant. The hypermultiplet content, derived in~\cite{Baykara:2023plc}, is
\begin{align}
\mathscr{H} = \quad &(\rep{1},\rep{2},\rep{1},\rep{1},\rep{56}) + (\rep{1},\rep{2},\rep{1},\rep{56},\rep{1}) + (\rep{2},\rep{1},\rep{1},\rep{1},\rep{56}) + (\rep{2},\rep{1},\rep{1},\rep{56},\rep{1})
\nonumber\\
+ \,&(\rep{2},\rep{2},\rep{8^v},\rep{1},\rep{1}) + (\rep{2},\rep{2},\rep{8^c},\rep{1},\rep{1}) + (\rep{2},\rep{2},\rep{8^s},\rep{1},\rep{1})~.
\label{eq:6dT=1asymorb1matter}
\end{align}
In particular, the spacetime massless spectrum is invariant under a $\Z_2\subset (\Z_2)^6$ center symmetry combining the centers of all $\mathrm{SU}(2)$ and $\mathrm{E}_7$ factors. To determine whether any massive supergravity state breaks this $\Z_2$ symmetry, one must analyze the heterotic partition function. In this perturbative setting, it can be obtained using modular orbits~\cite{Narain:1986qm}; we compute it in appendix~\ref{app:modularorbits}, and verify that the full heterotic spectrum is invariant under this $\Z_2$ center symmetry. It follows that the corresponding one-form symmetry is gauged, and that the global form of the gauge group is
\begin{align}
G = \left( \mathrm{SU}(2)\times\mathrm{SU}(2)\times\mathrm{E}_7\times\mathrm{E}_7 \right)/\Z_2 \times\mathrm{Spin}(8)~.
\end{align}
The worldsheet current algebra must therefore be extended by an integral spin simple current. Indeed, the current algebra admits a simple current in the primary representation
\begin{align}
(\rep{2},\rep{2},\rep{1},\rep{56},\rep{56})~,
\label{eq:simplecurrenta1a1e7e7}
\end{align}
with conformal dimension $h_\tL=2$, that precisely generates this $\Z_2$ center symmetry. This state appears in the heterotic spectrum as a GSO-even higher-spin current, confirming the enlargement of the left-moving algebra.

\subsubsection*{A model with mixed KMV extensions}
Let us now turn to a second supergravity model from~\cite{Baykara:2023plc}, which is obtained as an asymmetric orbifold of the same parent Narain theory as in the previous example, but with a left-moving action that now corresponds to a CHL-like involution exchanging the two $\mathrm{E}_8$ factors. The resulting six-dimensional $\mathcal{N}=(1,0)$ supergravity has gauge group and hypermultiplet content
\begin{align}
G  &=  \mathrm{Spin}(8)\times\mathrm{E}_8~,&
\mathscr{H} &= 2 \cdot (\rep{1},\rep{248}) + (\rep{8^v},\rep{1}) + (\rep{8^c},\rep{1}) + (\rep{8^s},\rep{1})~.
\label{eq:6dsugraD4E8}
\end{align}
The gauge group is simply-connected, as the $\Z_2\times\Z_2$ center symmetries of $\mathrm{Spin}(8)$ are already broken by the massless spectrum.

Despite the absence of a gauged one-form symmetry associated with the center of $G$, the left-moving algebra of the heterotic worldsheet CFT exhibits additional structure not immediately visible in the six-dimensional effective theory. The current algebra consists of $\mathfrak{so}(8)$ at level $1$ and $\mathfrak{e}_8$ at level $2$, whose Sugawara central charges are $4$ and $\frac{31}{2}$, respectively. Subtracting these contributions from the total left-moving central charge leaves a residual factor with $c_\tL=\frac{1}{2}$, which can be identified with the Ising model.
\begin{table}
\centering
{\renewcommand{\arraystretch}{1.2}
$\begin{array}{c|ccc}
\multicolumn{4}{c}{\mathfrak{e}_8}\\  \hline 
\rep{r} & \rep{1} & \rep{248} & \rep{3875} \\ \hline
h_\tL & 0 & \frac{15}{16} & \frac{3}{2} \\
\end{array}
\qquad\qquad
\begin{array}{c|ccc}
\multicolumn{4}{c}{\text{Ising}}\\  \hline 
\rep{r} & I & \sigma & \varepsilon\\ \hline
h_\tL & 0 & \frac{1}{16} & \frac{1}{2} \\
\end{array}$}
\caption{Primary representations of the $\mathfrak{e}_8$ current algebra at level $2$ and of the Ising model}
\label{tab:e8Isingcontent}
\end{table}
The $\mathfrak{e}_8$ KM algebra at level $2$ and the Ising model exhibit similar structure: both possess three primary operators (listed in table~\ref{tab:e8Isingcontent}), including a $\mathbb{Z}_2$ simple current.\footnote{The underlying reason is that the Ising model and the $\mathfrak{e}_8$ level $2$ algebra are complementary in the sense of~\cite{Schellekens:1990xy}: their characters combine into a single $c=16$ character, corresponding to a theory with a unique primary field, namely the $\mathfrak{e}_8 + \mathfrak{e}_8$ lattice CFT~\cite{Forgacs:1988iw}.} The $\mathfrak{e}_8$ simple current lies in the representation $\rep{3875}$; despite arising from a current algebra, it is not associated with a center symmetry, since $\mathrm{E}_8$ has trivial center. The Ising simple current corresponds to the primary $\varepsilon$ and generates the $\mathbb{Z}_2$ symmetry $\sigma \to -\sigma$. Their product, the operator in $\rep{3875}\otimes\varepsilon$, defines a $\Z_2$ simple current of the Kac--Moody--Virasoro (KMV) algebra with conformal dimension $h_\tL=2$. In the heterotic worldsheet CFT associated with~\eqref{eq:6dsugraD4E8}, the left-moving algebra is extended by this spin-$2$ current. This is reflected in the elliptic genus of the $c_\tL=20$, $c_\tR=6$ internal CFT, which takes the form
\begin{align}
\mathbb{E} = \quad &2 \left( \chi^{\mathfrak{e}_8}_\rep{1}\chi^{\text{I}}_{I} + \chi^{\mathfrak{e}_8}_\rep{3875}\chi^{\text{I}}_{\varepsilon} \right) \chi^{\mathfrak{d}_4}_\rep{1}
\nonumber \\
- & 2 \left( \chi^{\mathfrak{e}_8}_\rep{1}\chi^{\text{I}}_{\varepsilon} + \chi^{\mathfrak{e}_8}_\rep{3875}\chi^{\text{I}}_{I} \right) ( \chi^{\mathfrak{d}_4}_\rep{8^v} + \chi^{\mathfrak{d}_4}_\rep{8^c} + \chi^{\mathfrak{d}_4}_\rep{8^s} )
\nonumber \\
- & 4  \,\chi^{\mathfrak{e}_8}_\rep{248}\chi^{\text{I}}_{\sigma} \chi^{\mathfrak{d}_4}_\rep{1}~,
\end{align}
and depends only on characters of the extended KMV algebra.\footnote{This extended algebra has four primary representations: in terms of their KMV content, they consist of the identity $\rep{1}\otimes I+\rep{3875}\otimes\varepsilon$, one primary in the representation $\rep{1}\otimes \varepsilon+\rep{3875}\otimes I$, and two primaries in the representation $\rep{248}\otimes \sigma$.}\\

This simple current extension of the left-moving KMV algebra may appear more exotic than the current algebra extensions discussed earlier: it is not associated with any center one-form symmetry, and lacks a straightforward geometric interpretation due to the intrinsically non-geometric nature of the asymmetric orbifold. Nevertheless, the same spin-$2$ KMV current arises in a much more familiar setting: the nine-dimensional CHL string. Indeed, as shown in~\cite{Forgacs:1988iw} (see also~\cite{Collazuol:2024kzl} for a lucid presentation), the current algebra of the ten-dimensional $\mathrm{E}_8\times\mathrm{E}_8$ heterotic string can be decomposed with respect to the $\Z_2$ outer automorphism. A diagonal $\mathfrak{e}_8$ factor at level $2$ is invariant under this symmetry, together with a residual factor described by the coset
\begin{align}
\frac{\mathfrak{e}_{8,1}+\mathfrak{e}_{8,1}}{\mathfrak{e}_{8,2}}~,
\end{align}
which is isomorphic to the Ising model. The spin-$2$ simple current in $\rep{3875}\otimes\varepsilon$ survives the CHL orbifold, and gives rise to an extension of the $c_\tL=17$ left-moving algebra of the internal CFT. In this context, it is clear that the spin-$2$ simple current is a remnant of the outer automorphism symmetry of the heterotic $(\mathrm{E}_8\times\mathrm{E}_8)\rtimes\Z_2$ gauge group.

The preceding observations illustrate that, beyond the well-understood Kac--Moody simple currents emphasized in this paper, more general extensions of the left-moving algebras can arise in worldsheet CFT of probe strings, which may encode subtler discrete structures of the bulk theory. In particular, ``mixed'' KMV extensions---combining current algebras and minimal models---appear to be rather generic. For instance, at several maximally enhanced points of the eight-dimensional CHL string, the left-moving algebra decomposes into a current algebra associated with the gauge symmetry and a commuting unitary minimal model,\footnote{This was already noted in~\cite{Font:2021uyw}, see in particular their appendix 2, where the appearance of unitary minimal models is used as a consistency check of the CHL construction.} and may admit higher-spin simple currents intertwining the two sectors. Such KMV extensions have also recently appeared in six-dimensional supergravity theories without tensor multiplets, where certain probe BPS strings are described by rational CFTs exhibiting similar structures~\cite{Lockhart:2025lea,Lockhart:2026xml}. These KMV algebra extensions remain largely unexplored, and it would be very interesting to develop a clearer physical interpretation from the bulk perspective. One possible avenue may be to study their realization in dual F-theory constructions (for instance, for eight-dimensional CHL models discussed above), where the left-moving extension might be related to discrete geometric data of the underlying Calabi--Yau compactification. 

\section{Gauge group topology in 6d supergravity}
\label{sec:6dBPSstrings}
The heterotic compactifications discussed in the previous section provided a perturbative setting in which the correspondence between worldsheet simple current extensions and gauged one-form symmetries in the bulk could be verified explicitly. We now turn our focus to six-dimensional $\mathcal{N}=(1,0)$ supergravity theories, whose spectrum contains a variety of BPS strings which in suitable cases can serve as faithful probes of the gauge symmetry and provide information about the global structure of the gauge group.

We start in section~\ref{subsec:faithful6d} by reviewing general properties of 6d supergravity strings. We then study in section \ref{subsec:rational} a class of 6d theories without tensor multiplets identified in~\cite{Lockhart:2025lea,Lockhart:2026xml}, whose primitive BPS strings are described by a rational CFT. For these theories we will be able to determine the global form of the gauge group directly from the structure of the chiral algebra. Finally, in section~\ref{subsec:5dreduction} we study circle compactifications of six-dimensional theories with gauged center one-form symmetries. In this setting, we provide a new perspective on the consistency conditions introduced in~\cite{Kim:2024tdh} by pointing out a relation between the BPS particles considered there in order to modify the BPS cone in the resulting five-dimensional theory, and simple currents that extend the worldsheet CFT of faithful string probes.

\subsection{Faithful supergravity strings in 6d}
\label{subsec:faithful6d}
In six-dimensional $\mathcal{N}=(1,0)$ supergravity, BPS strings are extended objects charged under the two-form gauge fields of the $T$ tensor multiplets and of the gravity multiplet. Their charges $Q$ take values in the string charge lattice
\begin{align}
\Gamma_\BPS \subset \mathbb{R}^{1,T}~,
\end{align}
an integral self-dual lattice of signature $(1,T)$. 
For lattice vectors $x,y\in\Gamma_\BPS$, we write their inner product as $x\cdot y=\Omega_{\mu\nu}\, x^\mu y^\nu$, with $\Omega$ the Lorentzian lattice metric and $\mu,\nu=0,1,\dots,T$. The anomaly coefficients
\begin{align}
a,~b_1,~\dots,~b_N\quad\in\Gamma_\BPS~,
\end{align}
that characterize the Green--Schwarz coupling of the theory through~\eqref{eq:6danomalyX}, span an integral sulattice of $\Gamma_\BPS$~\cite{Kumar:2010ru, Seiberg:2011dr, Monnier:2017klz}, which is usually referred to as the anomaly lattice. At low energy, the worldsheet theories of BPS strings are captured by families of two-dimensional (0,4)  superconformal field theories. A distinguished class is given by supergravity strings, for which the $\mathfrak{su}(2)$ superconformal R-symmetry is identified with a subalgebra of the $\mathfrak{so}(4)= \mathfrak{su}(2)_\tL + \mathfrak{su}(2)_\tR$ rotation symmetry  transverse to the string. For these strings, the six-dimensional gauge symmetry and the above $\mathfrak{su}(2)_\tL$ algebra are expected to be realized on the worldsheet as left-moving current algebras~\cite{HaghighatMurthyVafaVandoren2015}. Anomaly inflow further determines the central charges of the string CFT in terms of the six-dimensional anomaly coefficients~\cite{Kim:2019vuc}: 
\begin{align}
c_\tL &= 3\,Q\cdot Q - 9\,a\cdot Q + 2~,&
c_\tR &= 3\,Q\cdot Q - 3\,a\cdot Q~,
\label{eq:centralchargessupergravitystring}
\end{align}
as well as the current algebra levels 
\begin{align}
k_\tL &= \tfrac{1}{2}\left(Q\cdot Q +a\cdot Q + 2\right)~,&
k_\alpha &= Q\cdot b_\alpha~.
\end{align}
Here, $k_\tL$ is the level of the affine $\mathfrak{su}(2)_\tL$ algebra while $k_\alpha$ is the level of the current algebra for the bulk gauge factor $\mathfrak{g}_\alpha$. In the expression for the central charges above we have subtracted the center of mass contribution, given by $c_\tL^{\text{com}}=4$, $c_\tR^{\text{com}}=6$, which corresponds to a free decoupled hypermultiplet.  These quantities are subject to unitarity constraints of the worldsheet theory. In particular, one requires
\begin{align}
c_\tL &\geq 0 ~,&
c_\tR &\geq 0 ~,&
k_\tL &\geq 0~,&
k_\alpha &\geq 0 ~,
\end{align}
and the gauge contribution $c_\mathfrak{g}$ to the left-moving central charge, obtained from the Sugawara construction, is constrained by unitarity to satisfy
\begin{equation}
c_\mathfrak{g} \leq c_\tL ~.
\end{equation}
Violations of the unitarity bound on the central charge signal inconsistencies in the corresponding bulk supergravity theory.  This perspective has proven particularly useful in relegating many anomaly-free six-dimensional supergravity models to the swampland~\cite{Kim:2019vuc,Lee:2019skh,Tarazi:2021duw,Hamada:2023zol,Hamada:2024oap}.

For a six-dimensional supergravity theory with semi-simple gauge group $G$, a supergravity string of charge $Q$ provides a faithful probe of the gauge symmetry when the induced current algebra on its worldsheet has strictly positive level for every simple factor $G_\alpha\subset G$. Equivalently, the charge vector must satisfy
\begin{align}
Q\cdot b_\alpha > 0 ~,
\label{eq:6dfaithfulstring}
\end{align}
for each simple component, where $b_\alpha$ denotes the corresponding anomaly coefficient. Such a string can then be used to probe the global structure of the gauge group: the appearance of a KM simple current in its worldsheet CFT signals that the corresponding center one-form symmetry is gauged in the six-dimensional bulk theory. More generally, useful information can still be extracted from strings that satisfy~\eqref{eq:6dfaithfulstring} only for a subset of the simple factors. Although such strings are not sensitive to the full global form of the gauge group, a simple current extension of their worldsheet theory still reflects the gauging of a one-form symmetry acting on those particular components.\\

Let us illustrate this more concretely by going back to the six-dimensional asymmetric orbifold model of section~\ref{subsec:6dasymmetric} with gauge symmetry $\mathfrak{g} = \mathfrak{su}(2)+\mathfrak{su}(2)+\mathfrak{so}(8)+\mathfrak{e}_7+\mathfrak{e}_7$ and matter content given in \eqref{eq:6dT=1asymorb1matter}. In $T=1$ supergravity theories, the string charge lattice $\Gamma_\BPS$ is a rank-$2$ self-dual lattice of signature $(1,1)$. Up to isometry, there are precisely two such inequivalent lattices, which we refer to as \emph{even} and \emph{odd}, whose Gram matrices are respectively
\begin{align}
\Omega_{\text{even}} &=
\begin{pmatrix}
0 & 1 \\
1 & 0
\end{pmatrix}~,&
\Omega_{\text{odd}}  &=
\begin{pmatrix}
1 & 0 \\
0 & -1
\end{pmatrix}~.
\end{align}
By explicit computation we find that the asymmetric orbifold model of section \ref{subsec:6dasymmetric} requires us to choose the even lattice. The $T=1$ anomaly cancellation condition for the gravitational anomaly coefficient $a$ fixes its norm to be $a\cdot a = 8$. Hence, it can be embedded in the even unimodular lattice as
\begin{align}
a =
\begin{pmatrix}
-2 \\
-2
\end{pmatrix}~.
\end{align}
With this realization, the string charge vector
\begin{align}
Q=
\begin{pmatrix}
1\\
0
\end{pmatrix}
\label{eq:6dchargeheterotic}
\end{align}
describes a heterotic string for which $Q\cdot Q = 0$ and $(c_\tL,c_\tR)=(20,6)$. The inner products between the anomaly coefficients are fixed by the matter representations, and determine
\begin{align}
b_{\mathfrak{su}(2)} &=
\begin{pmatrix}
12\\
1
\end{pmatrix}~,&
b_{\mathfrak{so}(8)} = b_{\mathfrak{e}_7}&=
\begin{pmatrix}
0\\
1
\end{pmatrix}~.
\label{eq:bb}
\end{align}
Here, $b_{\mathfrak{su}(2)}$ denotes the anomaly coefficient for each of the two $\mathfrak{su}(2)$ factors, and similarly $b_{\mathfrak{e}_7}$ denotes the coefficient for the two $\mathfrak{e}_7$ factors. From~\eqref{eq:6dchargeheterotic} and~\eqref{eq:bb} one recovers the correct levels $k_\alpha = 1$ for the Kac--Moody algebra arising on the heterotic string.

More generally, in this example we can consider different choices of supergravity strings labeled by a charge vector
\begin{align}
Q = \begin{pmatrix}
l_1 \\ l_2 \\
\end{pmatrix}~,
\label{eq:genericcharge}
\end{align}
with $l_1$ and $l_2$ non-negative coprime integers. The dynamics of such strings can be intricate: at low energies, the worldsheet theory may decompose into several interacting two-dimensional SCFTs corresponding to the separation of the classical moduli space of vacua into disconnected branches. In particular, besides the $\R^4$ center-of-mass multiplet, additional zero modes may arise from instanton moduli spaces associated with some of the gauge group factors. When the theory admits a non-instantonic branch, however, the resulting two-dimensional CFT may still serve as a useful probe of the six-dimensional bulk theory. Let us return to the $T=1$ model introduced above, whose gauge group is
\begin{align}
G = \left( \mathrm{SU}(2)\times\mathrm{SU}(2)\times\mathrm{E}_7\times\mathrm{E}_7 \right)/\Z_2 \times\mathrm{Spin}(8)~.
\end{align}
On the non-instantonic branch of the worldsheet theory for a string of charge~\eqref{eq:genericcharge}, the various gauge factors are realized as left-moving current algebras with levels
\begin{align}
k_{\mathfrak{su}(2)} &=l_1+12l_2~,&
k_{\mathfrak{so}(8)} = k_{\mathfrak{e}_7} &= l_1~.
\end{align}
Since the six-dimensional gauge group is not simply connected, we expect that on every faithful probe string the worldsheet chiral algebra is extended by the $\Z_2$ simple current
\begin{align}
J = J_{\mathfrak{su}(2)} J_{\mathfrak{su}(2)}' J_{\mathfrak{e}_7} J_{\mathfrak{e}_7}'~.
\end{align}
As a consistency check, this operator should appear with integral conformal weight for any faithful probe string. This is indeed the case. The individual simple currents have conformal weights
\begin{align}
h_\tL(J_{\mathfrak{su}(2)})=h_\tL(J_{\mathfrak{su}(2)}') &= \frac{l_1}{4}+3l_2~,&
h_\tL(J_{\mathfrak{e}_7})= h_\tL(J_{\mathfrak{e}_7})&= \frac{3l_1}{4}~,
\end{align}
and therefore $h_\tL(J)=2l_1+6l_2$ is always integer. 

It is also instructive to consider the string with charge vector
\begin{align}
Q = \begin{pmatrix}
0 \\ 1 \\
\end{pmatrix}~.
\end{align}
This choice of charge does not give rise to a faithful probe: at low energy, it is expected to decompose into disconnected CFTs, including  instantonic branches with zero modes associated to $\mathfrak{so}(8)$ and $\mathfrak{e}_7$ instantons, due to the identity $Q=b_{\mathfrak{so}(8)}=b_{\mathfrak{e}_7}$. The non-instantonic branch, described by a compact CFT, does not couple to the $\mathfrak{so}(8)$ or $\mathfrak{e}_7$ factors (the corresponding levels vanish), and its worldsheet current algebra reduces to two $\mathfrak{su}(2)$ factors at level $12$. Since it transforms trivially under $\mathrm{E}_7\times\mathrm{E}_7$, it is automatically neutral under the corresponding $\Z_2\times\Z_2$ center symmetries. Because, in the bulk supergravity, the one-form symmetry combining the centers of the two $\mathrm{SU}(2)$ and the two $\mathrm{E}_7$ factors is gauged, the string must also be neutral under the diagonal $\Z_2\subset\Z_2\times\Z_2$ center symmetry of $\mathrm{SU}(2)\times\mathrm{SU}(2)$. Accordingly, its worldsheet theory must admit a simple current extension by $J_{\mathfrak{su}(2)} J_{\mathfrak{su}(2)}'$. This is again consistent, since this operator appears in the worldsheet CFT of the string with integral conformal weight $h_\tL=6$. Hence, this unfaithful string still detects the presence of a gauged $\Z_2$ one-form symmetry in the bulk. However, because it is blind to the $\mathrm{Spin}(8)$ and $\mathrm{E}_7$ sectors, it cannot determine how this one-form symmetry is embedded into the full gauge group, and therefore, unlike what happens in the case of faithful strings, only encodes partial information about the global structure of $G$.

\subsection{Supergravity without tensors and the hyperplane string}
\label{subsec:rational}
We now focus on specific examples of faithful supergravity strings that arise in theories without tensor multiplets. We concentrate on a class of rational models introduced in~\cite{Lockhart:2025lea} for which we will infer the topology of the six-dimensional gauge group. Constraints on the landscape of six-dimensional $\mathcal{N}=(1,0)$ supergravity theories with at least one tensor multiplet can be obtained by analyzing the properties of a class of strings known as H-strings~\cite{Kim:2024eoa}. In theories without tensor multiplets, such strings are absent from the BPS spectrum;\footnote{In this sense, theories with $T=0$ are distinguished in the 6d supergravity landscape: theories with $T\geq 1$ always contain an H-string, apart from an isolated case with $T=9$~\cite{Kim:2024eoa}.} nevertheless, the theory still contains supergravity strings which provide alternative probes of the six-dimensional physics~\cite{Kim:2024eoa,Lockhart:2025lea,Lockhart:2026xml}.
In this case, the string charge lattice has rank one, $\Gamma_\BPS \simeq \mathbb Z$, and the minimal BPS string is a primitive object of charge $Q=1$. The worldsheet theory of this probe string
has central charges
\begin{align}
c_\tL &= 32~,&
c_\tR &= 12 ~,
\end{align}
and the gauge symmetry is realized as a left-moving current algebra with levels are given by the gauge anomaly coefficients, $k_\alpha=b_\alpha$.
In F-theory, six-dimensional theories without tensors arise through compactification on an elliptically fibered Calabi--Yau threefold over a base $\mathbb P^2$. The generator of the charge lattice is identified with the hyperplane class $[L]\in H_2(\mathbb{P}^2, \mathbb{Z})$, and the corresponding primitive BPS string is obtained from a D3-brane wrapped on $[L]$; we refer to it as the hyperplane string or L-string~\cite{Lockhart:2025lea,Lockhart:2026xml}. If the elliptic fiber does not degenerate along $[L]$, the L-string admits a large-volume description as a heterotic $(0,4)$ sigma model whose target space is a $T^4$-fibration over $\mathbb{P}^2$~\cite{HaghighatMurthyVafaVandoren2015,Kim:2016foj,Shimizu:2016lbw}. In this regime, the left-moving degrees of freedom associated with the gauge symmetry are expected to arise from chiral fermions valued in a holomorphic bundle over the hyper-Kähler target space. The hyperplane string provides a particularly effective probe of supergravity theories with $T=0$. Indeed, even in the absence of an explicit F-theory realization, evidence from five-dimensional compactification suggests that the L-string still exists in any consistent $T=0$ theory~\cite{Kim:2024eoa}, and it is expected to remain a reliable probe regardless of whether a specific model can be realized in terms of a conventional F-theory compactification on an elliptic threefold~\cite{Lockhart:2025lea}. 

It was argued in~\cite{Lockhart:2025lea} that the moduli space of the hyperplane string CFT is expected to probe the full set of six-dimensional $T=0$ supergravity theories---in particular, marginal deformations of the CFT can interpolate between theories whose gauge group are related by Higgsing. Within this moduli space, five distinguished points were identified in~\cite{Lockhart:2025lea}, at which the two-dimensional CFT living on the hyperplane string simplifies significantly. At these points, which are associated to five distinct anomaly-free $T=0$ supergravity models, the gauge symmetry is realized on the left-moving side of the hyperplane string CFT by the following current algebras:
\begin{align}
&\mathfrak{e}_{6,8}~,
&\mathfrak{a}_{7,8}~,&
&\mathfrak{a}_{5,6}+ \mathfrak{d}_{4,6}~,&
&\mathfrak{a}_{3,6}+\mathfrak{a}_{3,6}+ \mathfrak{e}_{7,2}~,&
&\mathfrak{a}_{3,10}+ \mathfrak{a}_{11,2}~,&
\label{eq:rationalmodels}
\end{align}
where the subscripts reflect the KM levels. The corresponding six-dimensional theories belong to the set of anomaly-free supergravity models identified in~\cite{Hamada:2023zol,Hamada:2024oap}. What singles out these theories is that, for all five examples, the residual left-moving central charge
\begin{equation}
c_\tL^{\rm res} = c_\tL - c_\mathfrak{g}~ ,
\end{equation}
obtained after subtracting the Sugawara contribution of the gauge current algebra, coincides with that of a unitary Virasoro minimal model. Hence, the CFT must be rational, with all possible left-moving excitations on the string encoded by the finite set of primary representations of the left-moving Kac--Moody--Virasoro algebra~\cite{Lockhart:2025lea, Lockhart:2026xml}. 

\subsubsection*{A supergravity model with $G=\mathrm{SU}(8)/\Z_4$}
Among the five rational points for the hyperplane string listed in~\eqref{eq:rationalmodels}, the second corresponds to a $T=0$ supergravity theory whose gauge group $G$ and hypermultiplet representation $\mathcal{H}$ are given---temporarily ignoring the precise global form of the gauge group---by
\begin{align}
G &\approx \mathrm{SU}(8)~,&
\mathscr{H} &= \rep{336}~.
\label{eq:6dSU(8)model}
\end{align}
This model, first identified in~\cite{Kumar:2010am}, satisfies all anomaly cancellation conditions; nevertheless, no string-theoretic realization is presently known. In fact, it appears to lack some of the basic features expected from an F-theory construction: it violates the Kodaira constraint required for a consistent elliptically fibered Calabi--Yau realization, and it contains no neutral hypermultiplet that could be interpreted as the universal hypermultiplet associated with the overall volume of the base of the elliptic fibration. Nevertheless, the model seems to satisfy all standard consistency requirements---in particular, the unitarity constraints arising from the hyperplane string CFT are satisfied: the $\mathfrak{su}(8)$ gauge symmetry is realized on the worldsheet as a current algebra at level $8$, while the residual left-moving chiral algebra has central charge
\begin{align}
c_\tL^{\mathrm{res}} = \frac{1}{2}~,
\end{align}
as already observed in~\cite{Kim:2019vuc}, which coincides with the central charge of the Ising model. Under the assumption that this model is consistent and admits a unitary CFT, primary representations of the left-moving chiral algebra on the string are labeled by their $\mathfrak{su}(8)$ representation together with a choice of Virasoro primary of the Ising model (see table~\ref{tab:e8Isingcontent} above).

The massless fields of the supergravity theory~\eqref{eq:6dSU(8)model} are neutral under a
\begin{align}
\Z_4\subset\mathrm{SU}(8)
\end{align}
subgroup of the center symmetries. Determining whether the corresponding one-form symmetry is gauged or broken, however, in principle requires substantially more information than the massless spectrum alone. This is a difficult task especially given that we cannot rely on any conventional string-theoretic realization to gain insight into this theory. Nevertheless, we will argue that the $\Z_4$ one-form symmetry is gauged by analyzing the chiral algebra of the hyperplane string CFT. As reviewed in section~\ref{subsec:KMcenter}, the $\mathfrak{su}(8)$ KM algebra at level $8$ admits a $\Z_8$ simple current symmetry, generated by the primary representation $\rep{6435}=[8000000]$. Its action on affine primary representations is described by the outer automorphism
\begin{align}
\text{g}\quad:\quad \rep{r} = [l_1l_2\cdots l_7]\quad\mapsto\quad \text{g}(\rep{r})=[l_0l_1\cdots l_6]
\end{align}
of the $\mathfrak{su}(8)$ extended Dynkin diagram, where the affine Dynkin label $l_0$ is defined by the relation $l_0+l_1+\dots+l_7=8$. This simple current has conformal weight
\begin{align}
h_\tL=\frac{7}{2}~.
\end{align}
The $\Z_4$ subroup of the center is generated instead by the primary representation $\rep{2147145}=[0800000]$, whose conformal weight is 
\begin{align}
h_\tL=6~.
\end{align}
Since this value is integral, the corresponding simple current is compatible with locality and can, in principle, extend the left-moving chiral algebra of the hyperplane string CFT. Establishing whether this extension is actually realized, however, requires more detailed information about the full CFT spectrum.\\

Fortunately, rationality of the hyperplane string CFT allows us to extract substantial information about its spectrum. In particular, the elliptic genus---which computes, in the Ramond--Ramond Hilbert space, the net number of antiholomorphic Ramond ground states---can be written entirely in terms of characters of the left-moving $c_\tL=32$ chiral algebra. Its explicit form was determined in~\cite{Lockhart:2026xml} by exploiting modularity: it is written as
\begin{align}
\mathbb{E}=\sum_{\rep{r}}\left(C_{\rep{r},I}\,\chi^{\text{I}}_{I}+C_{\rep{r},\sigma}\,\chi^{\text{I}}_{\sigma}+C_{\rep{r},\varepsilon}\,\chi^{\text{I}}_{\varepsilon}\right)\chi^{\mathfrak{su}(8)}_{\rep{r}}~,
\end{align}
where the sum runs over the $6435$ integrable primary representations of the $\mathfrak{su}(8)$ current algebra at level $8$, and the coefficients $C_{\rep{r},I}$, $C_{\rep{r},\sigma}$, and $C_{\rep{r},\varepsilon}$ are integers. Here, $\chi^{\mathfrak{su}(8)}_{\rep{r}}$ denotes the affine $\mathfrak{su}(8)$ character, while $\chi^{\text{I}}_I$, $\chi^{\text{I}}_\sigma$, $\chi^{\text{I}}_\varepsilon$ are the three Ising model characters. Only a small subset of $\mathfrak{su}(8)$ primaries appears with non-vanishing multiplicity. Remarkably, every such representation is neutral under the $\Z_4\subset\mathrm{SU}(8)$ center symmetry. Moreover, the multiplicities satisfy
\begin{align}
C_{\text{g}^2(\rep{r}),I} &= C_{\rep{r},I}~,&
C_{\text{g}^2(\rep{r}),\sigma} &= C_{\rep{r},\sigma}~,&
C_{\text{g}^2(\rep{r}),\varepsilon} &= C_{\rep{r},\varepsilon}~.
\end{align}
As a result, the elliptic genus naturally reorganizes into extended characters\footnote{The extended characters are defined for representations $\rep{r}$ neutral under the $\Z_4$ center symmetry. When $\rep{r}$ has a non-trivial stabilizer, i.e. when $\text{g}^2(\rep{r})=\rep{r}$ or $\text{g}^4(\rep{r})=\rep{r}$,  the definition of $\mathcal{X}^{\mathfrak{su}(8)}_{\rep{r}}$ includes an additional overall factor of $\frac{1}{4}$ or $\frac{1}{2}$, respectively.}
\begin{align}
\mathcal{X}^{\mathfrak{su}(8)}_{\rep{r}} = \chi^{\mathfrak{su}(8)}_{\rep{r}} + \chi^{\mathfrak{su}(8)}_{\text{g}^2(\rep{r})} + \chi^{\mathfrak{su}(8)}_{\text{g}^4(\rep{r})} + \chi^{\mathfrak{su}(8)}_{\text{g}^6(\rep{r})}~.
\end{align}
This strongly suggests that the left-moving chiral algebra is extended by the spin-$6$ simple current associated with the $\Z_4$ center symmetry, whose fusion product corresponds to $\text{g}^2$. From the spacetime perspective, this means that the corresponding $\Z_4$ one-form symmetry is gauged, so that the global form of the gauge group is
\begin{align}
G = \mathrm{SU}(8)/\Z_4~.
\end{align}
In fact, the worldsheet theory of the probe string appears to also contain more refined information which from the spacetime perspective is less straightforward to interpret. Namely, the Kac--Moody simple currents associated with the $\mathfrak{su}(8)$ factor are not the only simple currents of the theory: the Ising model contributing to the residual left-moving algebra also carries its own non-trivial $\Z_2$ simple current, corresponding to the primary field $\varepsilon$, whose conformal weight is $1/2$. Its action on the Ising primaries is particularly simple: it leaves $\sigma$ fixed while exchanging the identity $I$ and $\varepsilon$. From the intrinsic viewpoint of the CFT, this simple current is on exactly the same footing as the Kac--Moody simple currents, and it is therefore natural to ask whether the chiral algebra could be extended by simple currents of the full $c_\tL=32$ chiral algebra. Of course, the Ising simple current alone cannot define a consistent extension, since its conformal weight is half-integral. However, one can consider the composite simple current $J$, associated with the Kac--Moody--Virasoro representation
\begin{align}
\rep{6435}\otimes \varepsilon~,
\end{align}
whose conformal weight is instead
\begin{align}
h_\tL = 4~.
\end{align}
Since this is integral, $J$ can consistently appear as a local holomorphic operator and could, in principle, enlarge the chiral algebra. Interestingly, the elliptic genus suggests that such extension is indeed realized. In particular, the multiplicities appearing in the character decomposition satisfy
\begin{align}
C_{\text{g}(\rep{r}),\varepsilon} &= C_{\rep{r},I}~,&
C_{\text{g}(\rep{r}),\sigma} &= C_{\rep{r},\sigma}~,&
C_{\text{g}(\rep{r}),I} &= C_{\rep{r},\varepsilon}~,
\end{align}
as required in order for the spectrum to organize into orbits of the simple current $J$. This suggests that the left-moving algebra of the hyperplane string is indeed enhanced by this holomorphic spin-$4$ operator. Note that the action of $J$ generates a $\Z_8$ symmetry, and that its square acts trivially on the Ising sector and reproduces precisely the $\Z_4\subset\Z_8$ center symmetry of $\mathrm{SU}(8)$. In this sense, the gauged center symmetry of the bulk theory appears as only a substructure of a larger symmetry visible on the worldsheet.

It would be very interesting to understand the spacetime interpretation of this enhanced simple current structure. One possibility is that it reflects a more subtle bulk generalized symmetry, combining the ordinary $\Z_8$ center one-form symmetry of $\mathrm{SU}(8)$ with an additional discrete action not immediately visible from the gauge theory alone. The appearance of the Ising factor is suggestive in this respect: it closely resembles the structure encountered in CHL compactifications~\cite{Forgacs:1988iw}, where a residual $\Z_2$ symmetry can be understood as a vestige of the outer automorphism exchanging the two $\mathrm{E}_8$ factors of the heterotic string. It is therefore natural to ask whether a similar interpretation might be relevant here. Similar questions could also be investigated in less exotic heterotic settings, such as in CHL compactifications at points of maximal enhancement, where comparable mixing between current algebras and Virasoro minimal model can  occur~\cite{Font:2021uyw}.

\subsubsection*{Rational enhancement points for the hyperplane string}
The same type of rational structure that appears at the $\mathfrak{su}(8)$ enhancement point of the hyperplane string CFT also arises at the other rational points listed in~\eqref{eq:rationalmodels}. The elliptic genera of the corresponding (0,4) SCFTs were determined in~\cite{Lockhart:2025lea,Lockhart:2026xml}. These elliptic genera, given in terms of characters of the left-moving chiral algebra, provide direct information about possible extensions of the current algebra. In turn, these extensions allow us to determine the global structure of the six-dimensional gauge group $G$. In every case, we find that each center symmetry that acts trivially on the hypermultiplet representation $\mathscr{H}$ is realized on the worldsheet by a corresponding simple current extension, and is therefore interpreted as a gauged one-form symmetry in the bulk. The resulting global forms of the gauge groups are as follows.
\begin{itemize}
\item[-] At the $\mathfrak{e}_6$ enhancement point, there is no extension of the $\mathfrak{e}_6$ current algebra itself, and the six-dimensional theory has
\begin{align}
G &= \mathrm{E}_6~,&
\mathscr{H} &= \rep{351'}~.
\end{align}
In this case, the global structure is straightforward to determine directly from the matter content, since the hypermultiplets explicitly break the $\Z_3$ center symmetry of $\mathrm{E}_6$. Nevertheless, the CFT still exhibits a non-trivial extension, by a composite spin-$6$ current that combines the $\Z_3$ center symmetry of $\mathrm{E}_6$ with the $\Z_3$ symmetry of the residual Potts model.
\item[-] At the $\mathfrak{su}(8)$ enhancement point, the current algebra admits a $\Z_4$ simple current extension, leading to the refined identification
\begin{align}
G &= \mathrm{SU}(8)/\Z_4~,&
\mathscr{H} &= \rep{336}~.
\end{align}
In addition, the left-moving algebra is further extended by a composite spin-$4$ current that mixes the $\Z_8$ center symmetry of $\mathrm{SU}(8)$ with the $\Z_2$ symmetry of the residual Ising model.
\item[-] At the $\mathfrak{su}(6)+\mathfrak{so}(8)$ enhancement point, the current algebra admits a $\Z_3\times\Z_2\times\Z_2$ simple current extension, leading to the refined identification
\begin{align}
G &= \mathrm{SU}(6)/\Z_3\; \times\; \mathrm{Spin}(8)/(\Z_2\times\Z_2) ~,&
\mathscr{H} &= (\rep{56},\rep{1})+(\tfrac{1}{2}\rep{20},\rep{28}) ~.
\end{align}
The left-moving algebra is further extended by a composite spin-$3$ current that mixes the $\Z_6$ center symmetry of $\mathrm{SU}(6)$ with the $\Z_2$ symmetry of the residual Ising model.
\item[-] At the $\mathfrak{su}(4)+\mathfrak{su}(4)+\mathfrak{e}_7$ enhancement point, the current algebra admits a $\Z_4\times\Z_2$ simple current extension, leading to the refined identification
\begin{align}
G &= \left(\mathrm{SU}(4)/\Z_2 \times \mathrm{SU}(4)/\Z_2 \times \mathrm{E}_7\right)/\Z_2~,&
\nonumber\\
\mathscr{H} &= (\rep{10},\rep{10},\rep{1})+(\rep{1},\rep{6},\tfrac{1}{2}\rep{56})+(\rep{6},\rep{1},\tfrac{1}{2}\rep{56})~.
\end{align}
The left-moving algebra is further extended by a composite spin-$3$ current that mixes the $\Z_2$ center symmetry of $\mathrm{E}_7$ with the $\Z_2$ symmetry of the residual tricritical Ising model.
\item[-] At the $\mathfrak{su}(4)+\mathfrak{su}(12)$ enhancement point, the current algebra admits a $\Z_4\times\Z_2$ simple current extension, leading to the refined identification
\begin{align}
G &= \left(\mathrm{SU}(4)/\Z_2 \times \mathrm{SU}(12)/\Z_2\right)/\Z_2~,&
\mathscr{H} &= (\rep{35},\rep{1})+(\rep{6},\rep{66})~.
\end{align}
The left-moving algebra is further extended by a composite spin-$3$ current that mixes the $\Z_6\subset\Z_{12}$ center symmetry of $\mathrm{SU}(12)$ with the $\Z_3$ symmetry of the residual tricritical Potts model.
\end{itemize}
At each of the five rational points, the left-moving $c_\tL=32$ chiral algebra exhibits a mixed extension involving not only the KM algebra sector, but also a simple current from the residual minimal model factor. We stress that, from the intrinsic perspective of the probe string CFT, these composite simple currents stand on exactly the same footing as the Kac--Moody simple currents associated with gauged one-form symmetries. A clear interpretation in the six-dimensional bulk theory for these additional discrete structures is certainly desirable. A possible avenue to investigate this question is to compactify the six-dimensional theory on a circle. Upon reduction, both the KM simple currents and these more general composite simple currents give rise to BPS particles that appear on a democratic footing in the five-dimensional theory.

\subsection{Circle reduction and primitive 5d BPS particles}
\label{subsec:5dreduction}
Circle compactification of 6d $\mathcal{N}=(1,0)$ supergravity yields a minimal 5d supergravity theory, and there exists a well-studied interplay between the Green--Schwarz term in 6d and one-loop corrections to the Chern--Simons couplings in 5d~\cite{Bonetti:2011mw,Bonetti:2013ela,Grimm:2015zea,Corvilain:2017luj,Corvilain:2020tfb}. The reduced theory encodes non-trivial consistency conditions on its six-dimensional uplift, particularly through its BPS spectrum~\cite{Kim:2024tdh,Kim:2024eoa,Kaufmann:2024gqo,Cheng:2021zjh,Cheng:2025ikd,Duque:2025kaa}. In this section, we discuss the fate of one-form symmetries under this reduction by analyzing the 5d BPS particles arising from wrapped 6d supergravity strings. In particular, we relate the gauging of one-form symmetries in six dimensions to the presence of particular primitive particles in the 5d BPS spectrum, originating from simple currents in the worldsheet CFT of 6d supergravity strings wrapped on the circle. This perspective sheds further light on observations of~\cite{Kim:2024tdh} for some models with rank-one gauge algebras, and generalizes them to arbitrary gauge algebras.\\

Let us examine the circle reduction of a six-dimensional $\mathcal{N}=(1,0)$ supergravity theory with $T$ tensor multiplets and a (possibly non-simply connected) gauge group $G$, following \cite{Kim:2024tdh}. We consider the resulting five-dimensional minimal supergravity on its Coulomb branch, where vacuum expectation values of scalars in the vector multiplets break the five-dimensional gauge group to a product of $\mathrm{U}(1)$ factors. The abelian vector multiplets of the reduced theory consist of one Kaluza--Klein vector, $T+1$ vectors coming from the two-form fields of the 6d theory, and vectors valued in the Cartan subalgebra of $G$. We refer to the corresponding $\mathrm{U}(1)$ charges respectively as winding, tensor, and gauge charges, and denote them by $(Q_{\text{w}},Q_{\text{t}},Q_{\text{g}})$. The charge $Q_{\text{g}}$ is defined with respect to a maximal torus of the simply connected cover of $G$, and corresponds to a point in its weight lattice. The five-dimensional $\mathrm{U}(1)$ gauge connections are collectively denoted   by $A^I$, where the index runs over $I=0,1,\dots,T+\operatorname{rk}G+1$. The Coulomb branch moduli space is parameterized by scalar expectation values $t^I$. The trilinear CS couplings $C_{IJK}$ of the 5d theory appear in the prepotential 
\begin{align}
\mathcal{F} = \tfrac{1}{6}C_{IJK}t^I t^J t^K~,
\end{align}
which determines the scalar manifold through the hypersurface equation $\mathcal{F}=1$.

Several constraints on the prepotential follow from the positivity of the scalar metric and that of the tensions of probe BPS strings in the five-dimensional supergravity theory~\cite{Kim:2024tdh,Kaufmann:2024gqo}. These consistency conditions hold within the Kähler cone $\mathcal{K}$, the region of the Coulomb branch where the masses of all supergravity BPS states---electrically
charged particles and magnetically charged strings---are non-negative. Boundaries of the Kähler cone correspond to loci in the parameter space where certain BPS states become massless. In simple cases, $\mathcal{K}$ coincides with the cone $\mathcal{C}=\{t_I\geq0\}$. When this occurs, the CS coefficients must obey~\cite{Kim:2024tdh}
\begin{align}
C_{III} &\geq 0~,&
C_{IJJ} &\geq 0~,
\label{eq:CSpositivity}
\end{align}
for consistency of the 5d theory. In general, however, the positivity constraints are more subtle to analyze, as the Kähler cone may be non-simplicial. In such cases, $\mathcal{K}$ is strictly smaller than $\mathcal{C}$: it has additional boundaries where further primitive BPS states become massless. The positivity conditions~\eqref{eq:CSpositivity} then no longer apply directly. Nevertheless, one can introduce new Kähler parameters $\tilde{t}^I$ spanning a simplicial cone $\widetilde{\mathcal{C}}=\{\tilde{t}^I\geq 0\}$ contained inside the Kähler cone. Since the positivity conditions must hold within $\mathcal{K}$, they in particular hold in $\widetilde{\mathcal{C}}$. The resulting constraints read $\widetilde{C}_{III}\geq 0$ and $\widetilde{C}_{IJJ}\geq 0$, where $\widetilde{C}_{IJK}$ denote the coefficients of the prepotential expressed in the new basis.\\

The authors of~\cite{Kim:2024tdh} outline a strategy to properly identify the Kähler cone generators of the five-dimensional theory obtained by circle reduction of a six-dimensional supergravity model. This requires determining the primitive BPS states of the 5d theory. While some BPS particles descend directly from the massless spectrum of the six-dimensional theory, others arise from non-perturbative degrees of freedom, notably supergravity strings. Indeed, wrapped 6d BPS strings give rise to 5d BPS particles carrying non-trivial tensor charge. As shown in~\cite{Kim:2024tdh}, the ground state of a primitive BPS string of charge $Q$ yields a 5d BPS particles with charges
\begin{align}
(Q_{\text{w}},Q_{\text{t}},Q_{\text{g}}) = (\tfrac{1}{2}Q\cdot a,Q,0)~,
\end{align}
where $a$ is the 6d gravitational anomaly coefficient. Extra Kähler cone generators may arise from additional chiral primary states on the string. A purely left-moving KM primary state of conformal dimension $h_\tL$ produces BPS particles with
\begin{align}
Q_{\text{w}} &= h_\tL+\tfrac{1}{2}Q\cdot a~,&
Q_{\text{t}} &= Q~,
\end{align}
and charges $Q_{\text{g}}$ corresponding to the weights of the associated representation. Based on lower bounds on the left-moving conformal dimensions of KM primaries, the authors of~\cite{Kim:2024tdh} are able to identify potential primitive states, which if present alter structure of the Käler cone.
Our analysis of the worldsheet of faithful probes provides a criterion for establishing definitively the presence of additional primitive BPS states which arise in connection to the existence of gauged one-form symmetries. Specifically, when a primitive BPS state is associated to a KM simple current on the worldsheet, it is present in the 5d theory if and only if the corresponding one-form symmetry is gauged in the six-dimensional model. In particular, our results imply that whenever the gauge group of the 6d supergravity theory contains a non-simply connected non-abelian factor $G$, each element of $\pi_1(G)$ necessarily gives rise to a primitive 5d BPS particle upon circle reduction. The mass and $\mathrm{U}(1)$ charges of these particles are determined by the conformal dimension of the corresponding KM simple current and by the corresponding representation $\rep{r}$ of $G$. The state of minimal charge $Q_{\text{g}}$, associated to the lowest weight of $\rep{r}$, then yields an additional Kähler cone generator, since its charges cannot be obtained from a positive combination of charges of other 5d BPS particles.

\subsubsection*{Examples with rank-one gauge symmetry}
To illustrate these ideas, let us take a closer look at some examples discussed in~\cite{Kim:2024tdh} for gauge algebras of rank $1$. The first model we consider is a $T=0$ supergravity theory with $\mathfrak{su}(2)$ gauge algebra and $55$ charged hypermultiplets in the representation~$\rep{3}$. This model satisfies the anomaly cancelation conditions, with anomaly coefficients given by $a=-3$ and $b=12$. Since there is only adjoint matter, the gauge group could in principle be either $\mathrm{SU}(2)$ or $\mathrm{SO}(3)$. The version with $\mathrm{SO}(3)$ gauge group is known to admit a realization in F-theory as a compactification on an elliptically fibered Calabi–Yau threefold with $\Z_2$ Mordell--Weil torsion~\cite{Morrison:2021wuv}. 

This supergravity model admits a primitive BPS string of charge $Q=1$, namely the hyperplane string. On its worldsheet CFT, the six-dimensional gauge symmetry is realized as a left-moving $\mathfrak{su}(2)$ KM algebra at level $12$. In~\cite{Kim:2024tdh}, this supergravity string is used to test the consistency of the six-dimensional theory. The reasoning proceeds as follows. First, it is noticed that the naive prepotential has negative coefficients, suggesting the presence of additional BPS states arising from the BPS string spectrum. Assuming the presence of a primary state with $h_\tL=3$ in the representation $\rep{13}$, and supposing that the addition of this state is sufficient to generate the full BPS cone, one obtains the prepotential\footnote{The theory is considered after Higgsing a $\mathrm{U}(1)$ factor by giving an expectation value to a BPS state with charges $(Q_{\text{w}},Q_{\text{t}},Q_{\text{g}})=(1,0,-2)$, which corresponds to setting $t_0=0$ in the prepotential.}
\begin{align}
6\mathcal{F} = 6\tilde{t}_1^2\tilde{t}_2+36\tilde{t}_1\tilde{t}_2^2+72\tilde{t}_2^3~.
\end{align}
This prepotential appears to be inconsistent, since the corresponding scalar metric degenerates as $\tilde{t}_1\to0$. However, this issue arises only if the gauge group is $\mathrm{SU}(2)$. If instead the $\Z_2$ center one-form symmetry is gauged, the Kähler parameter must be rescaled as $\tilde{t}_2\to \tfrac{1}{2}\tilde{t}_2$ to account for the correct normalization of the $\mathrm{SO}(3)$ CS coupling. With this rescaling, the resulting prepotential is consistent.

In fact, these observations follow from the relationship between simple current extensions and gauging of center one-form symmetries. The $\mathfrak{su}(2)$ KM algebra at level $12$ supported by the hyperplane string CFT admits a non-trivial simple current associated with the $\Z_2$ center of $\mathrm{SU}(2)$. This current belongs to the KM primary representation $\rep{13}$ and has conformal weight $h_\tL=3$. Hence, it precisely coincides with the primary state postulated above. Its presence in the CFT spectrum indicates that the $\Z_2$ one-form symmetry is gauged, implying that the gauge group is $\mathrm{SO}(3)$. In particular, we see that some of the assumptions made in~\cite{Kim:2024tdh} about this $\mathfrak{su}(2)$ model are in fact related to each other: the presence of the KM primary state in $\rep{13}$ is inevitably tied to the gauging of the $\Z_2$ one-form symmetry. It is quite remarkable that independent arguments based on the five-dimensional prepotential also lead to the same conclusion, namely the necessity of this simple current extension.\\

As a second example, consider a $T=0$ supergravity theory with $\mathrm{U}(1)$ gauge group and $54$ triplets of hypermultiplets with charges
\begin{align}
&Q_1~,&
&Q_2~,&
Q_3 &= Q_1+Q_2~.
\end{align}
This six-dimensional model, introduced in~\cite{Taylor:2018khc} and further studied in~\cite{Kim:2024tdh}, satisfies the abelian anomaly cancellation conditions, with anomaly coefficient
\begin{align}
b = 6(Q_1^2+Q_2^2+Q_1Q_2) ~.
\end{align}
On the worldsheet CFT of the hyperplane string, the gauge symmetry is realized as a $\mathfrak{u}(1)$ current algebra with level $k=b/2$. Let
\begin{align}
n=\operatorname{gcd}(Q_1,Q_2)
\end{align}
denote the greatest common divisor of the charges. When $n>1$, the massless spectrum is invariant under a non-trivial $\Z_n\subset\mathrm{U}(1)$ one-form symmetry.

Upon circle reduction, it is argued in~\cite{Kim:2024tdh} that consistency of the resulting five-dimensional theory---specifically, positivity of the gauge kinetic terms and consistency of the CS couplings---requires the existence of an additional BPS state. This state arises from a primary operator in the worldsheet CFT of a six-dimensional BPS string, with conformal weight
\begin{align}
h_\tL = \frac{b}{2n^2}~.
\label{eq:candicateU1primary}
\end{align}
Including this contribution modifies the five-dimensional prepotential to
\begin{align}
6\mathcal{F} = 3n\,\tilde{t}_1^2\tilde{t}_2+9n^2\tilde{t}_1\tilde{t}_2^2+9n^3\tilde{t}_2^3~,
\end{align}
expressed in terms of suitably redefined Kähler parameters. However, this expression still leads to degeneracies unless the $\Z_n$ one-form symmetry is gauged, which effectively corresponds to a rescaling $\tilde{t}_2\to \tilde{t}_2/n$ of the Kähler parameter associated with the six-dimensional $\mathrm{U}(1)$ gauge field.

From the worldsheet perspective, the gauged $\Z_n$ one-form symmetry is naturally encoded in the presence of the primary state~\eqref{eq:candicateU1primary}. As reviewed in section~\ref{subsec:KMcenter}, the primaries of the $\mathfrak{u}(1)$ current algebra are labeled by $s\in\Z_{2k}$, with conformal dimensions given by~\eqref{eq:u(1)conformalweights}. The candidate state with label $s=b/n$ has precisely the conformal weight~\eqref{eq:candicateU1primary}, and its presence signals an extension of the $\mathfrak{u}(1)$ current algebra. The resulting extended algebra is simply the $\mathfrak{u}(1)$ current algebra at level $\widetilde{k}=k/n$. In spacetime, this extension corresponds to gauging the $\Z_n$ one-form symmetry, so that the gauge group is effectively $\mathrm{U}(1)/\Z_n$, and that all charges of matter fields are integral and primitive.

\section{Summary and open questions}
\label{sec:concl}
In this work we have studied center one-form symmetries from the perspective of strings charged under the two-form gauge fields of the bulk theory. Our focus was on faithful string probes: strings whose worldsheet CFT realizes the bulk gauge symmetry through a holomorphic current algebra. Such probes are sensitive to global properties of the spacetime gauge group. In particular, when the bulk theory has a non-simply connected non-abelian gauge group $G$, the left-moving chiral algebra on the string is extended by higher-spin currents associated with the non-trivial elements of $\pi_1(G)$. In the presence of abelian gauge factors, the worldsheet theory is likewise sensitive to gauged one-form symmetries mixing center symmetries of the non-abelian sector with discrete subgroups of $\mathrm{U}(1)$ symmetries: these again appear on the worldsheet as extensions of the KM algebra by integral spin simple currents. We argued that the appearance of such higher-spin chiral operators in the CFT spectrum of a faithful string probe provides a worldsheet explanation for the fact that the corresponding center one-form symmetry is gauged in the bulk theory.

Several observations support this correspondence. Most notably, the consistency condition for a simple current extension of the KM algebra precisely reproduces the known conditions for the absence of a mixed anomaly for the corresponding center one-form symmetry in spacetime. Our framework therefore unifies, from the worldsheet viewpoint, the consistency conditions derived in~\cite{Apruzzi:2020zot,Cvetic:2020kuw} for gauged one-form symmetries in six- and eight-dimensional supergravity theories and generalizes them to a broader class of theories in $d\geq 4$. We further tested this picture in theories obtained from compactification of the heterotic string, where both the spacetime spectrum and the worldsheet CFT are explicitly accessible. In a variety of examples in dimensions $d\geq 6$ with eight or sixteen supercharges, we found exact agreement between the spacetime and worldsheet determinations of the gauge group topology. In these constructions, where the faithful string CFT is solvable, our perspective reveals a common underlying structure for many well-studied models with non-simply connected gauge groups, such as CHL compactifications with enhanced gauge symmetry~\cite{Font:2021uyw}. The same worldsheet physics is at play for several six-dimensional supergravity theories without tensor multiplets, where a detailed analysis of faithful BPS strings leads to concrete predictions for the global form of the gauge group, despite the absence of any known string-theoretic realization.

Our results also fit naturally within broader swampland expectations for quantum gravity regarding the absence of global symmetries. From the worldsheet perspective, a center symmetry can only be realized in two ways: either it is broken by states of the CFT, or it appears through a simple current extension of the chiral algebra. In spacetime, these correspond respectively to broken or gauged center one-form symmetries. Importantly, this conclusion does not rely on assuming that the theory is coupled to gravity, but rather on the existence of faithful string probes themselves. Such probes appear naturally in various classes of supergravity theories, where in light of the completeness of the spectrum of charged branes one finds a plethora of BPS strings that are good candidates for being faithful probes.\\

Let us conclude by remarking on a number of open questions. A central lesson of our analysis is that faithful string probes capture genuinely global aspects of gauge symmetry. A natural direction is therefore to extend our discussion to theories with discrete gauge groups, which are connected to 3d Dijkgraaf--Witten TQFTs~\cite{Dijkgraaf:1989pz} and have been well studied from a F-theoretic perspective, starting with~\cite{Braun:2014oya}. It would be interesting to  determine whether simple current extensions on the worldsheet also play a role in this context. A complication worth noting is that, for non-abelian discrete gauge groups, an incomplete spectra need not imply the existence of a one-form symmetry~\cite{Harlow:2018tng}. 

Another open and possibly related question concerns the interpretation of the mixed extensions of the chiral algebra that we observed in several rational worldsheet CFTs associated to faithful probe strings. These extensions are generated by composite integral spin currents, that intertwine center symmetries with additional discrete symmetries associated with the residual left-moving chiral algebra. Their close parallel with ordinary KM simple current extensions suggests that they also admit a spacetime interpretation, possibly in terms of additional, non-center one-form symmetries mixing non-trivially with the gauge symmetry itself. This possibility is reinforced by the parallel with the CHL involution, where the decomposition of the $(\mathrm{E}_8\times\mathrm{E}_8)\rtimes\Z_2$ heterotic currents into an $\mathfrak{e}_8$ current algebra at level $2$ together with a residual Ising sector gives rise to a composite spin-$2$ current inherited from the $\Z_2$ outer automorphism exchanging the two $\mathrm{E}_8$ factors. It would be very interesting to determine whether similar enhancements appear more generally in CHL compactifications exhibiting analogous rational structures~\cite{Font:2021uyw}, and whether they admit a comparable spacetime origin.

A central ingredient in our analysis is modularity. Beyond its technical role in establishing the existence of a simple current extension, modularity is crucially tied to the fact that a faithful string probe couples to more than just the gauge sector of the bulk theory. This is already visible in the six-dimensional $\mathrm{SU}(8)$ supergravity model discussed in section~\ref{subsec:rational}. The primitive BPS string supports a $\mathfrak{su}(8)$ current algebra at level $8$, with thousands of integrable representations. Taken alone, these data do not define a modular object: modularity is restored only after combining the current algebra with an additional left-moving Ising sector and coupling to right-movers, which encode information about the landscape of supergravity vacua~\cite{Lockhart:2025lea}. In our view this provides satisfying conceptual explanation for the fact that the mere existence of faithful strings implies the absence of global center one-form symmetries expected in quantum gravity theory, and calls for a deeper study of faithful probe CFTs.

One may further ask whether the composite extensions observed in the CFTs of several faithful string probes encountered in this work are related to properties of the bulk background geometry. Indeed, in two-dimensional superconformal theories, enhancements of the superconformal algebra by additional (anti-)holomorphic operators are tied, in sigma model realizations, to the presence of covariantly constant tensor fields on the target space---the canonical example being the $\mathfrak{u}(1)$ R-current of the N=2 superconformal algebra. It would be interesting to understand whether the more exotic rational sectors encountered in this work, such as those involving an Ising model, admit a geometric interpretation, perhaps analogous to the extensions of N=1 superconformal algebras observed in~\cite{Shatashvili:1994zw} for supersymmetric sigma models on $7$- and $8$-manifolds with $\mathrm{G}_2$ and $\mathrm{Spin}(7)$ holonomy.\footnote{We thank B. Pioline for pointing out this parallel.} A related question is whether, in these rational models, the right-moving chiral algebra---corresponding to the supersymmetric sector---may likewise exhibit enhancements by higher-spin operators.

Another natural direction for further research is provided by F-theory compactifications on elliptically fibered Calabi--Yau manifolds with non-trivial Mordell--Weil torsion. In such models, D3-branes wrapped on curves in the base give rise to BPS strings, and suitable curve configurations may furnish faithful string probes. Our results then predict that the corresponding worldsheet theories must exhibit extensions of their chiral algebra by higher-spin currents associated with the torsional sections of $\mathrm{MW}(X)$. It would be very interesting to understand how these extended chiral operators are encoded geometrically. Moreover, in this framework, the elliptic genera of BPS strings are generating functions for enumerative invariants of the Calabi--Yau geometry and satisfy strong modular constraints~\cite{Haghighat:2014vxa,Huang:2015sta,DelZotto:2017mee,Oberdieck:2016nvt,Oberdieck:2017pqm,Duan:2020imo,Duque:2025kaa,Schimannek:2021pau,Cota:2019cjx}. Our perspective therefore leads to concrete predictions for Calabi--Yau manifolds with Mordell--Weil torsion: the relevant enumerative invariants should organize into extended characters reflecting the presence of the simple currents.

Finally, a very interesting question which we have left open is to extend our ideas to broader and less constrained settings. In this work we have focused on examples with at least eight supercharges, where the probe string worldsheet theory is sufficiently controlled to permit a precise comparison with the bulk physics. It would be particularly interesting to understand whether faithful string probes retain a useful role in analyzing the properties of theories with fewer supercharges such as four-dimensional $\mathcal{N}=1$ supergravity where dynamical aspects lead to much richer physics, or even in the absence of supersymmetry.

\section*{Acknowledgements}
We are grateful to M. Del Zotto, J. Dücker, C. Fierro Cota, D. Israël, A. Klemm, I. Melnikov, R. Minasian, B. Pioline, T. Rudelius, A. Sangiovanni, T. Schimannek and I. Zadeh for valuable discussions, and especially to K. Xu for ongoing collaboration on related topics. The work of GL, LN and YP has received funding from the European Research Council (ERC) under the Horizon Europe (grant agreement No. 101078365) research and innovation program.

\appendix
\section{Narain CFTs and the CHL orbifold}
\label{app:Narain}

In this appendix, we give a brief review of Narain CFTs and their CHL orbifolds. For two of the CHL models discussed in the main text, namely the $G= \mathrm{SU}(2)\times \mathrm{SU}(9)$ model in  $d=9$ and the $G= (\mathrm{SU}(2)\times \mathrm{SU}(3)\times \mathrm{E}_7)/\Z_2$ model in $d=8$, we then determine the low-lying spectrum of purely left-moving states and deduce whether the left-moving chiral algebra is extended by simple current operators.

\subsection{The Narain moduli space}
\label{app:Narainmoduli}
The Narain CFT with central charges $c_\tL=D+16$ and $c_\tR=D$ can be described in terms of free fields: $D$ compact bosons $X_\tL^I(z),X_\tR^I(\bar{z})$ associated to the torus degrees of freedom, together with sixteen chiral left-moving bosons valued in the Cartan torus of $\mathrm{E}_8\times\mathrm{E}_8$. The left- and right-moving $\mathfrak{u}(1)$ currents generate a current algebra, whose primary fields are the vertex operators $\mathcal{V}_\bp(z,\bar{z})$. The vertex operators organize into primary representations of this current algebra, and are labeled by points of a lattice $\Gamma$. The latter is an even self-dual lattice of signature $(D+16,D)$. Up to isometry, it can be written as\footnote{We designate Euclidean lattices as $\Lambda$ and keep the notation $\Gamma$ for lattices with a pseudo-definite norm.}
\begin{align}
\Gamma = \Gamma_{D,D}\oplus\Lambda_{\mathfrak{e}_8}\oplus\Lambda_{\mathfrak{e}_8}~.
\end{align}
Here, $\Gamma_{D,D}$ is spanned by null vectors $\boldsymbol{e}_I$ and $\boldsymbol{e}^{\ast I}$ satisfying $\boldsymbol{e}_I\cdot \boldsymbol{e}^{\ast J}=\delta_I^J$, while $\Lambda_{\mathfrak{e}_8}$ denotes the root lattice of $\mathfrak{e}_8$, the unique positive-definite even self-dual lattice of rank eight. Accordingly, we decompose lattice vectors as
\begin{align}
\boldsymbol{p} = \text{w}^I \boldsymbol{e}_I + \text{n}_I \boldsymbol{e}^{\ast I} +\boldsymbol{L}~,
\end{align}
where $\text{w}^I,\text{n}_I$ are integers and $\boldsymbol{L}=(\boldsymbol{\ell},\boldsymbol{\ell}')$ with $\boldsymbol{\ell}$ and $\boldsymbol{\ell}'$ two lattice vectors of $\Lambda_{\mathfrak{e}_8}$. We also fix an embedding
\begin{align}
\Gamma \subset \R^{D+16,D}
\end{align}
of the Narain lattice in the pseudo-Riemannian vector space $\R^{D+16,D}$.

The Narain moduli are fully specified by a choice of timelike $D$-plane $\Pi_\tR$ in $\R^{D+16,D}$, with orthogonal complement $\Pi_\tL = (\Pi_\tR)^\perp$. This choice determines a decomposition of lattice vectors
\begin{align}
\boldsymbol{p} = \boldsymbol{p}_\tL + \boldsymbol{p}_\tR~,
\end{align}
from which the left- and right-moving conformal weights follow as
\begin{align}
h_\tL(\boldsymbol{p}) &= \tfrac{1}{2}\boldsymbol{p}_\tL\cdot\boldsymbol{p}_\tL~,&
h_\tR(\boldsymbol{p}) &= -\tfrac{1}{2}\boldsymbol{p}_\tR\cdot\boldsymbol{p}_\tR~.
\end{align}
By construction, states have integral spin
\begin{align}
h_\tL(\boldsymbol{p}) - h_\tR(\boldsymbol{p}) = \text{n}_I \text{w}^I + \tfrac{1}{2}\boldsymbol{L}\cdot\boldsymbol{L}~,
\end{align}
which only depends on the lattice vector $\boldsymbol{p}$. The heterotic moduli are identified as follows. The plane $\Pi_\tR$ is characterized by a choice of $D$ independent timelike vectors $\widetilde{\boldsymbol{\pi}}_I$. A convenient parametrization is obtained as
\begin{align}
\widetilde{\boldsymbol{\pi}}_I = \boldsymbol{e}_I - (\mathcal{G}_{IJ}-\mathcal{B}_{IJ}+\tfrac{1}{2}A_I\cdot A_J)\boldsymbol{e}^{\ast J} - A_I~,
\end{align}
which satisfy $\widetilde{\boldsymbol{\pi}}_I\cdot\widetilde{\boldsymbol{\pi}}_J = -2\mathcal{G}_{IJ}$. This naturally identifies the geometric data of the toroidal compactification: the metric $\mathcal{G}_{IJ}$ and two-form field $\mathcal{B}_{IJ}$ on $T^D$, together with the Wilson line parameters $A_I$. These furnish a local description of the Narain moduli space, which must be further quotiented by the discrete identifications induced by Narain T-dualities. It is also convenient to define
\begin{align}
\mathcal{E}_{IJ}=\mathcal{G}_{IJ}-\mathcal{B}_{IJ}+\tfrac{1}{2}A_I\cdot A_J~.
\end{align}

Specializing to the case of a single circle, the heterotic moduli reduce to a radius $r$ and a single Wilson line parameter $A$. In this case, the right-moving conformal weight becomes
\begin{align}
h_\tR(\boldsymbol{p}) = \frac{1}{4r^2}\left( \text{n} -\left(r^2+\tfrac{1}{2}A\cdot A\right)\text{w}-A\cdot\boldsymbol{L}\right)^2~.
\end{align}

\subsection{Spectrum of the CHL string}
\label{app:CHLspectrum}
The internal CFT of the $(10-D)$-dimensional CHL string may be obtained as a $\Z_2$ orbifold of the Narain theory with $c_\tL=16+D$ and $c_\tR=D$. The orbifold is characterized by the following shift vector and lattice automorphism:
\begin{align}
s_g &= \frac{\be_1}{2}~,&
\varphi_g(\bp) &=\tw^I\be_I+\tn_I\be^{\ast I} +(\bell',\bell)~.
\end{align}
It is a symmetry of the Narain CFT on the loci in moduli space with Wilson line parameters of the form $A_I=(a_I,a_I)$. We denote the (unprojected) untwisted and twisted Hilbert spaces of the orbifold CFT by $\mathcal{H}_{\text{ut}}$ and $\mathcal{H}_{\text{t}}$ respectively. In order to describe the untwisted Hilbert space, it is useful to introduce the states
\begin{align}
\ket{\bp} = \lim_{z,\bar{z}\to 0}\mathcal{V}_\bp(z,\bar{z})\ket{0}~,
\end{align}
associated to vertex operators. The CHL orbifold acts on those as 
\begin{align}
g\ket{\bp} = \ee^{2\pi\ii s_g\cdot\bp}\ket{\varphi_g(\bp)}~.
\end{align}
We also denote by
\begin{align}
&\mathcal{J}^I(z)~,&
&\mathcal{J}_\pm^i(z)~,&
&\bar{\mathcal{J}}^I(\bar{z})~,
\end{align}
the left- and right-moving $\mathfrak{u}(1)$ currents, with $I=1,\dots,D$ and $i=1,\dots,8$. They transform under the CHL symmetry as 
\begin{align}
g\, \mathcal{J}^I g^{-1} &= \mathcal{J}^I~,&
g\, \mathcal{J}^i_\pm \,g^{-1} &= \pm\mathcal{J}^i_\pm~,&
g\, \bar{\mathcal{J}}^I g^{-1} &= \bar{\mathcal{J}}^I~.
\label{eq:CHLoscillators}
\end{align}
In the untwisted sector, the $\mathfrak{u}(1)$ currents all have integer modes. We schematically denote combinations of oscillators by $\mathcal{J}_{\text{osc}}$ (that is, $\mathcal{J}_{\text{osc}}$ is a product of negative modes of $\mathcal{J}^I$, $\mathcal{J}^i_\pm$ and $\bar{\mathcal{J}}^I$). States of the untwisted Hilbert space that survive the CHL projection may all be written as linear combinations of states of the form
\begin{align}
\mathcal{J}_{\text{osc}}\ket{\bp} + \ee^{2\pi\ii s_g\cdot\bp} g\,\mathcal{J}_{\text{osc}}\,g^{-1} \ket{\varphi_g(\bp)}~,
\end{align}
where the transformation $g\,\mathcal{J}_{\text{osc}}\,g^{-1}$ of oscillators is straightforwardly derived from~\eqref{eq:CHLoscillators}.

In the twisted sector, states are labeled by
\begin{align}
\widetilde{\bp} = \widetilde{\text{w}}^I\be_I + \widetilde{\text{n}}_I\be^{\ast I} + (\widetilde{\bell},\widetilde{\bell})~,
\end{align}
in terms of the half-integer winding number $\widetilde{\text{w}}^1\in\Z+\frac{1}{2}$, the integer winding and momentum numbers $\widetilde{\text{w}}^2,\dots,\widetilde{\text{w}}^D,\widetilde{\text{n}}_1,\dots,\widetilde{\text{n}}_D\in\Z$, and of $\widetilde{\bell}$, an element of the $\mathfrak{e}_8$ root lattice. The twisted state $\ket{\widetilde{\bp}}_{\text{t}}$ has spin equal to
\begin{align}
h_\tL- h_\tR = \widetilde{\text{n}}_I\widetilde{\text{w}}^I + \tfrac{1}{4}\widetilde{\bell}\cdot\widetilde{\bell} + \tfrac{1}{2} ~,
\end{align}
and right-moving conformal dimension given by
\begin{align}
h_\tR = \tfrac{1}{4}\mathcal{G}^{IJ}\left(\widetilde{\text{n}}_I-\mathcal{E}_{IK}\widetilde{\text{w}}^K-a_I\cdot\widetilde{\bell}\right)\left(\widetilde{\text{n}}_J-\mathcal{E}_{JL}\widetilde{\text{w}}^L-a_J\cdot\widetilde{\bell}\right)  ~.
\end{align}
Under the CHL symmetry, it transforms according to
\begin{align}
g \ket{\widetilde{\bp}}_{\text{t}} = (-1)^{\widetilde{\text{n}}_1+\frac{1}{2}\widetilde{\bell}\cdot\widetilde{\bell}+1}\ket{\widetilde{\bp}}_{\text{t}}~.
\end{align}
States of the twisted Hilbert space $\mathcal{H}_{\text{t}}$ are obtained by acting on $\ket{\widetilde{\bp}}_{\text{t}}$ with a combination of oscillators---note that in the twisted sector, the currents $\mathcal{J}^i_-$ have a half-integer mode expansion.

\subsubsection*{Purely left-moving spectrum of the $G=\mathrm{SU}(2)\times\mathrm{SU}(9)$ model}
Consider the nine-dimensional CHL string at the following point in moduli space, specified by the radius and Wilson line
\begin{align}
r &=\frac{3}{2\sqrt{2}}~,&
a &= \frac{\boldsymbol{\omega}}{4}~,
\end{align}
where $\boldsymbol{\omega}$ is a vector in the $\mathfrak{e}_8$ lattice satisfying $\boldsymbol{\omega}\cdot\boldsymbol{\omega}=14$. At this point, the current algebra is enhanced to $\mathfrak{su}(2)+\mathfrak{su}(9)$, with both factors at level $2$. Indeed, the Wilson line breaks $\mathfrak{e}_8$ to $\mathfrak{su}(2)+\mathfrak{su}(8)$, and the twisted sector contains $16$ roots which enhance $\mathfrak{su}(8)$ to $\mathfrak{su}(9)$. The purely left-moving spectrum can be determined at fixed conformal weight $h_\tL$ using the general formulas described above; the result is summarized in table~\ref{tab:holomorphicspectrumSU2SU9}. In particular, the orbifold CFT contains $14580$ purely left-moving states at $h_\tL=2$. They lie in the $\mathfrak{su}(2)$ representation
\begin{align}
10447\cdot\rep{1}+1376\cdot\rep{3}+1\cdot\rep{5}~,
\end{align}
which can be extracted from the charges of states under the Cartan $\mathfrak{u}(1)\subset\mathfrak{su}(2)$ current, linearly realized in the untwisted sector.

\begin{table}[h!]
\centering
{\renewcommand{\arraystretch}{1.2}
$\begin{array}{c|ccccc}
\multicolumn{6}{c}{\text{$\mathfrak{u}(1)^{17}$ KM primaries}} \\\hline
h_\tL & 0 & \frac{1}{2}  & 1 & \frac{3}{2} & 2 \\ \hline
\mathcal{H}_{\text{ut}}^+ & 1 & 0 & 58 & 0 & 7936 \\
\mathcal{H}_{\text{ut}}^- & 0 & 0 & 58 & 0 & 7792 \\
\mathcal{H}_{\text{t}}^+ & 0 & 0 & 16 & 0 & 1904 \\
\mathcal{H}_{\text{t}}^- & 0 & 0 & 0 & 368 & 0 \\
\end{array}$
\qquad
$\begin{array}{c|ccccc}
\multicolumn{6}{c}{\text{All states}} \\\hline
h_\tL & 0 & \frac{1}{2}  & 1 & \frac{3}{2} & 2 \\ \hline
\mathcal{H}_{\text{ut}}^+ & 1 & 0 & 67 & 0 & 9012 \\
\mathcal{H}_{\text{ut}}^- & 0 & 0 & 66 & 0 & 8858 \\
\mathcal{H}_{\text{t}}^+ & 0 & 0 & 16 & 0 & 5568 \\
\mathcal{H}_{\text{t}}^- & 0 & 0 & 0 & 496 & 0 \\
\end{array}$}
\caption{Number of purely left-moving spectrum of the $G=\mathrm{SU}(2)\times\mathrm{SU}(9)$ model, organized with respect to their CHL parity}
\label{tab:holomorphicspectrumSU2SU9}
\end{table}

Let us also note that the Sugawara contribution of the current algebra does not saturate the left-moving central charge. The residual sector has $c_\tL=\frac{21}{22}$ and is therefore identified with the $10$th unitary Virasoro minimal model. This model has $55$ primary representations, labeled by pairs ${r,s}$ with $1\leq s\leq r\leq 10$, and conformal dimensions
\begin{align}
h_\tL = \frac{(12r-11s)^2-1}{528}~.
\end{align}
Purely left-moving states organize into representations of the KMV algebra generated by $\mathfrak{su}(2)+\mathfrak{su}(9)$ together with this Virasoro factor. Among the states with $h_\tL=2$, there are $3570$ KMV descendants. Subtracting these descendants from the total leaves $11010$ states, transforming as
\begin{align}
7125\cdot\rep{1}+1295\cdot\rep{3}~,
\end{align}
which must assemble into KMV primaries. There are $13$ such primary representations at $h_\tL=2$, namely
\begin{align}
&(\rep{1},\rep{80})\otimes\{3,1\}~,&
&(\rep{1},\rep{1050})\otimes\{6,5\}~,&
&(\rep{1},\repconj{1050})\otimes\{6,5\}~,&
&(\rep{1},\rep{1215})\otimes\{6,5\}~,&\nonumber\\
&(\rep{1},\rep{2520})\otimes\{1,1\}~,&
&(\rep{1},\repconj{2520})\otimes\{1,1\}~,&
&(\rep{1},\rep{5760})\otimes\{7,7\}~,&
&(\rep{3},\rep{80})\otimes\{8,7\}~,&\nonumber\\
&(\rep{3},\rep{240})\otimes\{2,1\}~,&
&(\rep{3},\repconj{240})\otimes\{2,1\}~,&
&(\rep{3},\rep{1050})\otimes\{5,5\}~,&
&(\rep{3},\repconj{1050})\otimes\{5,5\}~,&\nonumber\\
&(\rep{3},\rep{1215})\otimes\{5,5\}~.
\end{align}
In particular, the representations $(\rep{1},\rep{2520})$ and $(\rep{1},\repconj{2520})$ correspond to the simple currents of the $\mathfrak{su}(9)$ current algebra associated with the $\Z_3\subset\Z_9$ center.

Inspecting the $\mathfrak{su}(2)$ decomposition of the spectrum, one finds that the $1295$ states transforming in the adjoint $\rep{3}$ can be uniquely accounted for by the representation $\rep{80}+\rep{1215}$. For the $7125$ singlets, several decompositions are a priori possible; however, none of them includes the representation $\rep{2520}+\repconj{2520}$. We therefore conclude that the $\Z_3\subset\Z_9$ simple currents of $\mathfrak{su}(9)$ are absent from the CHL worldsheet theory. Correspondingly, the associated one-form symmetry of $\mathrm{SU}(9)$ is broken.

\subsubsection*{Purely left-moving spectrum of the $G=(\mathrm{SU}(2)\times\mathrm{SU}(3)\times\mathrm{E}_7)/\Z_2$ model}
Consider the eight-dimensional CHL string at the following point in moduli space, specified by the metric, B-field and Wilson lines
\begin{align}
\mathcal{G}_{IJ}-\mathcal{B}_{IJ} &= \begin{pmatrix}
2 & -1 \\
0 & \frac{1}{2} \\
\end{pmatrix}~,&
a_1 &= 0~,&
a_2 &= \frac{\boldsymbol{\omega}}{2}~,
\end{align}
where $\boldsymbol{\omega}$ is a root of the $\mathfrak{e}_8$ lattice satisfying $\boldsymbol{\omega}\cdot\boldsymbol{\omega}=2$. At this point, the current algebra is enhanced to $\mathfrak{su}(2)+\mathfrak{su}(3)+\mathfrak{e}_7$, with each factor at level $2$. Indeed, the Wilson lines break $\mathfrak{e}_8$ to $\mathfrak{su}(2)+\mathfrak{e}_7$, while the untwisted currents associated to $T^2$ are preserved by the CHL orbifold: they give rise to an $\mathfrak{u}(1)+\mathfrak{su}(2)$ algebra, which is further enhanced to $\mathfrak{su}(3)$ by $4$ roots in the twisted sector. The purely left-moving spectrum is presented in table~\ref{tab:holomorphicspectrumSU2SU3E7}. 

\begin{table}[h!]
\centering
{\renewcommand{\arraystretch}{1.2}
$\begin{array}{c|ccccc}
\multicolumn{6}{c}{\text{$\mathfrak{u}(1)^{18}$ KM primaries}} \\\hline
h_\tL & 0 & \frac{1}{2}  & 1 & \frac{3}{2} & 2 \\ \hline
\mathcal{H}_{\text{ut}}^+ & 1 & 0 & 130 & 0 & 19688 \\
\mathcal{H}_{\text{ut}}^- & 0 & 0 & 134 & 0 & 19558 \\
\mathcal{H}_{\text{t}}^+ & 0 & 0 & 4 & 0 & 4036 \\
\mathcal{H}_{\text{t}}^- & 0 & 0 & 0 & 512 & 0 \\
\end{array}$
\qquad
$\begin{array}{c|ccccc}
\multicolumn{6}{c}{\text{All states}} \\\hline
h_\tL & 0 & \frac{1}{2}  & 1 & \frac{3}{2} & 2 \\ \hline
\mathcal{H}_{\text{ut}}^+ & 1 & 0 & 140 & 0 & 22161 \\
\mathcal{H}_{\text{ut}}^- & 0 & 0 & 142 & 0 & 22026 \\
\mathcal{H}_{\text{t}}^+ & 0 & 0 & 4 & 0 & 8316 \\
\mathcal{H}_{\text{t}}^- & 0 & 0 & 0 & 544 & 0 \\
\end{array}$}
\caption{Number of purely left-moving spectrum of the $G=(\mathrm{SU}(2)\times\mathrm{SU}(3)\times\mathrm{E}_7)/\Z_2$ model, organized with respect to their CHL parity}
\label{tab:holomorphicspectrumSU2SU3E7}
\end{table}

The orbifold CFT contains $30477$ purely left-moving states at $h_\tL=2$. As discussed in the main text, these states can be entirely characterized in terms of the current algebra, whose Sugawara central charge saturates $c_\tL=18$. Among them, $10584$ are descendants, leaving $19893$ KM primary states at $h_\tL=2$. Their representation content is uniquely fixed and given by
\begin{align}
(\rep{3},\rep{1},\rep{1463})+(\rep{3},\rep{8},\rep{133})+(\rep{1},\rep{8},\rep{1539})~.
\end{align}
In particular, this spectrum contains the spin-$2$ simple current associated with the gauged $\Z_2$ one-form symmetry of the model.

\section{Asymmetric orbifolds and modular orbits}
\label{app:modularorbits}
In this appendix we explicitly construct the partition function of the heterotic CFT associated to the 6d supergravity model with $G=(\mathrm{SU}(2)\times\mathrm{SU}(2)\times\mathrm{E}_7\times\mathrm{E}_7)/\Z_2\times\mathrm{Spin}(8)$ discussed in the main text. We focus on the internal sector of the heterotic worldsheet CFT, formulated in the RNS formalism, with central charges $c_\tL=20$ and $c_\tR=6$. The contribution of the Narain sector of the parent theory is given by 
\begin{align}
Z_{\Gamma} &= \frac{1}{\eta^{20}\bar{\eta}^4}\sum_{\boldsymbol{p}\in\Gamma}q^{h_\tL(\boldsymbol{p})}\bar{q}^{h_\tR(\boldsymbol{p})} 
\nonumber\\
&= \chi^{\mathfrak{e}_8}_\rep{1}\chi^{\mathfrak{e}_8}_\rep{1}\left( \chi^{\mathfrak{d}_4}_\rep{1}\bar{\chi}^{\mathfrak{d}_4}_\rep{1} + \chi^{\mathfrak{d}_4}_\rep{8^v}\bar{\chi}^{\mathfrak{d}_4}_\rep{8^v} + \chi^{\mathfrak{d}_4}_\rep{8^s}\bar{\chi}^{\mathfrak{d}_4}_\rep{8^s} + \chi^{\mathfrak{d}_4}_\rep{8^c}\bar{\chi}^{\mathfrak{d}_4}_\rep{8^c} \right)~,
\end{align}
where the second line follows from the specific choice of Narain moduli. This expression is modular
by itself; more precisely, it transforms under $\mathrm{SL}(2,\Z)$ as
\begin{align}
Z_\Gamma(\tau+1,\bar{\tau}+1) &= \ee^{\frac{2\pi\ii}{3}} Z_\Gamma(\tau,\bar{\tau})~,&
Z_\Gamma(-1/\tau,-1/\bar{\tau}) &= Z_\Gamma(\tau,\bar{\tau})~.
\end{align}
The partition function $Z^{0,0}$ of the unorbifolded theory also includes contributions from the right-moving superpartners of the torus degrees of freedom. In the Neveu--Schwarz ($\NS$) and Ramond ($\Ram$) sectors, it takes the form
\begin{align}
Z^{0,0}_\NS(\bar{z}) &= Z_{\Gamma}\frac{\bar{\vartheta}_3(\bar{z})^2}{\bar{\eta}^2}~,&
Z^{0,0}_\Ram(\bar{z}) &= Z_{\Gamma}\frac{\bar{\vartheta}_2(\bar{z})^2}{\bar{\eta}^2}~,
\nonumber\\
Z^{0,0}_\tNS(\bar{z}) &= Z_{\Gamma}\frac{\bar{\vartheta}_4(\bar{z})^2}{\bar{\eta}^2}~,&
Z^{0,0}_\tRam(\bar{z}) &= Z_{\Gamma}\frac{\bar{\vartheta}_1(\bar{z})^2}{\bar{\eta}^2}~,
\end{align}
where the dependence on the fugacity $\bar{z}$ keeps track of the $\mathfrak{u}(1)$ R-symmetry of the right-moving N=2 superconformal algebra. Here, $\eta$ denotes the Dedekind eta function, and
\begin{align}
\vartheta_1(z)&=-\ii\sum_{m\in\Z}(-1)^mq^{\frac{1}{2}(m+\frac{1}{2})^2}y^{m+\frac{1}{2}} ~,&
\vartheta_2(z)&=\sum_{m\in\Z}q^{\frac{1}{2}(m+\frac{1}{2})^2}y^{m+\frac{1}{2}}~,\nonumber\\
\vartheta_3(z)&=\sum_{m\in\Z}q^{\frac{1}{2}m^2}y^{m}~,&
\vartheta_4(z)&=\sum_{m\in\Z}(-1)^mq^{\frac{1}{2}m^2}y^{m}~,
\end{align}
are the Jacobi theta functions. The sectors labeled by $\tNS$ and $\tRam$ include an additional insertion of the fermion number.

The partition function $Z_\NSRam(\bar{z})$ of the orbifold theory, in a sector $\NSRam\in\{\NS,\Ram,\tNS,\tRam\}$, decomposes as
\begin{align}
Z_\NSRam(\bar{z}) = \frac{1}{2}\left( Z^{0,0}_\NSRam(\bar{z}) + Z^{1,0}_\NSRam(\bar{z}) + Z^{0,1}_\NSRam(\bar{z}) + Z^{1,1}_\NSRam(\bar{z}) \right)~,
\end{align}
where $Z^{\ell,\ell'}_\NSRam(\bar{z})$ denotes the contribution of the $\ell'$-twisted sector with an insertion of $g^\ell$. Under modular transformations, these sectors are permuted according to
\begin{align}
Z^{\ell,\ell'}\left(\frac{a\tau+b}{c\tau+d},\frac{a\tau+b}{c\bar{\tau}+d},\frac{\bar{z}}{c\bar{\tau}+d}\right) \sim Z^{a\ell+b\ell',c\ell+d\ell'}(\tau,\bar{\tau},\bar{z})~.
\end{align}
Modular transformations also mix spin structures; for example,
\begin{align}
Z^{1,0}_\tNS(-1/{\tau},-1/\bar{\tau},\bar{z}/\bar{\tau}) 
= \ee^{-2\pi\ii\bar{z}^2/\bar{\tau}} Z^{0,1}_\Ram(\tau,\bar{\tau},\bar{z})~.
\end{align}
The orbifold partition function can be  obtained by first determining $Z^{1,0}_\NSRam$, the trace with an insertion of the orbifold generator, and then reconstructing the remaining sectors by modular covariance. 

To describe the action of the orbifold on the left-moving $\mathrm{E}_8\times\mathrm{E}_8$ degrees of freedom, it is convenient to decompose the $\mathfrak{e}_8$ characters as
\begin{align}
\chi^{\mathfrak{e}_8}_\rep{1} = \chi^{\mathfrak{a}_1}_\rep{1}\chi^{\mathfrak{e}_7}_\rep{1} + \chi^{\mathfrak{a}_1}_\rep{2}\chi^{\mathfrak{e}_7}_\rep{56}~.
\end{align}
Under the $\Z_2$ action, states in $(\rep{1},\rep{1})$ are even, whereas those in $(\rep{2},\rep{56})$ are odd. The modular orbit then yields
\begin{align}
Z_{\NS}^{1,0}(\bar{z}) &= (\chi^{\mathfrak{e}_7}_\rep{1}\chi^{\mathfrak{a}_1}_\rep{1} - \chi^{\mathfrak{e}_7}_\rep{56}\chi^{\mathfrak{a}_1}_\rep{2})^2 \chi^{\mathfrak{d}_4}_\rep{1} 4\frac{\bar{\vartheta}_4(\bar{z})^2}{\bar{\vartheta}_2(0)^2}~,
\nonumber\\
Z_{\NS}^{0,1}(\bar{z}) &= (\chi^{\mathfrak{e}_7}_\rep{1}\chi^{\mathfrak{a}_1}_\rep{2} + \chi^{\mathfrak{e}_7}_\rep{56}\chi^{\mathfrak{a}_1}_\rep{1})^2 (\chi^{\mathfrak{d}_4}_\rep{1} +\chi^{\mathfrak{d}_4}_\rep{8^v} +\chi^{\mathfrak{d}_4}_\rep{8^c} +\chi^{\mathfrak{d}_4}_\rep{8^s} )2\frac{\bar{\vartheta}_2(\bar{z})^2}{\bar{\vartheta}_4(0)^2}~,
\nonumber\\
Z_{\NS}^{1,1}(\bar{z}) &= (\chi^{\mathfrak{e}_7}_\rep{1}\chi^{\mathfrak{a}_1}_\rep{2} - \chi^{\mathfrak{e}_7}_\rep{56}\chi^{\mathfrak{a}_1}_\rep{1})^2 (\chi^{\mathfrak{d}_4}_\rep{1} -\chi^{\mathfrak{d}_4}_\rep{8^v} -\chi^{\mathfrak{d}_4}_\rep{8^c} -\chi^{\mathfrak{d}_4}_\rep{8^s} )2\frac{\bar{\vartheta}_1(\bar{z})^2}{\bar{\vartheta}_3(0)^2}~.
\end{align}
Several consistency checks can be performed. Expanding the unrefined partition function,
\begin{align}
q^{\frac{20}{24}}\bar{q}^{\frac{6}{24}} Z_\NS = 1+ 300\, q +18 \,\bar{q} +112\, q^{\frac{1}{2}}\bar{q}^{\frac{1}{2}}+\dots~,
\end{align}
one identifies the $300$ holomorphic spin-$1$ currents associated with the gauge algebra. A similar analysis in the Ramond sector yields the elliptic genus
\begin{align}
\mathbb{E} =\quad &2\left( \chi^{\mathfrak{e}_7}_\rep{1} \chi^{\mathfrak{e}_7}_\rep{1} \chi^{\mathfrak{a}_1}_\rep{1} \chi^{\mathfrak{a}_1}_\rep{1} + \chi^{\mathfrak{e}_7}_\rep{56} \chi^{\mathfrak{e}_7}_\rep{56} \chi^{\mathfrak{a}_1}_\rep{2} \chi^{\mathfrak{a}_1}_\rep{2} \right)\chi^{\mathfrak{d}_4}_\rep{1}
\nonumber\\
- &2\left( \chi^{\mathfrak{e}_7}_\rep{1} \chi^{\mathfrak{e}_7}_\rep{1} \chi^{\mathfrak{a}_1}_\rep{2} \chi^{\mathfrak{a}_1}_\rep{2} + \chi^{\mathfrak{e}_7}_\rep{56} \chi^{\mathfrak{e}_7}_\rep{56} \chi^{\mathfrak{a}_1}_\rep{1} \chi^{\mathfrak{a}_1}_\rep{1} \right)(\chi^{\mathfrak{d}_4}_\rep{8^v}+\chi^{\mathfrak{d}_4}_\rep{8^c}+\chi^{\mathfrak{d}_4}_\rep{8^s})
\nonumber\\
- &2\left( \chi^{\mathfrak{e}_7}_\rep{1} \chi^{\mathfrak{e}_7}_\rep{56} \chi^{\mathfrak{a}_1}_\rep{1} \chi^{\mathfrak{a}_1}_\rep{2} + \chi^{\mathfrak{e}_7}_\rep{56} \chi^{\mathfrak{e}_7}_\rep{1} \chi^{\mathfrak{a}_1}_\rep{2} \chi^{\mathfrak{a}_1}_\rep{1} \right)\chi^{\mathfrak{d}_4}_\rep{1}
\nonumber\\
- &2\left( \chi^{\mathfrak{e}_7}_\rep{1} \chi^{\mathfrak{e}_7}_\rep{56} \chi^{\mathfrak{a}_1}_\rep{2} \chi^{\mathfrak{a}_1}_\rep{1} + \chi^{\mathfrak{e}_7}_\rep{56} \chi^{\mathfrak{e}_7}_\rep{1} \chi^{\mathfrak{a}_1}_\rep{1} \chi^{\mathfrak{a}_1}_\rep{2} \right)\chi^{\mathfrak{d}_4}_\rep{1}~,
\end{align}
which is modular invariant, as expected.

Finally, one can directly identify the spin-$2$ simple current of the left-moving current algebra in the spectrum. This $\Z_2$ simple current lies in the representation $(\rep{2},\rep{2},\rep{1},\rep{56},\rep{56})$, which appears in the following contribution:
\begin{align}
Z_\NS(\bar{z}) \supset\quad \chi^{\mathfrak{a}_1}_\rep{2}\chi^{\mathfrak{a}_1}_\rep{2}\chi^{\mathfrak{d}_4}_\rep{1}\chi^{\mathfrak{e}_7}_\rep{56}\chi^{\mathfrak{e}_7}_\rep{56}\, \tfrac{1}{2}\left(\frac{\bar{\vartheta}_4(\bar{z})^2}{(\tfrac{1}{2}\bar{\vartheta}_2)^2}+\bar{\chi}^{\mathfrak{d}_4}_\rep{1}\frac{\bar{\vartheta}_3(\bar{z})^2}{\bar{\eta}^2}\right)~.
\end{align}
The right-moving conformal weights of the corresponding states can be extracted by expanding the partition function in powers of $\bar{q}$:
\begin{align}
\bar{q}^{\frac{6}{24}}Z_\NS \supset\quad \chi^{\mathfrak{a}_1}_\rep{2}\chi^{\mathfrak{a}_1}_\rep{2}\chi^{\mathfrak{d}_4}_\rep{1}\chi^{\mathfrak{e}_7}_\rep{56}\chi^{\mathfrak{e}_7}_\rep{56}\left(1+ 18\bar{q}+64\bar{q}^{\frac{3}{2}}+70\bar{q}^2+\dots\right)~,
\end{align}
where we have included the standard vacuum energy shift associated with the central charge $c_\tR=6$. This expansion shows that the spectrum indeed contains purely left-moving states in the required representation. This confirms that the left-moving current algebra is extended by the corresponding spin-$2$ operator, in agreement with the gauging of the $\Z_2$ one-form symmetry of $(\mathrm{SU}(2)\times\mathrm{SU}(2)\times\mathrm{E}_7\times\mathrm{E}_7)/\Z_2$. The reader can easily verify that the partition function of the orbifold theory can be expressed in terms of characters of the extended algebra, which consist of the four combinations
\begin{align}
\mathcal{X}_0 &=\chi^{\mathfrak{a}_1}_\rep{1} \chi^{\mathfrak{a}_1}_\rep{1} \chi^{\mathfrak{e}_7}_\rep{1} \chi^{\mathfrak{e}_7}_\rep{1}  + \chi^{\mathfrak{a}_1}_\rep{2} \chi^{\mathfrak{a}_1}_\rep{2} \chi^{\mathfrak{e}_7}_\rep{56} \chi^{\mathfrak{e}_7}_\rep{56} ~,&
\mathcal{X}_1 &= \chi^{\mathfrak{a}_1}_\rep{2} \chi^{\mathfrak{a}_1}_\rep{2} \chi^{\mathfrak{e}_7}_\rep{1} \chi^{\mathfrak{e}_7}_\rep{1}  + \chi^{\mathfrak{a}_1}_\rep{1} \chi^{\mathfrak{a}_1}_\rep{1} \chi^{\mathfrak{e}_7}_\rep{56} \chi^{\mathfrak{e}_7}_\rep{56} ~,
\nonumber\\
\mathcal{X}_2 &= \chi^{\mathfrak{a}_1}_\rep{1} \chi^{\mathfrak{a}_1}_\rep{2} \chi^{\mathfrak{e}_7}_\rep{1} \chi^{\mathfrak{e}_7}_\rep{56}  +\chi^{\mathfrak{a}_1}_\rep{2} \chi^{\mathfrak{a}_1}_\rep{1}  \chi^{\mathfrak{e}_7}_\rep{56} \chi^{\mathfrak{e}_7}_\rep{1} ~,&
\mathcal{X}_3 &= \chi^{\mathfrak{a}_1}_\rep{2} \chi^{\mathfrak{a}_1}_\rep{1} \chi^{\mathfrak{e}_7}_\rep{1} \chi^{\mathfrak{e}_7}_\rep{56}  + \chi^{\mathfrak{a}_1}_\rep{1} \chi^{\mathfrak{a}_1}_\rep{2} \chi^{\mathfrak{e}_7}_\rep{56} \chi^{\mathfrak{e}_7}_\rep{1} ~.
\end{align}

\bibliographystyle{JHEP}
\bibliography{refs} 

\end{document}